\pdfoutput=1
\documentclass[11pt,a4paper,twoside,openright]{report}
\usepackage[english]{ETHDAsfs}

\usepackage{pdfpages}
\usepackage{amsbsy}
\usepackage{amssymb}
\usepackage{graphicx}
\usepackage{subcaption} 
\usepackage[longnamesfirst]{natbib}
\usepackage{texab}
\usepackage{amsmath}
\usepackage{enumerate}
\usepackage{relsize}
\usepackage{color} 
\usepackage{listings}
\usepackage{xfrac} 
\usepackage[toc,page, titletoc]{appendix}
\usepackage{float} 
\usepackage{bbm} 
\usepackage{mathtools} 

\definecolor{Mygrey}{gray}{0.75}
\definecolor{Cgrey}{gray}{0.4}
\newtheorem{definition}{Definition}[subsection]

\newtheorem{proposition}[definition]{Proposition}

\theoremstyle{definition} 
\newtheorem{example}[definition]{Example}

\DeclareMathOperator{\Span}{span}
\DeclareMathOperator{\dse}{d-sep}
\DeclareMathOperator{\dco}{d-conn}
\DeclareMathOperator{\by}{by}
\DeclareMathOperator{\rank}{rank}
\DeclarePairedDelimiter{\ceil}{\lceil}{\rceil}

\newcommand{\dsep}[3]{{#1} \dse_\mathcal{G} {#2} \by {#3}}
\newcommand{\dcon}[3]{{#1} \dco_\mathcal{G} {#2} \by {#3}}

\numberwithin{equation}{subsection}

\graphicspath{{Pictures/}}

\def\pr{\mathbb{P}}
\def\ex{\mathbb{E}}
\def\PA{PA}
\def\DE{DE}
\def\ND{ND}

\def\realnumbers{\mathbb{R}}

\makeatletter
\newcommand*{\indep}{%
  \mathbin{%
    \mathpalette{\@indep}{}%
  }%
}
\newcommand*{\nindep}{%
  \mathbin{
    \mathpalette{\@indep}{\not}
  }%
}
\newcommand*{\@indep}[2]{%
  \sbox0{$#1\perp\m@th$}
  \sbox2{$#1=$}
  \sbox4{$#1\vcenter{}$}
  \rlap{\copy0}
  \dimen@=\dimexpr\ht2-\ht4-.2pt\relax
  \kern\dimen@
  {#2}%
  \kern\dimen@
  \copy0 
} 
\makeatother

\begin{document}
\bibliographystyle{chicago}

\pagenumbering{roman}

\period{Autumn 2015}
\dasatype{Semester Project}
\students{Emiliano Díaz}
\mainreaderprefix{Adviser:}
\mainreader{Prof.\ Dr.\ Marloes Maathuis}
\submissiondate{July 8 2016}
\title{Causality and Surrogate Variable Analysis}

\maketitle
\cleardoublepage

\markright{}
\vspace*{\stretch{1}}
\begin{center}
    I would like to express my gratitude to Prof. Dr. Marloes Maathuis for all her patience and guidance. 
\end{center}
\vspace*{\stretch{2}}


\newpage
\markboth{Abstract}{Abstract}
\chapter*{Abstract}

Gene expression depends on thousands of factors and we usually only have access to tens or hundreds of observations of gene expression levels meaning we are in a high-dimensional setting. Additionally we don't always observe or care about all the factors. However, many different gene expression levels depend on a set of common factors. By observing the joint variance of the gene expression levels together with the observed \emph{primary variables} (those we care about) Surrogate Variable Analysis (SVA) seeks to estimate the remaining unobserved factors. The ultimate goal is to assess whether the primary variable (or vector) has a significant effect on the different gene expression levels, but without estimating unobserved factors first the various regression models and hypothesis tests are dependent which complicates significance analysis. 

In this work we define a class of additive gene expression structural equation models (SEMs) which are convenient for modeling gene expression data and which provides a useful framework to understand the various steps of the SVA methodology. We justify the use of this class from a modeling viewpoint but also from a causality viewpoint by exploring the independence and causality properties of this class and comparing to the biologically driven data assumptions. For this we use some of the theory that has been developed elsewhere on graphical models and causality. We then give a detailed description of the SVA methodology and its implementation in the \Rp package \verb|sva| referring each step to different parts of the additive gene expression SEM defined previously. 

Given the possible dependency of the primary variable with unobserved factors, recovering these factors accurately presents several issues. SVA tries to solve some of them by obtaining unobserved factors that are \emph{allowed} to be correlated to the primary variable, however the exact form of this dependency is not \emph{modeled} and so is not necessarily accurate. We perform simulation experiments and sensibility analysis to assess the performance of SVA, comparing it to other methods designed for the same purpose, and to identify for which parameters of the additive gene expression SEM, SVA performs better. In general, we found SVA performs comparably well at estimating unobserved factors to SVDR and much better than all other methods at capturing the dependence of these factors to the primary variable. Additionally the significance analysis performed were shown to be much closer to being valid for SVA than those for other methods. However, for the base scenario simulated this did not translate into significant improvement in the accuracy of the estimation of the effect of the primary variable on gene expression levels. Using sensitivity analysis we found that, assuming that performance measures vary independently with respect to the different parameters,  for gene expression SEMS with low number of unobserved factors, high level of gene expression variables, high sparsity (low number of edges in the SEM), a lower variance for unobserved factors than for the primary variable and low complexity for the signals from the primary variable to the unobserved factors and to the gene expression level variables,  SVA has superior performance to the other methods considered, including in terms of the accuracy of the estimation of the effect of the primary variable on gene expression levels.


\newpage
\tableofcontents
\newpage
\listoffigures


\cleardoublepage
\pagenumbering{arabic}

\chapter{Introduction} \label{ch:intro}

\section{Structure}

The report is organized as follows. In Section \ref{SVA} we describe the problem \emph{Surrogate Variable Analysis} (SVA) tries to solve in its original \emph{gene expression} context. 

Chapter \ref{ch:causality} introduces the basic \emph{graphical models}, \emph{structural equation models} (SEMs) and \emph{causality} concepts and results necessary to frame the gene expression modeling problem in the context of causality. We also define a class of structural equation models, which we call \emph{additive gene expression} SEMs. This class of SEMs can be used to model gene expression and forms the basis for the SVA methodology. We conclude the section by deriving some of the properties of this class.

In Chapter \ref{ch:SVA} the SVA methodology is described in detail. Section \ref{sec:overview} gives a general overview of the methodology which consists of three main steps: estimation of the span of the $c_l$ and $h_k$ variables, respectively, from the additive gene expression SEM, and estimation of effects $f_{x_j}$ of $y$ on $x_j$. Sections \ref{sec:cl} and \ref{sec:hk} give a detailed description of the first two steps. 

In Chapter \ref{ch:sims} we evaluate the performance of the SVA methodology by applying it to simulated data. Section \ref{sec:methods} describes other methods that can be used to estimate the gene expression SEM. In Sections \ref{sec:lowdim} and \ref{sec:highdim} we we will apply these methods, together with SVA,  in order to provide benchmarks for the SVA methodology. Section \ref{sec:lowdim} includes the results of \emph{low dimensional} experiments, where the number of gene expression variables is less than 10. The results of \emph{high dimensional} experiments, where the number of gene expression variables in the order of 1000, are included in Section \ref{sec:highdim}. This Section also includes sensitivity analysis on certain simulation parameters to gauge the efectiveness of SVA in different data environments. 

In Chapter \ref{s:Summary} we summarize the key findings and list possible future lines of investigation.

\section{Surrogate Variable Analysis} \label{SVA}

Gene expression is the biological process by which the information contained in a gene is used to produce material such as proteins or RNA. The degree to which this process occurs can be measured on a continuous scale and we call this measurement the gene expression level.

It has been shown that genetic, environmental, demographic and other factors have an effect on gene expression levels. Often it is of interest to study the effect of one, or  a group, of \emph{primary} variables on the expression levels of a group of genes. Other \emph{unmodeled} variables which also have an effect on gene expression may not be studied explicitly because measurements are not available or because it is inconvenient to do so - the relationship between the unmodeled variable and gene expression maybe too complex or limited sample size may restrict the number of variables that can be used as predictors. The following model can be used to describe this situation:

\begin{align} \label{eq:geneModel}
	x_j = f_{x_j}(y) + \sum_{l=1}^L \gamma_l g_l + N_{x_j}
\end{align}	

Where, 

\begin{itemize}
	\item $j \in \{1,...,J\}$, 
	\item $x_1,...,x_J$ are the gene expression levels of the $J$ genes of interest,
	\item $y$ is the primary variable (which may be a vector) of interest, the variable whose effect on the gene expression levels $\underline{x}=(x_1,...,x_J)^T$ we want to study, 
	\item $g_l := \sum_{p=1}^P g_{lp}(w_p)$ and $w_1,...,w_P$ are the \emph{unmodeled} variables, and
	\item $N_{x_j}$ is a white noise random process. 
\end{itemize}	

The choice of an additive model is quite general as it has been shown in \cite{additive} that an appropriate choice of non-linear basis, i.e. the right choice of functions $f_{x_j},g_{11},...,g_{lp}$, can be used to accurately represent even complicated non-linear functions of the variables $y,w_1,...,w_p$. Notice that the above model is a multivariate regression model since there are $J$ response variables involved. \\

\begin{example}[Disease state and age] \label{ex1}
	As a simple illustrative example take the hypothetical human expression study proposed in \cite{SVA}. The disease state of a certain tissue is the primary variable. Additionally, changes in expression are also influenced by the age of the individuals. The expression level of some genes depends on the disease state, the expression level of other genes depends on age and for others still, expression level depends on both variables. We are only interested in the influence of disease state on expression level. The age of the individuals corresponding to the collected samples is unknown to us. The corresponding model is: 
	\begin{align}
		x_j = f_{x_j}(y) + \sum_{l=1}^L \gamma_l g_l + N_{x_j}
	\end{align}	

	Where, 

	\begin{itemize}
		\item $j \in \{1,...,J\}$, 
		\item $x_1,...,x_J$ are the gene expression levels of the $J$ genes of interest,
		\item $y \in \{0,1\}$ is the primary variable, disease state, 
		\item $g_l := g_{l}(w)$ and $w$ is the age of the individual, and
		\item $N_{x_j}$ is a white noise random process.  
	\end{itemize}
	
\end{example}

Since we can't observe $w_1,...,w_p$ normal regression techniques can't be used to estimate the model.  We also know from multiple regression that if we omit the variables $w_1,...,w_p$ from the model, i.e. if we estimate the simplified model:

\begin{align}
	x_j = f_{x_j}(y) + \epsilon_j
\end{align}

we will obtain a model with systematic error such that $\ex[\epsilon_{ij}] \neq 0$. Additionally, since we are dealing with a multivariate regression model, omitting the variables $w_1,...,w_p$ leads to correlation among residuals $\epsilon_i$ and $\epsilon_j$, for $i \neq j$ which complicates significance analysis. 

To avoid this we need not estimate $g_1,...,g_L$ individually, it suffices to estimate $\sum_{l=1}^L \gamma_l g_l$. We could, for example, estimate variables $h_1,...,h_K$ as long as there exist $\beta_1,...,\beta_K$ such that $\sum_{k=1}^K \beta_k h_k=\sum_{l=1}^L \gamma_l g_l$. This is what the \emph{Surrogate Variable Analysis} methodology (SVA) proposed in \cite{SVA} tries to do. It aims to estimate \emph{surrogate variables} $h_1,...,h_K$ which generate the same linear space of the unmodeled factors  $g_1,...,g_L$ so that we can then accurately estimate $f_{x_1},...,f_{x_J}$ and produce valid significance analysis. 

The variables $h_1,...,h_K$ must include signal from other sources than the primary variable $y$, i.e. from certain unobserved $w_1,...,w_p$. However, the linear space generated by $h_1,...,h_K$ need not be orthogonal to $f_{x_j}(y)$ so the estimation of the surrogate variables must allow for potential \emph{overlap} in signal with the primary variable. The model to estimate is then:

\begin{align} \label{eq:model}
	x_j = f_{x_j}(y) + \sum_{k=1}^K \beta_k h_k + N_{x_j}
\end{align}	

Where, 

\begin{itemize}
	\item $j \in \{1,...,J\}$, 
	\item $x_1,...,x_J$ are the gene expression levels of the $J$ genes of interest,
	\item $y$ is the primary variable (which may be a vector) of interest, the variable whose effect on the gene expression levels $\underline{x}=(x_1,...,x_J)^T$ we want to study, 
	\item $N_{x_j}$ is a white noise random process, 
	\item $h_k$  are surrogate variables such that $\sum_{k=1}^K \beta_k h_k=\sum_{l=1}^L \gamma_l g_l$, to be estimated using SVA, 
	\item $h_k$ can't be modelled as $h_k = f(y) + N$, where $N$ is a white noise process, i.e. it must include signal from \emph{unmodeled} variables, and
	\item typically $Cor(f_{x_j}(y),h_k)\neq 0$ for some $k$.
\end{itemize}	

There are an estimated 20,000-25,000 human protein-coding genes while typically, the cost of measuring gene expression levels means we only have available in the order of tens or hundreds of samples with the gene expression level and primary variable measurements. This means we are in a high-dimensional setting where $J >> n$, where $n$ is the sample size. Any estimation technique designed to estimate model \ref{eq:model} must take this into account, as does SVA. 

In SVA as in techniques such as \emph{Canonical Correlation Analysis} (CCA) and \emph{Reduced Rank Regression} (RRR) response variables (in this case gene expression levels) are considered simultaneously. Where as in CCA and RRR the characterization of the space spanned by the response variables is used to estimate model parameters, in the case of SVA it is used to recover the effects of unmodeled variables so as to produce an analysis that essentially includes all relevant variables.

\chapter{Causality} \label{ch:causality}

In \cite{SVA} the relationship between, on the one hand, the primary variable $y$ and unmodeled variables $w_1,...,w_p$ and, on the other, the gene expression levels $x_1,...,x_J$ is not explicitly described as \emph{causal} in nature. Nevertheless, emphasis is made on measuring the \emph{effect} of the first group of variables on the second, suggesting an underlying causal structure. Additionally, the SVA methodology can be succintly described and justified within the framework of \emph{Causality} if we assume these relationships to be \emph{causal}. We give some definitions and results necessary to frame the problem in this setting. The rest of this section follows \cite{jonas} closely.

\section{DAGs and SEMs} \label{DAGsSEMs}

Causal relations are defined in the context of structural equation models (SEMs). Every SEM $\mathcal{M}$ has an associated graph $\mathcal{G}$ that summarizes the functional dependencies of the SEM. As we will see each SEM induces a unique joint probability distribution $\pr^\mathbb{X}$ for the variables in the SEM. We will be interested in SEMs whose graph is a directed acyclic graph (DAG). In this case the graph also represents, through the d-separation relation, the independence relations of $\pr^\mathbb{X}$ and can be used to refute claims about causal relationships between variables. 

\hspace{1mm}

\begin{definition}[Directed acyclic graph (DAG)]
	A directed acyclic graph (DAG) is a tuple $\mathcal{G}:=(\mathcal{V}, \mathcal{E})$ where $\mathcal{V}$ is a finite set of nodes and the set of edges $\mathcal{E} \subseteq \mathcal{V} \times \mathcal{V}$ is such that if $v,w \in \mathcal{V}$ and $v \neq w$ then $(v,v) \notin \mathcal{E}$ and if $(v,w) \in \mathcal{E}$ then $(w,v) \notin \mathcal{E}$. 
\end{definition}

\hspace{1mm}

\begin{definition}[D-separation]
	 Let $\mathcal{G}=(\mathcal{V}, \mathcal{E})$ be a DAG, with $\mathcal{V}=\{x_1,...,x_p\}$. Then:
	 \begin{enumerate}[i.]
		 \item \textbf{Notation.} We may refer to node $x_{i_k}$ or a sequence of nodes $x_{i_1},...,x_{i_n} \in \mathcal{V}$ by their indices, e.g. node $i_k$ or nodes $i_1,...,i_n$. If there is an edge between node $i$ and $j$, i.e. if $(i,j) \in \mathcal{E}$, we write $i \rightarrow j$ (or $j \leftarrow i$). If $\mathcal{V}$ includes nodes labeled with different letters we don't use the indices to refer to them as, for example, when $\mathcal{V}=\{y,h_1,...,h_K,x_1,...,x_J\}$
		 \item \textbf{Parent.} A node $i$ is a parent of $j$ in DAG $\mathcal{G}$ if $i \rightarrow j$. The set of parents of node $j$ is denoted $\PA_j^\mathcal{G}$.
		 \item \textbf{Path.} There is a path between nodes $i_1$ and $i_n$ if there exists a sequence of nodes  $i_1,i_2,...,i_n \in \mathcal{V}$ such that either $i_k \rightarrow i_{k+1}$ or $i_{k+1} \rightarrow i_{k}$ for $k=1,...,n-1$.
		 \item \textbf{Directed path.} There is a directed path between nodes $i_1$ and $i_n$ if there exists a sequence of nodes $i_1,i_2,...,i_n \in \mathcal{V}$ such that  $i_k \rightarrow i_{k+1}$ for $k=1,...,n-1$.
		 \item \textbf{Descendent.} We say node $i_k$ is a descendant of node $i_1$ if there is a directed path from $i_1$ to $i_k$. The sets of descendent and non-descendant nodes to node $i$ in DAG $\mathcal{G}$ are denoted $\DE_i^\mathcal{G}$ and $\ND_i^\mathcal{G}$ respectively. 
		 \item \textbf{Collider.} A node $i_k$ is a collider in a path $i_1,...,i_n$ if $i_{k-1} \rightarrow i_{k} \leftarrow i_{k+1}$. The structure $i_{k-1} \rightarrow i_k \leftarrow i_{k+1}$ is referred to as a v-structure.
		 \item \textbf{Blocking set.} The set $S$ blocks the path from $i_1$ to $i_n$ if either: 
		 	\begin{itemize}
				\item $i_k \in S$ and $i_{k-1} \rightarrow i_{k} \rightarrow i_{k+1}$, $i_{k-1} \leftarrow i_{k} \leftarrow i_{k+1}$ or $i_{k-1} \leftarrow i_{k} \rightarrow i_{k+1}$, or
				\item $i_k \notin S$, $\DE_{i_k}^\mathcal{G} \cap S = \emptyset$ and $i_{k-1} \rightarrow i_k \leftarrow i_{k+1}$ (i.e. $i_k$ is a collider). 
			\end{itemize}
		 \item \textbf{D-separation.} Let $A,B,S \subseteq \mathcal{V}$ disjoint. $A$ is d-separated from $B$ by $S$, if $S$ blocks all paths between nodes in $A$ and nodes in $B$. We denote this as $\dsep{A}{B}{S}$.
		 \item \textbf{D-connection.} We say that $A$ is d-connected from $B$ by $S$, if $A$ is not d-separated from $B$ by $S$. We denote this as $\dcon{A}{B}{S}$.
	 \end{enumerate}	 
\end{definition}

\hspace{1mm}

\begin{definition}[Structural equation model (SEM)] \label{SEM}
	 A structural equation model (SEM) for variables $\mathbb{X}=\{x_1,...,x_p\}$ is defined as a tuple $\mathcal{M}:=(\mathcal{S},\pr^{\mathbb{N}})$ where $\mathcal{S}=\{S_1,...,S_p\}$, a collection of $p$ equations,  and  $\pr^{\mathbb{N}}$, a joint probability distribution, are such that:
	 \begin{align}
		 S_j: x_j &= f_j(x_{PA_j},N_j)\\
		 \pr^{\mathbb{N}} &= \pr^{N_1,...,N_p} = \pr^{N_1}...\pr^{N_p}
	 \end{align}	 
	 Where, 
	 \begin{itemize}
		 \item $j \in \{1,...,p\}$,
		 \item $x_{\PA_j}$ is the set of variables upon which $x_j$ functionally depends, where $\PA_j \subseteq \{1,...,p\}$ denotes the indices of those variables; we require the functions $f$ to \emph{really} depend on the set of parent variables $x_{\PA_j}$, i.e. for all $x_i \in x_{\PA_j}$ and for all values of $x_{PA_j \setminus i}$ and $N_j$ there must exist values $x_i^1$ and $x_i^2$ such that $f_j(x_i^1,x_{PA_j \setminus i},N_j) \neq f_j(x_i^2, x_{PA_j \setminus i},N_j)$. If this is not the case we may simply redefine $x_{PA_j} \leftarrow x_{PA_j \setminus i}$,
		 \item $N_j$ are random noise variables, and
		 \item $\pr^{\mathbb{N}}$ is the joint distribution of noise variables, which we require to be jointly independent.
	 \end{itemize}	 
\end{definition}

\hspace{1mm}

\begin{definition}[Graph of a SEM]
	 The graph of a SEM for variables $\mathbb{X}=\{x_1,...,x_p\}$ is $\mathcal{G}=(\mathcal{V}, \mathcal{E})$ where $\mathcal{V}=\{1,...,p\}$ and $i \rightarrow j$ if and only if $i \in PA_j$. If the resulting graph $\mathcal{G}$ is a DAG then $PA_j^\mathcal{G}=PA_j$ $ \forall j$. From now on we only consider SEMs that induce a DAG graph.  
\end{definition}

\hspace{1mm}

\begin{example}[A gene expression SEM] \label{exGESEM}
As a an example of a SEM consider $\mathcal{M}=(\mathcal{S},\pr^\mathbb{N})$ where $\mathbb{X}=\{y,h_1,...,h_K,x_1,...,x_J\}$ and the joint noise distribution is a multivariate normal composed of independent standard normal variables $\mathbb{N} \sim \mathcal{N}_p(0,I_p)$ with $p=K+J+1$ , $j \in \{1,...,J\}$, $k \in \{1,...,K\}$ and $\mathcal{S}$ is such that:
	\begin{align}
		S_{y} &: y=N_y\\
		S_{h_k} &: h_k=f_{h_k}(y,N_{h_k})\\
		S_{x_j} &: x_j=f_{x_j}(y,h_1,...,h_k,N_{x_j})
	\end{align}	

If $J=3$ and $K=2$ the DAG $\mathcal{G}$ corresponding to this SEM is: 

	\begin{figure}[H]
	  \centering
	  \includegraphics[width=.5\textwidth]{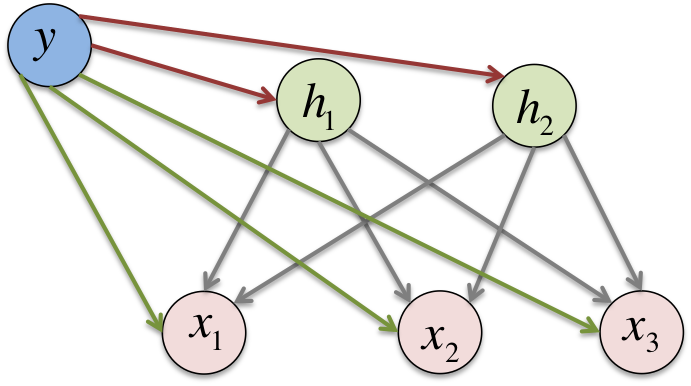} 
	  \caption[Gene expression SEM]
	  {Gene expression SEM}
	  \label{fig:exGESEM}
	\end{figure}

\end{example}

\hspace{1mm}

\begin{proposition}[Distribution of a SEM]
	 A SEM for variables $\mathbb{X}=\{x_1,...,x_p\}$, with an acyclic structure (i.e. a DAG graph), defines a unique distribution $\pr^{\mathbb{X}}$ over these variables. 
	 \begin{proof}
	   See Proposition 2.1.2 in \cite{jonas}.
	 \end{proof}
\end{proposition}

\hspace{1mm}
 This means that by specifying the SEM we implicitly specify the probability distribution $\pr^{\mathbb{X}}$. However, as we will see the SEM also includes additional information that allows us to determine which dependence relations between the variables in $\mathbb{X}$ are causal (and incidentally, other information that allows us to make \emph{counterfactual} statements).\\ \\
 
We have seen that a SEM $\mathcal{M}$ induces a unique graph $\mathcal{G}$ and joint probability distribution $\pr^\mathbb{X}$. We now study the relationship between the DAG of a SEM and its joint probability distribution to see to what degree the DAG structure accurately describes the independence relations encoded in the joint probability distribution.

\hspace{1mm}

\begin{definition}[Markov properties] \label{Markov}
	We say that the joint distribution $\pr^{\mathbb{X}}$ (which we require to have a density $p$ with respect to some product measure) has the following Markov properties with respect to the  DAG $\mathcal{G}$ over variables $\mathbb{X}$ if it satisfies the corresponding conditions:
	\begin{enumerate}[i.]
		\item the global Markov property if $\dsep{A}{B}{C} \Rightarrow A \indep B | C$ where $A,B,C \in \mathcal{V}$ are disjoint. 
		\item the local Markov property if $x_j \indep ND_j^\mathcal{G}|PA_j^\mathcal{G}$ $\forall j \in \mathcal{V}$.
		\item the Markov factorization property if $p(\mathbb{X})=\prod_{j=1}^p p(x_j|x_{PA_j^\mathcal{G}})$ in which case we say that $\pr^{\mathbb{X}}$ factorizes over $\mathcal{G}$.
	\end{enumerate}
	 
\end{definition}

\hspace{1mm}

Notice that the global Markov property implies that:

\begin{align}
	\dcon{A}{B}{C} \Leftarrow A \nindep B | C
\end{align}

\begin{proposition}[Equivalence of Markov properties]
	 If $\pr^\mathbb{X}$ has a density $p$ (with respect to a product measure), then all Markov properties of \ref{Markov} are equivalent. 
	 \begin{proof}
		 See Theorem 3.27 in \cite{lauritzen}.
	\end{proof} 
\end{proposition}

\hspace{1mm}

Let $\mathcal{I}(\mathcal{G})$ denote the set of all independencies implied by the d-separations in $\mathcal{G}$  when we equate $\dsep{A}{B}{C}$ to $A \indep B| C$. Then $\mathcal{G}$ is an I-map for a set of independencies $\mathcal{I}$ if: 

\begin{align}
	\mathcal{I}(\mathcal{G}) \subseteq \mathcal{I}
\end{align}	

A DAG $\mathcal{G}$ is said to be an I-map for a joint probability distribution $\pr^\mathbb{X}$ if and only if $\pr^\mathbb{X}$ is Markov with respect to $\mathcal{G}$. This is represented as:

\begin{align}
	\mathcal{I}(\mathcal{G}) \subseteq \mathcal{I}(\pr^\mathbb{X})
\end{align}	

Where:

\begin{itemize}
	\item $\mathcal{I}(\pr^\mathbb{X})$ denotes all independencies in $\pr^\mathbb{X}$.
\end{itemize}	

So if $\pr^\mathbb{X}$ is Markov with respect to a DAG $\mathcal{G}$  it means d-separation on $\mathcal{G}$ is a \emph{sound} procedure for obtaining the independencies in $\pr^\mathbb{X}$: we can \emph{read-off} independencies from the graph. \\

\begin{proposition}[The joint distribution of a SEM is Markov] \label{SEMarkov}
	If $\pr^{\mathbb{X}}$ and $\mathcal{G}$ are induced by a SEM $\mathcal{M}$ then $\pr^{\mathbb{X}}$ is Markov with respect to $\mathcal{G}$. 
	\begin{proof}
		See Theorem 1.4.1 in \cite{pearl}.
	\end{proof}	
\end{proposition}

This means that the DAG $\mathcal{G}$ of any SEM $\mathcal{M}$ is an I-map for its joint distribution $\pr^{\mathbb{X}}$. Using that figure \ref{fig:exGESEM} is an I-map for the SEM of example \ref{exGESEM} we can now deduce, for example, that $y \indep h_1 |x_1,x_2,x_3$ and $x_2 \indep x_3 | h_1, h_2, y$.  The \emph{faithfulness} property of a DAG $\mathcal{G}$ will allow us to check if d-separation is a \emph{complete} procedure for finding the indpendencies of $\pr^{\mathbb{X}}$.\\

\begin{definition}[Faithfulness]
	We say that the the joint probability distribution $\pr^{\mathbb{X}}$ is faithful with respect to DAG $\mathcal{G}$ over variables $\mathbb{X}$ if it satisfies the following condition:
	\begin{align}
		\dsep{A}{B}{C} \Leftarrow A \indep B | C
	\end{align}
\end{definition}

Notice that the faithfulness property implies that:

\begin{align}
	\dcon{A}{B}{C} \Rightarrow A \nindep B | C
\end{align}	

If a joint distribution $\pr^{\mathbb{X}}$  is faithful with respect to DAG $\mathcal{G}$  we write:

\begin{align}
	\mathcal{I}(\mathcal{G}) \supseteq \mathcal{I}(\pr^\mathbb{X})
\end{align}	

This means that if a joint distribution $\pr^\mathbb{X}$  is faithful with respect to DAG $\mathcal{G}$ we are able to \emph{read-off} dependencies from the graph. If a joint distribution $\pr^\mathbb{X}$ is Markov and faithful with respect to a DAG $\mathcal{G}$ we say $\mathcal{G}$ is a perfect I-map for $\pr^\mathbb{X}$ and write: 

\begin{align}
	\mathcal{I}(\mathcal{G}) = \mathcal{I}(\pr^\mathbb{X})
\end{align}	

In this case d-separation on $\mathcal{G}$ is a sound and complete procedure for obtaining the independencies in $\pr^{\mathbb{X}}$ and we can read-off dependencies and independencies from the graph. Unfortunately, neither the fact that a joint distribution $\pr^{\mathbb{X}}$ factorizes over $\mathcal{G}$ nor the fact that a DAG $\mathcal{G}$ and a joint distribution $\pr^{\mathbb{X}}$ are induced by the same SEM $\mathcal{M}$ guarantees that $\pr^{\mathbb{X}}$ is faithful with respect $\mathcal{G}$. We can verify this using a simplified version of example \ref{exGESEM}:\\

\begin{example}[A simple SEM] \label{exSEMsmpl}
Consider the  SEMs $\mathcal{M}=(\mathcal{S},\pr^\mathbb{N})$ where $\mathbb{X}=\{y,h,x\}$, 
the joint noise distribution is a multivariate normal composed of independent standard normal variables $\mathbb{N} \sim \mathcal{N}_3(0,I_3)$ and $\mathcal{S}$ is such that :
	\begin{align} \label{smplSEM}
		S_{y} &: y=N_y\\
		S_{h} &: h=f_{h}(y,N_{h})=ay+N_h\\
		S_{x} &: x=f_{x}(y,h,N_{x})=by+ch+N_x
	\end{align}	

The DAG $\mathcal{G}$ corresponding to this SEM is: 

	\begin{figure}[H]
	  \centering
	  \includegraphics[width=.3\textwidth]{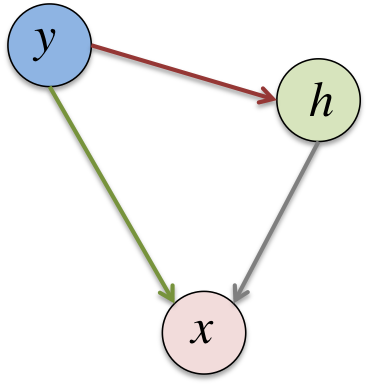} 
	  \caption[A simple SEM]
	  {A simple  SEM}
	  \label{fig:simpleSEM}
	\end{figure}

First notice that $\mathbb{X} \sim \mathcal{N}_3(0,\Sigma)$ where the covariance matrix $\Sigma$ depends on $a$, $b$ and $c$. 	
We may rexpress the equation $S_{x}$ as:

\begin{align}
	x=by+ch+N_x = bN_y + c(aN_y+N_h) = (b+ca)N_y + cN_h
\end{align}	

From which we can calculate the covariance between $y$ and $x$:

\begin{align}
	Cov(y,x)= Cov(N_y,(b+ca)N_y+cN_h) = (b+ca)Var(N_y)=b+ca
\end{align}	

Where we have used the bi-linear properties of the covariance and the fact that the noise variables $N_y$ and $N_h$ are independent. Now if $a$, $b$ and $c$ are such that $b+ca=0$, $y$ and $x$ are not correlated and since $\mathbb{X}$ has a multivariate normal distribution this means that $y$ is independent of $x$. This in turn implies that if $a$, $b$ and $c$ are such that $b+ca=0$ then $\mathcal{G}$ is not faithful to $\pr^{\mathbb{X}}$. 

\end{example}

\hspace{1mm}

The above SEM can actually be considered to be a \emph{class} of SEMs parametrized by $a$, $b$ and $c$. Although for many classes of SEMs, such as the one above, faithfulness cannot be guaranteed, a somewhat weaker condition called \emph{causal minimality} can sometimes be guaranteed. \\

\begin{definition}[Causal minimality]
 	We say that a joint probability distribution $\pr^{\mathbb{X}}$ satisfies causal miniimality with respect to the DAG $\mathcal{G}$ over variables $\mathbb{X}$ if it is Markov with respect to $\mathcal{G}$, but not with respect to $\mathcal{G}'$, where $\mathcal{G}'$ is a proper subgraph of $\mathcal{G}$. 
\end{definition}

\hspace{1mm}
We say that a DAG $\mathcal{G}$ is a minimal I-map of $\pr^\mathbb{X}$ if the following two conditions hold:

\begin{align}
	\mathcal{I}(\mathcal{G}) &\subseteq \mathcal{I}(\pr^\mathbb{X})\\
	\mathcal{I}(\mathcal{G}') &\not\subseteq \mathcal{I}(\pr^\mathbb{X})
\end{align}	

where $\mathcal{G}'$ is a proper subgraph of $\mathcal{G}$. The idea is that, if $\mathcal{G}$ satisfies causal minimality, while a d-connection in the graph does not imply the corresponding dependency, if we take off any of the edges of DAG $\mathcal{G}$, to obtain the proper subgraph $\mathcal{G}'$, then we create new independencies which are not in $\pr^\mathbb{X}$ i.e. all the edges in $\mathcal{G}$ contribute to the Markov property of $\pr^\mathbb{X}$ with respect to $\mathcal{G}$. We now present a result that will help to discern if the joint distribution of a given SEM satisfies causal minimality with respect to its DAG. \\

\begin{proposition}[Causal minimality condition for SEMs] \label{minimCond}
	 Let the joint distribution $\pr^\mathbb{X}$  be Markov with respect to a DAG $\mathcal{G}$ over variables $\mathbb{X}=\{y,x_1,...,x_p\}$ as is the case with the distribution and graph of a SEM. Assume the joint distribution $\pr^{\mathbb{X}}$ has a density with respect to some product measure. Then:

		 $\pr^\mathbb{X}$ satisfies causal minimality with respect to $\mathcal{G}$  $\iff$ $\forall x_j$ and $\forall y \in PA_j^\mathcal{G}$  $x_j \nindep y |PA_j^\mathcal{G} \setminus \{y\}$ 
	\begin{proof}
		See Appendix A.2.5 in \cite{jonas}.
	\end{proof}	
	  
\end{proposition}

\hspace{1mm}

We now define general a class of SEMs which we assume can be used to accurately model the gene expression \emph{causal} process. We don't assume this class of SEMs represents the actual data generating process as we know this is not likely to be additive, for example. \\

\begin{definition}[Class of gene expression SEMs] \label{GESEM}
	 We define the class of gene expression SEMs as the collection of SEMs for variables $\mathbb{X} \cup A =\{y,h_1,...,h_K,x_1,...,x_J\} \cup A$ such that:
	 \begin{enumerate}
		 \item $S_{x_j}: x_j = f_{x_j}(y) + \sum_{k=1}^K \beta_{kj}h_k + N_{x_j}$ $\forall j \in \{1,...,J\}$
		 \item $\pr^\mathbb{X}$ is such that $y \nindep h_k$ $\forall k \in \{1,...,K\}$
		 \item $A$ is the set of variables that are not parents of $x_1,...,x_J$ but form part of the SEM. 
	\end{enumerate}	 
\end{definition}

\hspace{1mm}

The variables in $A$ represent the part of the SEM which we are not so interested in but which we might need to estimate to understand the rest of the SEM. Specifically we are interested in estimating the functions $f_{x_j}$, which represent the the \emph{direct} effect that $y$ has on $x_j$. In this work we have assumed that the gene expression level $x_j$ does not depend on any other expression level $x_k$ with $k \in \{1,...,J\} \setminus j$, however this is not implicit in the SVA methodology. It could be that a more appropriate SEM model for the $x_j$ variables has the  equations:

\begin{align}
	S_{x_j}: x_j = f_{j}(x_{\setminus j}) + f_{x_j}(y) + \sum_{k=1}^K \beta_{kj}h_k + N_{x_j} 
\end{align}	 

$\forall j \in \{1,...,J\}$ where $x_{\setminus j}$ is a vector which includes all $x_k$ variables such that $k \in \{1,...,J\} \setminus j$. In this case the SVA methodology still works but does not estimate the \emph{direct} effect of $y$ on $x_j$ rather the \emph{filtered} effect of $y$ on $x_j$: the effect of $y$ that \emph{passes through} the unobserved vatiables $h_k$ has been filtered out.   

 Notice that since the joint probability distributions $\pr^\mathbb{X}$ of all SEMs are Markov with respect their DAGs $\mathcal{G}$,  if $\dcon{A}{B}{C}$ then  $A \nindep B | C$: we can satisfy condition 2 above by assuring that $\dcon{y}{h_k}{\emptyset}$. We give some examples of DAGs that correspond to SEMs belonging to class \ref{GESEM}. \\

 \begin{figure}[H]
  \begin{subfigure}{.4\textwidth}
    \centering
	\includegraphics[width=1\textwidth]{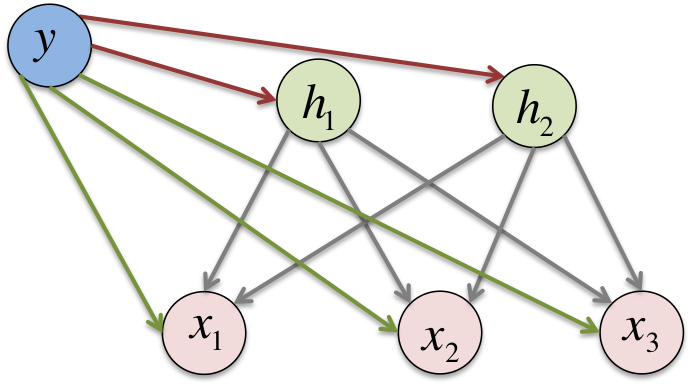} 
    \caption{}
    \label{fig:GESEM1}
  \end{subfigure}%
  \begin{subfigure}{.4\textwidth}
    \centering
    \includegraphics[width=1\textwidth]{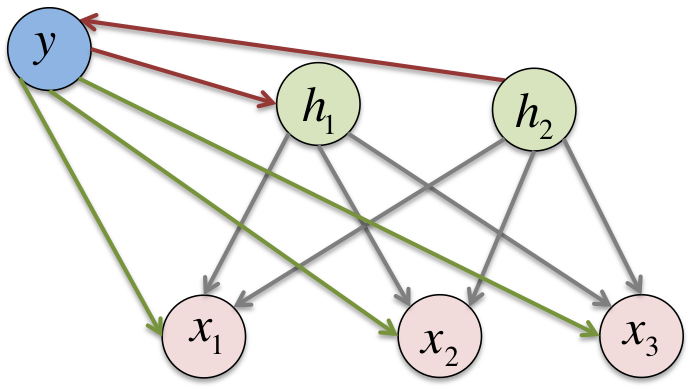}
    \caption{}
    \label{fig:GESEM2}
  \end{subfigure}
  \begin{subfigure}{.4\textwidth}
    \centering
    \includegraphics[width=1\textwidth]{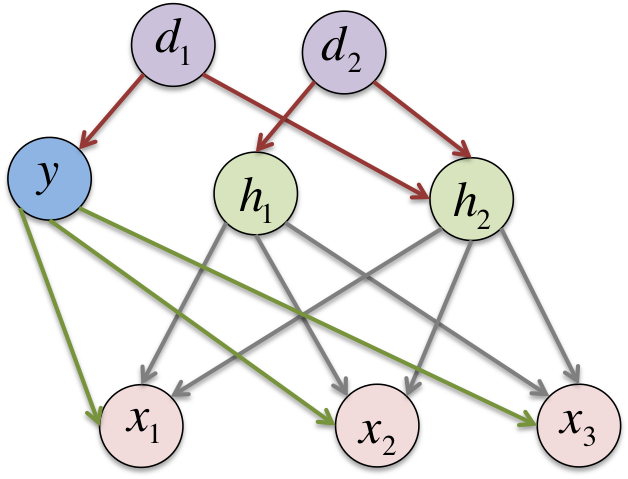}
    \caption{}
    \label{fig:GESEM3}
  \end{subfigure}
  \begin{subfigure}{.4\textwidth}
    \centering
    \includegraphics[width=1\textwidth]{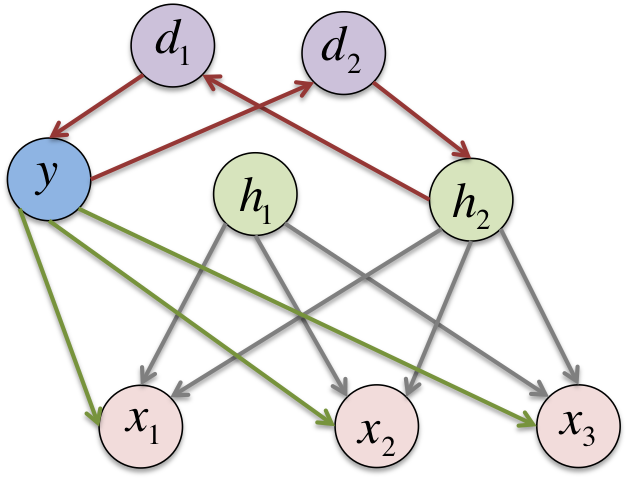}
    \caption{}
    \label{fig:GESEM4}
  \end{subfigure}
  \caption[DAGs $\mathcal{G}$ in gene expression class]
  {DAGs $\mathcal{G}$ in gene expression class}
  \label{fig:GESEM}
\end{figure}

\hspace{1mm}

Notice that in all the examples it holds that $\dcon{y}{h_k}{\emptyset}$, so for a SEM with any of these DAGs, $\mathcal{G}$ is an I-map of $\pr^\mathbb{X}$. The above class of SEMs is very large: the number of nodes is not parametrized and the functional form of the equations $S_{h_k}$ is not determined. This leads to an identifiability problem which we discuss here. 

Before we can estimate the parameters or functions of a SEM we must specify its structure, which means specifying the arguments of each equation in $\mathcal{S}$ or equivalently estimating the induced graph $\mathcal{G}$.

 One way to approach the problem is not to assume any specific form of SEM, first learn the graph structure and then model each equation using observations from $\pr^\mathbb{X}$. The way we can learn the graph structure is by performing some form of independence test. Assume we have an infinite number of observations from $\pr^\mathbb{X}$, then we can estimate the right set of independencies $\mathcal{I}(\pr^\mathbb{X})$ every time, which corresponds to having an independence oracle. Even in this situation, identifying most graph structures is impossible because different graphs can represent the same set of independencies: there exist $\pr^\mathbb{X}$ and $\mathcal{G}_1 \neq \mathcal{G}_2$ such that $\mathcal{I}(\mathcal{G}_1) \subseteq \mathcal{I}(\pr^\mathbb{X})$ and $\mathcal{I}(\mathcal{G}_2) \subseteq \mathcal{I}(\pr^\mathbb{X})$.\\
 
 \begin{example}[Non-identifiability] \label{nonIdent}
 As an example consider the variables $x$, $y$ and $z$ and suppose we want to find the DAG that is an I-map for $\mathcal{I}=\{x \indep y | z\}$. Even for such a simple example there are 3 graphs that are I-maps for $\mathcal{I}$:
\begin{figure}[H]
  \centering
  \includegraphics[width=.3\textwidth]{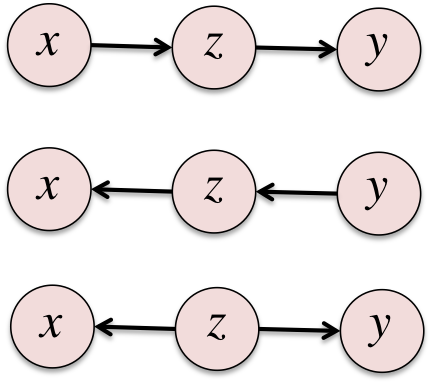} 
  \caption[Non-identifiability]
  {Non-identifiability}
  \label{fig:nonIdent}
\end{figure}
\end{example}
 
 Something similar occurs in the example of figure \ref{fig:GESEM} where the four DAGs are I-maps for sets of dependencies all of which include $y \nindep h_k$. This means that a SEM with any of the four DAGs can induce a $\pr^\mathbb{X}$ where $y \nindep h_k$.\\
 
 \begin{definition}[Markov equivalence class]
		 The Markov equivalence class of DAGs, $\mathcal{K}$, for a set of independencies $\mathcal{I}$ is the set of DAGs $\mathcal{G}$ over variables $\mathbb{X}$, such that $\mathcal{I}(\mathcal{G})=\mathcal{I}$. 
 \end{definition}	 
 
 The Markov equivalence class of a DAGs, $\mathcal{K}$, can be represented using a completed directed acyclic graph (CPDAG) where:
 
 \begin{enumerate}[1.]
	 \item The set of nodes of the graph is $\mathbb{X}$,
	 \item Directed edges correspond to nodes $i,j$ which in all DAGs $\mathcal{G} \in \mathcal{K}$ are joined either with a $i \rightarrow j$  edge or a $i \leftarrow j$ edge, and
	 \item Undirected edges correspond to nodes $i,j$ which in some DAGs $\mathcal{G} \in \mathcal{K}$ are joined with a $i \rightarrow j$ edge and in others with a$i \leftarrow j$ edge.
 \end{enumerate}	 
 
 \hspace{1mm}
	 
 \begin{example}[CPDAG of a Markov equivalence class] \label{PDAG}
	 The CPDAG representing the Markov equivalence class which contains the three DAGs shown in figure \label{fig:nonIdent} is:
	 \begin{figure}[H]
	   \centering
	   \includegraphics[width=.3\textwidth]{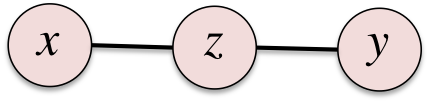} 
	   \caption[CPDAG of a Markov equivalence class]
	   {CPDAG of a Markov equivalence class}
	   \label{fig:CPDAG}
	 \end{figure}
\end{example} 

We can obtain all the DAGs $\mathcal{G}$ in Markov equivalence class $\mathcal{K}$ by directing undirected edges in all possible ways that don't create cycles in the graph. 
  
  \section{Causality} \label{sec:causality}
  
 Using the observational distribution $\pr^\mathbb{X}$ of an unknown SEM $\mathcal{M}$ we can only identify the correct DAG $\mathcal{G}$ up to its Markov equivalence class. This is why we deal with DAGs in the context of SEMs, instead of as models in themselves, because they allow us to define \emph{interventional distributions}. Through intervention distributions we are able to define causal relationships. This gives us another tool in establishing the directionality of edges and thus the possibility of identifying the correct DAG structure. We don't always have access to observations from the interventional distribution-  in fact SVA works on the assumption that we don't - however sometimes we can express the interventional distribution $\pr^{\mathbb{X}}_{\widetilde{\mathcal{M}}}$ in terms of the observational distribution $\pr^\mathbb{X}$ (for example through \emph{instrumental variables} or by \emph{adjusting}) meaning we can establish causal relationships. 
 
 A second approach to establishing the causal structure is to assume that the SEM belongs to a more restricted class. Depending on the class of SEM assumed there are results available establishing whether the identifiability of the DAG is possible or not, and techniques for actually finding it (see Chapter 4 in \cite{jonas} for an overview). The approach of SVA is more akin to this second approach. In fact, the assumptions that will be made about the underlying SEM in definition \ref{adGESEM} almost completely determine the DAG structure. 

We now study intervention distributions and causality so that we may give a causal interpretation of SVA. We are also interested in defining causal relationships so as to verify that the proposed SEM classes satisfy our assumption about the causal nature of the relationship between primary and unmodeled variables and the gene expression level variables. 

\hspace{1mm}

\begin{definition}[Intervention distribution]
	 Consider a distribution $\pr^{\mathbb{X}}$ induced by a SEM $\mathcal{M}=(\mathcal{S},\pr^{\mathbb{N}})$. If we replace one (or more) of the equations $S_i \in \mathcal{S}$ we obtain a new SEM $\widetilde{\mathcal{M}}$, a new joint distribution $\pr^{\mathbb{X}}_{\widetilde{\mathcal{M}}}$, and new marginals $\pr^{x_i}_{\widetilde{\mathcal{M}}}$  which we call intervention distributions and denote:
	 
	 \begin{align}
		 \pr^{\mathbb{X}}_{\widetilde{\mathcal{M}}} = \pr^{\mathbb{X}|do(x_j=\tilde{f}(\widetilde{PA}_j,\widetilde{N}_j))}_{\mathcal{M}}
	 \end{align}	 
	 Where: 
	 \begin{itemize}
		 \item Some noise variables have been replaced but $(\widetilde{N}_1,...,\widetilde{N}_p)$ is still mutually independent,
		 \item $\widetilde{PA}_j$ can be any new set of nodes in $\mathcal{V}$ as long as there are no cycles in $\widetilde{\mathcal{G}}$ but usually $\widetilde{PA}_j=\{\}$ ($x_j$ is deterministic or just noise) or $\widetilde{PA}_j=PA_j$ (the DAG $\mathcal{G}$ stays the same but we change the noise or the function $f$)
		 \item When $\tilde{f}(\widetilde{PA}_j, \widetilde{N}_j)$ puts a point mass on a real value $b$ we simply write $\pr^{\mathbb{X}}_{\widetilde{\mathcal{M}}} = \pr^{\mathbb{X}|do(x_j=b)}_{\mathcal{M}}$
	 \end{itemize}	 
\end{definition}

\hspace{1mm}

If we already know the SEM $\mathcal{M}$ and the induced distribution $\pr^\mathbb{X}$ we can obtain the interventional distribution $\pr^{\mathbb{X}}_{\widetilde{\mathcal{M}}}$ from the above definition. If we don't know the SEM, are learning the DAG structure, and have access to data generating mechanism, we can perform interventions.  This allows us to obtain observations from the corresponding interventional distributions and obtain the independence relations therein by performing independence tests. If we know the set of independencies in the interventional distribution,  $\mathcal{I}(\pr^{\mathbb{X}}_{\widetilde{\mathcal{M}}})$,  we can establishing causal relations between variables through definition \ref{causalEffec} and proposition \ref{causalEquivalence}. \\

\begin{definition}[Total causal effect] \label{causalEffec}
	 Given a SEM $\mathcal{M}$ there is a (total) causal effect from $x$ to $y$ if and only if $x \nindep y$  in $\pr^{\mathbb{X}|do(x=\widetilde{N}_x)}_{\mathcal{M}}$ for some variable $\widetilde{N}_x$.
\end{definition}

\hspace{1mm}

\begin{proposition}[Equivalent causality queries] \label{causalEquivalence}
	 Given a SEM $\mathcal{M}$, the following are equivalent:
	 \begin{enumerate}
		 \item There is a causal effect from $x$ to $y$.
		 \item $\exists x_1,x_2:x_1 \neq x_2$ such that $\pr^{y|do(x=x_1)}_{\mathcal{M}} \neq \pr^{y|do(x=x_2)}_{\mathcal{M}}$.
		 \item $\exists x_1$ such that $\pr^{y|do(x=x_1)}_{\mathcal{M}} \neq \pr^{y}_{\mathcal{M}}$.
		 \item  $x \nindep y$  in $\pr^{\mathbb{X}|do(x=\widetilde{N}_x)}_{\mathcal{M}}$ for all $\widetilde{N}_x$ such that $\widetilde{p}(x)>0$  $\forall x$. i.e. the pdf of $\widetilde{N}_x$ has full support. 
	\end{enumerate}	 
	\begin{proof}
		See Appendix A.2.1 in \cite{jonas}.
	\end{proof}	
\end{proposition}

\hspace{1mm}

Assume we have learned the Markov equivalence class $\mathcal{K}$ of DAGs that are I-maps to $\pr^\mathbb{X}$. The following result can help us to identify the DAG structure, from within $\mathcal{K}$, by establishing the directionality of undirected edges in the CPDAG.\\

\begin{proposition}[Directed paths and causality] 
	The following statements relate the DAG $\mathcal{G}$ of a SEM $\mathcal{M}$ to whether the relationships between its variables are causal or not.
	 \begin{enumerate}[i.]
		 \item If there is no directed path from $x$ to $y$, then there is no causal effect.
		 \item If there is a directed path from $x$ to $y$, there may not be a causal effect. 
	\end{enumerate} 
	\begin{proof}
		See Appendix A.2.2 in \cite{jonas}.
	\end{proof}	
\end{proposition}

\hspace{1mm}

	Notice that an equivalent form of statement i. is that if there is a causal effect from $y$ to $x$ then there must be a directed path from $y$ to $x$. Suppose we have identified the DAG $\mathcal{G}$ up to the CPDAG that represents $\mathcal{K}$ using observations from $\pr^\mathbb{X}$. Using observations from  $\pr^{\mathbb{X}}_{\widetilde{\mathcal{M}}}$ we can establish if variable $y$ causes variable $x$. If $y$ causes $x$ we can then use the above proposition to direct undirected edges by making sure there is at least one directed path from $y$ to $x$ in $\mathcal{G}$.


\section{Additive Gene Expression SEMs} \label{sec:addGESEM}

We now define a class of gene expression SEMs that is more restricted than the one defined in \ref{GESEM}. Below we argue that we may use this SEM to model the variables $\mathbb{X}=\{y, h_1,...,h_K,x_1,...,x_J\}$ since the distribution $\pr^{\mathbb{X}}$ that both SEMs induce is essentially the same: in this case we use non-identifiability in our favour to model a set of variables with the simpler SEM. \\

\begin{definition}[Class of additive gene expression SEMs] \label{adGESEM}
	 We define the class of additive gene expression SEMs as the collection of SEMs for variables $\mathbb{X} \cup C=\{y,c_1,...,c_L, h_1,...,h_K,x_1,...,x_J\}$ such that:
	 \begin{enumerate}
		 \item $S_{h_k}: h_k = f_{h_k}(y) + \sum_{l=1}^L \gamma_{lk}c_l + N_{h_k}$ $\forall k \in \{1,...,K\}$
		 \item $S_{x_j}: x_j = f_{x_j}(y) + \sum_{k=1}^K \beta_{kj}h_k + N_{x_j}$ $\forall j \in \{1,...,J\}$
	\end{enumerate}	
	
	\textbf{Remark:} Since we really care about modeling $x_j$ and the term $f_{x_j}(y)$ can be made to include any constant term, we assume without loss of generality that 
	\begin{itemize}
		\item $\mathbb{E}[c_l]=\mathbb{E}[f_{h_k}(y)]=\mathbb{E}[h_k]=\mathbb{E}[N_{h_k}]=\mathbb{E}[N_{x_j}]=0$ and
		\item $\mathbb{V}[c_l]=\mathbb{V}[h_k]=1$
	\end{itemize}	
\end{definition}

\hspace{1mm}
Clearly the class of additive gene expression SEMs is a subset of the class of gene expression SEMs of definition \ref{GESEM} since $h_k$ depends on $y$. Since we assume that the larger class of gene expression SEMs can be used to accurately model the gene expression causal process why can we use this reduced class to model the data? We again use the justification, from \cite{additive}, proposed for the use of model \ref{eq:geneModel} for modeling gene expression data. Since $y$ and $h_k$ must be dependent by condition ii.) of Definition \ref{GESEM} we know that for the right choice of non-linear basis $\{c_l\}_{\{l=1,...,L\}}$ we can model $h_k$ as:

\begin{align} \label{eq:geneModel}
	h_k = f_{h_k}(y) + \sum_{l=1}^L \gamma_l c_l + N_{h_k}
\end{align}	

Where, 

\begin{itemize}
	\item $k \in \{1,...,K\}$,  
	\item $c_l := \sum_{p=1}^P c_{lp}(z_p)$ and $z_1,...,z_P$ are \emph{unmodeled} variables upon which $h_k$ also depends, and
	\item $N_{h_k}$ is a white noise random process. 
\end{itemize}	

This means we can replace condition ii.) in \ref{GESEM} with equation $S_{h_k}$ of Definition \ref{GESEM}.

The DAG corresponding to a SEM from the above class where $L=3$, $K=2$ and $J=3$, and where $\gamma_{lk},\beta_{kj}>0$ $\forall l,k,j$ is: 

\begin{figure}[H]
  \centering
  \includegraphics[width=.5\textwidth]{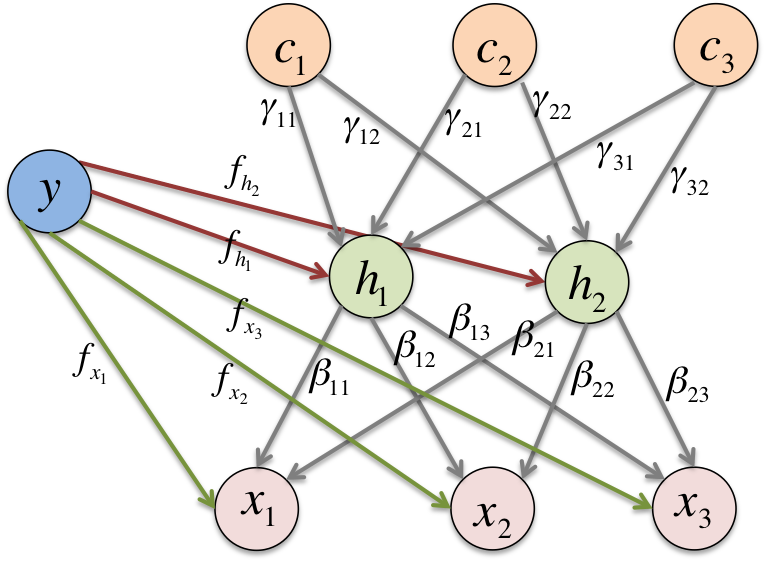} 
  \caption[Additive gene expression DAG]
  {Additive gene expression DAG}
  \label{fig:addGESEM}
\end{figure}

The $f_{h_k}$, $f_{x_j}$, $\gamma_{lk}$ and $\beta_{kj}$ edge labels do not form part of the DAG and are included to represent the SEM graphically. 

Notice that the DAG $\mathcal{G}$ of any SEM in the class above will be the same as that shown in the figure up to choice of $L$, $K$ and $J$, and deletion of edges (in the case that $\gamma_{lk}=0$ or $\beta_{kj}=0$ for any given $l$, $k$ and $j$, or that any of the $f_{x_j}$ or $f_{h_k}$ functions are constant in $y$). This means that we have essentially identified the causal structure of the underlying SEM by assumming it belongs to the above class.\\ 

Also notice that if we use an alternative definition for a SEM, where the joint noise distribution is not necessarily mutually independent then we can define a class of additive gene expression SEMs without the need for the $c_l$ variables, and instead allow the $N_{h_k}$ variables to be mutually dependent. 

Using the definitions and results presented in this Section we can establish the following properties for the additive gene expression class. \\

\begin{proposition}[Properties of class of additive gene expression SEMs] \label{propertiesAddSEM}
Suppose the SEM $\mathcal{M}$ belongs to the class of additive gene expression SEMs and that $\mathcal{G}$ and $\pr^{\mathbb{X}\cup C}$ are its induced DAG and joint distribution, respectively. Then:

\begin{enumerate}[1.]
	
\item $y \indep c_l$ and $c_l \indep c_m$ for all $l,m \in \{1,...,L\}$.
	
\item $\pr^{\mathbb{X} \cup C}$ satisfies causal minimality with respect to $\mathcal{G}$.

\item If $\beta_{kj} \neq 0$ then $h_k$  is a cause of $x_j$.

\item If $f_{x_j}(y) + \sum_{k=1}^K \beta_{kj}f_{h_k}(y) \neq 0$  then $y$ is a cause of $x_j$.


\begin{proof} 
	See Appendix \ref{app:proofs}.
\end{proof}	

\end{enumerate}

\end{proposition}

As we can see the class of additive gene expression SEMs satisfy our assumption that $y$ and $h_k$ are causes of $x_j$. The causal effect of $y$ on $x_j$ can be described in terms of the interventional distribution $\pr^{x_j|do(y=y_1)}_{\mathcal{M}}$. Although we have assumed that in the context of SVA we don't have access to observations from this distribution, we can express the interventional distribution in terms of the observational distribution if we can find a valid $\emph{adjustment set}$:
	\begin{align}
		p_{\mathcal{M},do(y=y_1)}(x_j)= \int_\mathcal{Z} p_\mathcal{M}(x_j|y,\mathcal{Z})p_\mathcal{M}(\mathcal{Z}) d\mathcal{Z}
	\end{align}
	
where $\mathcal{Z}$ is a valid adjustment set of variables. Although we don't study why \emph{adjustment} works or what is a criterion for a set of variables constituting a valid $\emph{adjustment set}$, we mention that the $c_l$ variables in the additive gene expression SEMs constitute a valid adjustment set for determining $p_{\mathcal{M},do(y=y_1)}(x_j)$. In Section 3.1.2 of \cite{jonas} the adjustment principle is explained and valid adjustment criterions derived. Using the adjustment set $\mathcal{Z}=\{c_1,...,c_L\}$ we can obtain $p_{\mathcal{M},do(y=y_1)}(x_j)$ from the observational distributions $p_\mathcal{M}(x_j|y,c_1,...,c_L)$ and $p_\mathcal{M}(c_1,...,c_L)$. This means we don't need to estimate the $h_k$ variables to obtain the causal effect of $y$ on $x_j$. However, as we mentioned following Definition \ref{GESEM}  we are only interested in the part of the effect that $y$ has on $x_j$ which does not go through the unmodeled factors $h_k$. For this we must first estimate the variables $h_k$ so that we may filter out the part of the effect that $y$ has on $x_j$ that goes through $h_k$. The SVA methodology developed by \citeauthor{SVA} provides a way to estimate the variables $h_k$.

\chapter{SVA Methodology} \label{ch:SVA}

\section{Overview} \label{sec:overview}

Given $n$ i.i.d. observations of $y$ and $x_j$ for $j \in \{1,...,J\}$ the SVA methodology uses the model class \ref{adGESEM} to fit the data. Although the goal is to estimate the effects $f_{x_j}$ of $y$ on $x_j$, to do this the variables  $\{h_k\}_{k = 1,...,K}$, or rather their span, must be estimated so that their effect on $y$ can be filtered out. To estimate the span of $\{h_k\}_{k=1,...,K}$ it is necessary to first estimate the span of $\{c_l\}_{l=1,...,L}$ since the $c_l$ variables are causes of the $h_k$ variables. The basic steps of the SVA methodology are thus:

\begin{enumerate}[1.]
	\item Estimate the span of $\{c_l\}_{l=1,...,L}$,
	\item Estimate the span of $\{h_k\}_{k=1,...,K}$, and
	\item Fit $S_{x_j}$ equations from \ref{adGESEM} to estimate $f_{x_j}$.
\end{enumerate}	

	 \begin{figure}[H]
	   \centering
	   \includegraphics[width=.7\textwidth]{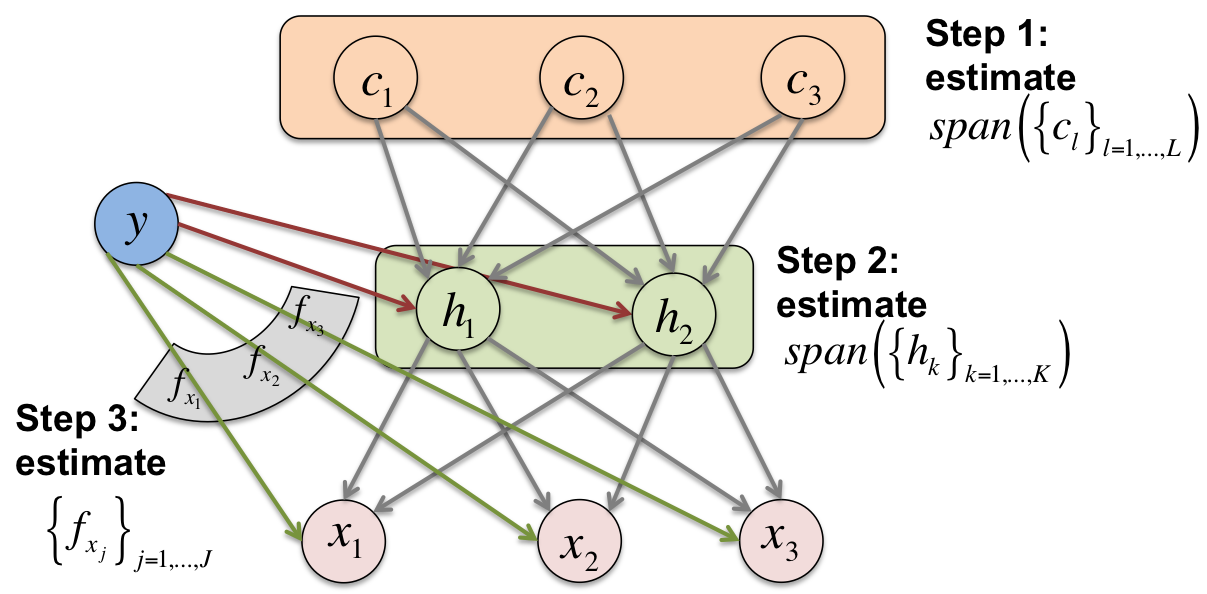} 
	   \caption[3 steps of SVA estimation]
	   {3 steps of SVA estimation}
	   \label{fig:steps}
	 \end{figure}

\section{Estimation of $\Span(\{c_l\})$} \label{sec:cl}


\subsection{Estimation procedure} \label{SVAest}

If we combine equations $S_{x_j}$ and $S_{h_k}$ from \ref{adGESEM} we have 

\begin{align}
  x_j &= f_{x_j}(y) + \sum_{k=1}^K \beta_{kj}h_k + N_{x_j}\\
  x_j &= f_{x_j}(y) + \sum_{k=1}^K \beta_{kj}(f_{h_k}(y) + \sum_{l=1}^L \gamma_{lk}c_l + N_{h_k}) + N_{x_j}\\
  x_j &= f_{x_j}(y) + \sum_{k=1}^K \beta_{kj}f_{h_k}(y) + \sum_{k=1}^K \beta_{kj}\sum_{l=1}^L \gamma_{lk}c_l + \sum_{k=1}^K \beta_{kj}N_{h_k} + N_{x_j}\\
  x_j &= (f_{x_j}(y) + \sum_{k=1}^K \beta_{kj}f_{h_k}(y)) + \sum_{l=1}^L c_l (\sum_{k=1}^K \beta_{kj} \gamma_{lk}) + (\sum_{k=1}^K \beta_{kj}N_{h_k} + N_{x_j})\\
  x_j &= f_j(y) + \sum_{l=1}^L \alpha_{lj} c_l  + N_j \label{errors}
\end{align}

Where,
\begin{itemize}
	\item $f_j(y) := f_{x_j}(y) + \sum_{k=1}^K \beta_{kj}f_{h_k}(y)$,
	\item $\alpha_{lj}:= \sum_{k=1}^K \beta_{kj} \gamma_{lk}$, and 
	\item $N_j := \sum_{k=1}^K \beta_{kj}N_{h_k} + N_{x_j}$
\end{itemize}		

The last expression suggests a reduced equation model class for variables $y$, $c_k$ and $x_j$ with the following set of equations:

\begin{align}
	S_{x_j}: x_j &= f_j(y) + \sum_{l=1}^L \alpha_{lj} c_l  + N_j
\end{align}	

and with the following associated DAG

	 \begin{figure}[H]
	   \centering
	   \includegraphics[width=.7\textwidth]{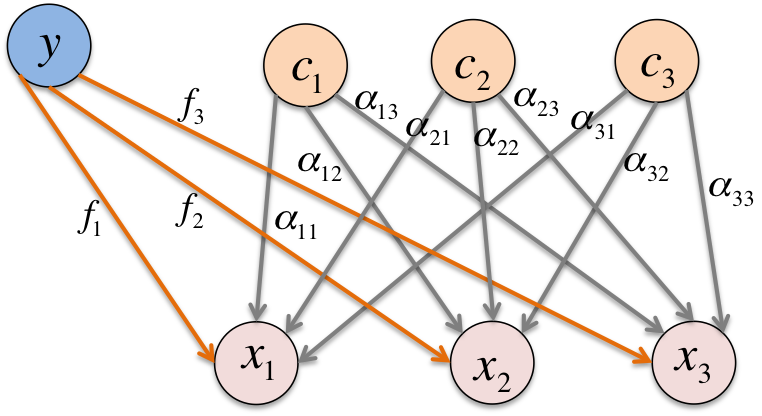} 
	   \caption[Reduced equation model DAG]
	   {Reduced equation model DAG}
	   \label{fig:reduced}
	 \end{figure}

Notice that since $N_j = \sum_{k=1}^K \beta_{kj}N_{h_k} + N_{x_j}$ the noise variables $N_j$ are not independent which means the above equation model is not a SEM as defined in \ref{SEM}, and we have, with the theory shown here, no guarantee that the above DAG is an I-map of $\pr^\mathbb{X}$. However for the same noise realizations of $N_y$ and $N_{c_k}$, and for the above definition of the $\alpha_{lj}$ parameters, the above derivation shows that the reduced model  would generate exactly the same realizations of $x_j$. This means that, as in the additive gene expression SEMs, $y \indep c_l$  for all $l \in \{1,...,L\}$ which implies that for the model:

\begin{align} \label{fullModel}
	x_j &= f_j(y) + \sum_{l=1}^L \alpha_{lj} c_l  + N_j
\end{align}	

we can estimate $f_j$ separately from $\alpha_{lj}$ by fitting the model:

\begin{align} \label{marginalModel}
	x_j &= f_j(y) + \epsilon_j
\end{align}	

and we can expect to have an equivalent estimate for $f_j$, in terms of bias and variance, by fitting model \ref{marginalModel} as if we had fit model \ref{fullModel}. If we define the \emph{residuals} $r_j$ of model \ref{fullModel} as:

\begin{align}
	r_j := x_j - f_j(y) = \sum_{l=1}^L \alpha_{lj} c_l  + N_j
\end{align}	

or in vector form 

\begin{align} \label{resVec}
	r := x - f(y) =  A^T c  + N
\end{align}	

where

\begin{itemize} 
	\item $r = (r_1,...,r_J)^T$,
	\item $x = (x_1,...,x_J)^T$,
    \item $f(y) = (f_1(y),...,f_J(y))^T$
	\item $N = (N_1,...,N_J)^T$,
	\item $c = (c_1,...,c_L)^T$, and
	\item $A \in \realnumbers^{L \times J}$ with $A_{lj}=\alpha_{lj}$. 
\end{itemize}

then we can estimate the span of $\{c_l\}_{l=1,...,L}$ by factorizing the data matrix $R$:

\begin{align}
	R := X - F
\end{align}	 

where 

\begin{itemize}
	\item $X \in \realnumbers^{n \times J}$ and $X_{ij}$ is the i-th observation of variable $x_j$,
	\item $Y \in \realnumbers^n$ and $Y_i$ is the i-th observation of variable $y$,
	\item $F \in \realnumbers^{n \times J}$, $F_{ij}$ is the i-th observation of $\hat{f}_j(y)$, that is  $\hat{f}_j$ applied to $Y_i$, where $\hat{f}_j$ is the estimation of $f_j$ obtained by fitting model \ref{marginalModel}.
\end{itemize}	

In Section \ref{subsec:basisModels} we explore the way the functions $f_j$ are estimated in the SVA methodology of \citeauthor{SVA} . Once the matrix $F$ and then $R$ are obtained we factorize $R$ as:

\begin{align} \label{resMat}
	R = C\Lambda + E 
\end{align}	

where 

\begin{itemize}
	\item $C \in \realnumbers^{n \times L}$ and $C_{il}$ is the estimate for i-th unobserved realization of variable $c_l$,
	\item $\Lambda \in \realnumbers^{L \times J}$ is the estimate of $A$ from \ref{resVec} and
	\item $E \in \realnumbers^{n \times J}$ and $E_{ij}$ is the estimate for the i-th unobserved realization of variable $N_j$.
\end{itemize}	

The chosen factorization should reflect the fact that the set of $c_l$ and $N_j$ variables are mutually independent. 

In SVA this factorization is obtained by applying singular value decomposition (SVD) to $R$ as is detailed in Section \ref{subsec:SVD}. Before applying SVD, the number of $c_l$ variables, $L$, must be estimated. In SVA this is done by using \emph{parallel analysis} as described in Section \ref{subsec:PA}.

\subsection{Estimation of basis function model} \label{subsec:basisModels}

To estimate the functions $f_j$ the SVA methodology prescribes using standard basis models. The task is then to choose a feature mapping $\phi$:

\begin{align}
& \phi: \realnumbers \rightarrow \realnumbers^p \\
& y \mapsto (\phi_1(y),...,\phi_p(y))^T
\end{align}

assuming that $y$ is univariate so that:  

\begin{align}
f_j(y) = \mathbb{E}[x_j|y] := \sum_{k=1}^p \beta_{jk}\phi_k(y)
\end{align}

and estimating $f_j$ for $j \in \{1,...,J\}$ reduces to estimating the matrix $\beta \in \realnumbers^{J \times p}$. Least squares estimation is then applied to the regression model

\begin{align}
	x_j = \sum_{k=1}^p \beta_{jk}\phi_k(y) + \epsilon
\end{align}	

or in vector form 

\begin{align} \label{linMod}
	X_j =  \Phi \beta_j + \epsilon
\end{align}	

Where 

\begin{itemize}
	\item $X_j \in \realnumbers^n$ is the j-th column of the observation matrix $X$,
	\item $\beta_j \in \realnumbers^p$ is the j-th row of $\beta$ as a column vector,
	\item $\Phi \in \realnumbers^{n \times p}$ and $\Phi_{ik}=\phi_k(y_i)$.
\end{itemize}	

The well known least squares solution for $\beta_j$ is

\begin{align}
	\beta_j = (\Phi^T\Phi)^{-1}\Phi^T X_j
\end{align}	

and the fitted values $F$ are 

\begin{align}
	F = HX = \Phi(\Phi^T\Phi)^{-1}\Phi^T X
\end{align}	

where $H$ is the hat matrix. Notice that since we choose one mapping $\phi$ to define all $f_j$ then the hat matrix $H$, which only depends on $\Phi$, is the same for all the models to be fitted. In the function \verb|sva| of the \Rp package \verb|sva|, which implements the SVA methodology, the hat matrix $H$ must be provided as input so that the choice of basis defined by $\phi$ and the estimation of functions $f_j$  is done independently by the user. The matrix $F$ is then calculated by the function \verb|sva| as $F=HX$. Notice that this means that no regularization is applied to any of the $J$ regression models to avoid overfitting. This is because since $J$ is large, each regression model would need its separate regularization parameter $\lambda$ which would be unfeasible - for example because cross validation would need to be performed for each model. This means that we must regularize by choosing a \emph{simple} mapping $\phi$. This is especially true since the number of observations $n$ will usually be in the order of 10 meaning a complex $f_j$ will tend to overfit the data.

\subsection{Factorization of residuals using SVD} \label{subsec:SVD}

Suppose $J>n$. If we apply SVD to the residual matrix $R$ we have:

\begin{align}
	R = QDS^T
\end{align}	

Where 

\begin{itemize}
	\item $D \in \realnumbers^{n x J}$ and $D = \diag\{d_1,...,d_n\}$ where $d_1 \geq d_2 \geq...\geq d_n \geq 0$, $d_i$ are the singular values of $R$ and $d_i^2$ are the eigenvalues of $R^TR$ and $RR^T$,
	\item $Q \in \realnumbers^{n \times n}$ and its columns $\{q_i\}_{i=1,...,n}$ are an orthonormal basis for $\realnumbers^n$ and the eigenvectors of $RR^T=QDS^TSDQ^T=QD^2Q^T$ corresponding to the eigenvalues $\{d_i^2\}_{i=1,...,n}$,
	\item $S \in \realnumbers^{J \times J}$ and its columns $\{s_i\}_{i=1,...,J}$ are an orthonormal basis for $\realnumbers^J$ and the first $n$ are the eigenvectors of $R^TR=SDQ^TQDS^T=SD^2S^T$ corresponding to the eigenvalues $\{d_i^2\}_{i=1,...,n}$.
\end{itemize}	

In the following we suppose that we know exactly $L \leq n$ of the singular values $d_i$ are significantly different from zero. 

\subsubsection{Left SVD}

Let $W=Q^TR$ then

\begin{align} \label{factData}
R = QW = \left( \begin{array}{cc}
Q_1 & Q_2 
\end{array} \right)
\left( \begin{array}{c}
W_1  \\
W_2
\end{array} \right)
= Q_1 W_1 + Q_2 W_2
\end{align}	

Where 
\begin{itemize}
\item $Q \in \realnumbers^{n \times n}$, $Q_1 \in \realnumbers^{n \times L}$, $Q_2 \in \realnumbers^{n \times (n-L)}$ and
\item $W \in \realnumbers^{n \times J}$, $W_1 \in \realnumbers^{L \times J}$, $W_2 \in \realnumbers^{(n-L) \times J}$.
\end{itemize}

So if we make $C:=Q_1$, $\Lambda:=W_1$ and $E:= Q_2 W_2$ we have an estimate for the factorization of the observed residuals $R$ in terms of the estimated realizations of $c_l$ and $N_j$ variables as in \ref{resVec} and \ref{resMat}. Some of the properties of this factorization are:

\begin{enumerate}[1.]
	\item Since the unobserved realizations of $c_l$ are etimated as the l-th left eigenvector of $R$, which is orthonormal, the estimated realizations have a sample variance of 1 which is consistent with definiton \ref{adGESEM}. 
	\item Since the columns of $C$ are the left eigenvectors of $R$ and we assume the expected value of $c_l$ to be zero, the estimated realizations of $c_l$ have zero sample correlation with the estimated realizations of $c_m$ for $l \neq m$:
	
	\begin{align}
		\widehat{\Cor}(c_l, c_m) = q_l^Tq_m = 0
	\end{align}

	 This is consistent with property 1. of \ref{propertiesAddSEM} however it is not sufficient to satisfy it since zero correlation does not imply independence except in the case of normally distributed $c_l$ variables. 
	 
	\item The estimated realizations of $N_j$ are given by the j-th column of $E:= Q_2 W_2$ which is $Q_2w_{2j}$ where $w_{2j}$ is the j-th column of $W_2$.  Since the columns of $C$ are the left eigenvectors of $R$ and we assume the expected value of $c_l$ to be zero, the estimated realizations of $c_l$ have zero sample correlation with the estimated realizations of $N_j$:
	
	\begin{align}
		\widehat{\Cor}(c_l, N_j) = q_l^T(Q_2w_{2j}) = (q_l^TQ_2)w_{2j}= 0
	\end{align}

	 This is consistent with definition \ref{adGESEM} however it is not sufficient to satisfy it since zero correlation does not imply independence except in the case of normally distributed $c_l$ and $N_j$ variables. 
	 
\end{enumerate}

\subsubsection{Right SVD}

Let $Y=RS$ then:

\begin{align} \label{factData}
R = YS^T = \left( \begin{array}{cc}
Y_1 & Y_2 
\end{array} \right)
\left( \begin{array}{c}
S_1^T  \\
S_2^T
\end{array} \right)
= Y_1 S_1^T + Y_2 S_2^T
\end{align}	

Where 
\begin{itemize}
\item $Y \in \realnumbers^{n \times J}$, $Y_1 \in \realnumbers^{n \times L}$, $Y_2 \in \realnumbers^{n \times (J-L)}$ and
\item $S \in \realnumbers^{J \times J}$, $S_1 \in \realnumbers^{J \times L}$, $S_2 \in \realnumbers^{J \times (J-L)}$.
\end{itemize}

So if we make $C:=Y_1$, $\Lambda:=S_1$ and $E:= Y_2 S_2$ we have an estimate for the factorization of the observed residuals $R$ in terms of the estimated realizations of $c_l$ and $N_j$ variables as in \ref{resVec} and \ref{resMat}. Notice that the matrix $Y$ contains the principal components of the residuals $R$. Some of the properties of this factorization are:

\begin{enumerate}[1.]
	\item Since the unobserved realizations of $c_l$ are etimated as the l-th principal component of $R$, the estimated realizations have a sample variance of $d_l^2$.  This is inconsistent with definiton \ref{adGESEM} which assumes that the $c_l$ variables have a variance of 1, but we may easily \emph{fix} this by using $Z= YD^{-1}=RSD^{-1}$ so that $C:=Z_1$ instead of $Y_1$. 
	\item Since the columns of $C$ are the first $L$ principal components of $R$ and we assume the expected value of $c_l$ to be zero, the estimated realizations of $c_l$ have zero sample correlation with the estimated realizations of $c_m$ for $l \neq m$:
	
	\begin{align}
		\widehat{\Cor}(c_l, c_m) = (Rs_l)^T(Rs_m) = s_l^TR^TRs_m = s_l^TSD^2S^T s_m= e_l^TD^2e_m = 0
	\end{align}

	 Where $e_l$ is the l-th canonical vector. This is consistent with property 1. of \ref{propertiesAddSEM} however it is not sufficient to satisfy it since zero correlation does not imply independence except in the case of normally distributed $c_l$ variables. 
	 
	\item The estimated realizations of $N_j$ are given by the j-th column of $E:= Y_2 S_2^T$ which is $Y_2\tilde{s}_{2j}$ where $\tilde{s}_{2j}$ is the j-th row of $S_2$ as a column vector.  Since the columns of $C$ are the first $L$ principal components of $R$ and we assume the expected value of $c_l$ to be zero, the estimated realizations of $c_l$ have zero sample correlation with the estimated realizations of $N_j$:
	
	\begin{align}
		\widehat{\Cor}(c_l, N_j) &= y_l^T(Y_2\tilde{s}_{2j})   = (Rs_l)^T(Y_2\tilde{s}_{2j}) \\
								 &= s_l^TR^TRS_2\tilde{s}_{2j} = s_l^TSD^2S^TS_2\tilde{s}_{2j} \\
								 &= e_lD^2(S^TS_2)\tilde{s}_{2j} = d_l^2 e_l \left( 
								 	\begin{array}{c}
											0_{L \times (J-L)}\\
											I_{(J-L) \times (J-L)} 
									\end{array} \right)\tilde{s}_{2j} = 0
	\end{align}

	 Where $y_l$ is the l-th column of $Y_1$ and $l \in \{1,...,L\}$. This is consistent with definition \ref{adGESEM} however it is not sufficient to satisfy it since zero correlation does not imply independence except in the case of normally distributed $c_l$ and $N_j$ variables. 
	 
\end{enumerate}	

Now from the SVD factorization of $R$ and from left and right SVD factorizations above we have that: 

\begin{align}
R  = YS^T = QW = QDS^T
\end{align}

So that

\begin{align}
Y=RS=QWS=QD \\
YD^{-1} = Q
\end{align}

So we see that performing right and left SVD is equivalent up to standardization of the estimated $c_l$ variables. Since $Y$ is the matrix of principal components of $R$ we also see that this factorization corresponds to obtaining the principal components and then standardizing them. 

Equations \ref{resVec} and \ref{resMat} suggest that  \emph{factor analysis} may be a more suitable technique for extracting factors from the matrix $R$. However since in our setting $J>n$ factor analysis is not possible since it relies on the spectral decomposition of the covariance matrix $R^TR=SDQ^TQDS^T=SD^2S^T$ which in this case is not positive definite. 

\subsection{Parallel Analysis} \label{subsec:PA}

The SVA methodology uses a parallel anlaysis method based on \cite{pa} to estimate the number factors $L$ needed to approximate the residual matrix as: 
\begin{align} \label{resMatApprox}
	R \approx C\Lambda
\end{align}	

where 

\begin{itemize}
	\item $C \in \realnumbers^{n \times L}$ and 
	\item $\Lambda \in \realnumbers^{L \times J}$ is the estimate of $A$ from \ref{resVec}.
\end{itemize}	

 We assume that the $\{r_j\}_{j=1,...,J}$ residual variables are mutually independent irrespective of the particular marginal distribuition of each $r_j$ variable. Under this assumption a realization of $r_j$ is not linked to any particular realization of $r_k$ for $j \neq k$ meaning we can permute observations of the columns of $R$ without altering the underlying joint distribution of $\{r_j\}_{j=1,...,J}$ . This assumption also implies that the population singular values for $R$ (i.e. if $n \rightarrow \infty$) are  $d_j = \sigma_{\psi(j)}$ where $\sigma_j$ is the  standard deviation of $r_j$ and $\psi(j)$ is such that:

\begin{align}
	\psi(i) < \psi(j) \Rightarrow \sigma_{\psi(i)} \geq \sigma_{\psi(j)}
\end{align}	

One method to choose the number of factors, for example in the context of PCA, is to choose the number of singular values such that:

\begin{align}
	\hat{d}_j^2 \geq \hat{\sigma}_{\psi(j)}^2
\end{align}	

Where $\hat{d}_j$ is the estimate of the population singular value $d_j$ made by applying SVD to $R$ and $\hat{\sigma}_{\psi(j)}^2$ is the sample variance of $r_{\psi(j)}$ . In the context of PCA this corresponds to choosing the number of principal components such that their corresponding eigenvalue is greater than one in the case that the data matrix has been standardized. We choose only the principal components that account for more of the variance than we would expect under the assumption of independence, or in other words we choose only those principal components that summarize the information of more than one variable. However, for limited data settings, the variance of $\hat{d}_j$ and $\hat{\sigma}_{\psi(j)}^2$ means that it is possible for $\hat{d}_j^2 \geq \hat{\sigma}_{\psi(j)}^2$ even when $d_j^2 < \sigma_{\psi(j)}^2$. To take this variance into account parallel analysis prescribes permuting the values of each column of $R$ several times so that we obtain a distribution for each singular value under the independence assumption. For each column $j$, $B$ permutations $\pi_{ij}(k)$ are generated where $i \in \{1,...,B\}$ indicates the permutation number, $j \in \{1,...,J\}$ the column to which it will be applied and $k \in \{1,...,n\}$ is the row index argument:

\begin{align}
& \pi_{ij}: \{1,...,n\} \rightarrow \{1,...,n\} \\
& (1,...,n) \mapsto (\pi_{ij}(1),...,\pi_{ij}(n))
\end{align}

In this way we obtain $B$ realizations of the residual matrix $R$ which we denote $R^b_{i}$. For each $R^b_{i}$ we apply SVD so that we have $B$ estimates $d^b_{ij}$ of $d_j$ for $i \in \{1,...,B\}$ and $j \in \{1,...,J\}$ under the assumption of independence. We then compare the estimated proportion of variance explained by each principal component under the assumption of independence to the estimated proportion of variance explained without this assumption, i.e. with the non-permuted $R$ matrix:

\begin{align}
	\hat{\nu}_j &= \frac{\hat{d}_j^2}{\sum_{j=1}^J\hat{d}_j^2}\\
	\hat{\nu}_{ij}^b &= \frac{(\hat{d}_{ij}^b)^2}{\sum_{j=1}^J(\hat{d}_{ij}^b)^2}
\end{align}	

If $d_j^2 \geq \sigma_{\psi(j)}^2$ then we expect that for \emph{most} realizations $i \in \{1,...,B\}$:

\begin{align}
	\hat{\nu}_j \geq \hat{\nu}_{ij}^b
\end{align}	

If this is the case we say the corresponding singular value $d_j$ is significant. To obtain an estimate for $L$ we count the number of significant singular values. 

Other considerations taken into account by \citeauthor{SVA} in their implementation of the parallel analysis algorithm suggested by \cite{pa} are:

\begin{enumerate}[1.]
	\item \textbf{Degrees of freedom of model}: We know that:
	\begin{align}
		R=X-HX
	\end{align}
	so that 
	\begin{align}
		\rank(R)=\rank(X)-\rank(HX)
	\end{align}
	and if we assume that $n<J$ and $X$ has full rank then
	\begin{align}
		\rank(R)=n-\rank(H) = n - \trace(H)
	\end{align}
	since $H$ is an idempotent matrix. So we only consider  $M := n - \ceil{\trace(H)} \geq L$ non-zero singular values as potentially significant.

	\item \textbf{Orthogonal residuals}: Since least squares is used to fit models \ref{linMod} the residuals should be, by construction, orthogonal to the predictor variables: each column of $R$ should be orthogonal to the linear span of $\Phi$. However, by permuting the residuals in each column of $R$ to create $R_i^b$ we create a linear dependence beteween the columns of $R_i^b$ and the linear span of $\Phi$. This is fixed by projecting the permuted residuals $R_i^b$ on the linear span of $\Phi$ using the hat matrix to create the \emph{corrected} residuals $\tilde{R}^b_i$:
		\begin{align}
			\tilde{R}^b_i := R^b_i - HR^b_i
		\end{align}	
		
	\item \textbf{Quasi hypothesis test}: To decide if a singular value $d_j$ is significant  the proportion $p^b_j$ of boostrap estimates $\hat{\nu}_{ij}^b$ such that $\hat{\nu}_{ij}^b \geq \hat{\nu}_j$ is calculated:
	
	\begin{align}
		p^b_j := \frac{\sum_{i=1}^B \mathbbm{1}_{\{\hat{\nu}_{ij}^b \geq \hat{\nu}_j\}} }{B}
	\end{align}	
	
	and  the null hypothesis that $d^b_j$ is not significant is rejected if $p^b_j < 0.1$. The cutoff value 0.1 is arbitrarily chosen and since  the distribution of the test statistic $p^b_j$ is unknown the significance or power of this test is unknown.
	
	\item \textbf{Non-increasing singular values}: Singular values $d_j$ are non-increasing, meaning $v_j$ should be non-increasing and, for a large enough $B$, $p^b_j$ non-decreasing.  However because of the randomness of the boostrap estimates and the limited data setting, the $p^b_j$ values may not be non-decreasing. A corrected value $\tilde{p}^b_j$ is calculated:

	\begin{align}
		\tilde{p}^b_j := \max(p^b_{j-1},p^b_j)
	\end{align}	
	
	for $j \in \{2,...,M\}$.
	
\end{enumerate}	

With these consideration in mind we present the pseudo code for the parallel analysis algorithm described in \cite{SVA} and implemented as the default method in the function \verb|num.sv| of the \Rp package \verb|sva|. 

\subsubsection{Parallel analysis algorithm of \cite{SVA}} \label{subsec:PA}

\begin{enumerate}[1.]
	\item Estimate matrices $X$, $F$, $R$ and the rank $M$ of $R$:
		\begin{align}
			 H &= \Phi(\Phi^T\Phi)^{-1}\Phi^T  \\
			 F &= HX \\
			 R &= X - F \\
			 M &= n - \ceil{\trace{H}}
		\end{align}	
	
	\item Estimate singular values $d_j$  by performing SVD on $R$:
	
	\begin{align}
		R = QDS^T
	\end{align}	
	
	where $D = \diag\{\hat{d}_1,...,\hat{d}_n\}$. 
	
	\item Estimate the proportion of variance explained by first $M$ principal components:
	
	\begin{align}
		\hat{\nu}_j &= \frac{\hat{d}_j^2}{\sum_{j=1}^J\hat{d}_j^2}
	\end{align}	
	
	for $j \in \{1,...,M\}$.
	
	\item Calculate boostrap estimates $\hat{d}_{ij}^b$ for $i \in \{1,...,B\}$ and $j \in \{1,...,M\}$. 
	
		 For $i=1,...,B$
		 \begin{enumerate}[a.]
			 \item Generate $J$ permutations $\pi_1,...,\pi_J$ one for each column of $R$.
			 \item Permute columns of $R$ according to these permutations to create $R_i^b$:
			 
			 \begin{align}
				 R_i^b  = \left( \begin{matrix}
				 	R_{\pi_1(1),1} & R_{\pi_2(1),2}   & \ldots & R_{\pi_J(1),J}\\
					R_{\pi_1(2),1} & R_{\pi_2(2),2}   & \ldots & R_{\pi_J(2),J}\\
					\vdots & \vdots & \ddots & \vdots\\
					R_{\pi_1(n),1} & R_{\pi_2(n),2}   & \ldots & R_{\pi_J(n),J}
							\end{matrix} \right)
			 \end{align}	 
			 
			 \item Project each column of $R_i^b$ onto the columns of $\Phi$ and calculate adjusted residuals:
			 
			 \begin{align}
				 \tilde{R}_i^b = R_i^b - HR_i^b
			 \end{align} 
			 
			 \item Apply SVD to $\tilde{R}_i^b$ to obtain estimates $\hat{d}_{ij}^b$ under independence assumption.
		 	
			\begin{align}
		 		\tilde{R}_i^b = Q_i^bD_i^b(S_i^b)^T
		 	\end{align}
			
			where $D_i^b = \diag\{\hat{d}_1^b,...,\hat{d}_n^b\}$. 
			 
		 	\item Estimate the proportion of variance explained by first $M$ principal components: 
	
		 	\begin{align}
		 		\hat{\nu}_{ij}^b &= \frac{(\hat{d}_{ij}^b)^2}{\sum_{j=1}^J(\hat{d}_{ij}^b)^2}
		 	\end{align}	
	
		 	for $j \in \{1,...,M\}$.
		 \end{enumerate} 
		 
	\item Calculate test statistics: For each singular value $d_j$, $j \in \{1,...,M\}$ calculate the proportion of bootstrap estimates for which the variance explained is higher than the non-permuted estimate:
	
		\begin{align}
			p^b_j := \frac{\sum_{i=1}^B \mathbbm{1}_{\{\hat{\nu}_{ij}^b \geq \hat{\nu}_j\}} }{B}
		\end{align}
	
	\item Correct test statistics $\{p^b_1,...,p^b_M\}$ to ensure they are non-decreasing:
	
	\begin{align}
		\tilde{p}^b_j := \max(p^b_{j-1},p^b_j)
	\end{align}
	
	\item Calculate the number of of singular values for which the corresponding corrected proportion of boostrap estimates $p^b_j$ is lower than 0.1.
	\begin{align}
		L = \sum_{j=1}^M \mathbbm{1}_{\{\tilde{p}^b_j \leq 0.1 \}}
	\end{align}	
\end{enumerate}	

\section{Estimation of $\Span(\{h_k\})$} \label{sec:hk}

\subsection{General description} \label{sigMeth}

We can observe the $x_j$ variables and have estimated the span of the $c_l$ variables. We now want to estimate the span of the $h_k$ variables. The $x_j$ variables depend on three quantities - $c_l$, $f_{h_k}(y)$ and $f_{x_j}(y)$ - while the $h_k$ variables only depend on $c_l$ and $f_{h_k}(y)$. If we can somehow filter out the $f_{x_j}(y)$ \emph{signature} from the $x_j$ variables we could then reconstruct $h_k$ from these filtered variables.  The problem is it is impossible to separate the $f_{x_j}(y)$ and $f_{h_k}(y)$ \emph{signals} since they both depend on $y$. The strategy will be to build an $h_k$ variable for every $c_l$ variable which contains the signal from $c_l$ but which is also allowed to contain signal from $y$ even if it is not exactly the $f_{h_k}(y)$ signal we would like it to have. This is important since not including the \emph{overlap} in signal between $y$ and $h_k$ can lead to biased estimate of $h_k$ as is pointed out in the third paragraph of the section titled \emph{comparison with existing methods} of \cite{SVA}. The basic algorithm has the following steps, which we will elaborate on in the subsequent subsections:

For $l \in \{1,...,L\}$
\begin{enumerate}
	\item Identify the subset $\mathcal{S} \subset \{1,...,J\}$ such that $x_j$ for $j \in \mathcal{S}$ includes the \emph{signature} of $c_l$. 
	\item Form an \emph{enriched} matrix $X^\mathcal{S} \in \realnumbers^{n \times |\mathcal{S}|}$ which includes the observations of the $x_j$ variables such that $j \in \mathcal{S}$.
	\item Factorize enriched matrix $X^\mathcal{S}$ into $M$ factors using left SVD of section \ref{subsec:SVD} and the parallel analysis method of section \ref{subsec:PA}.
	\item Set estimate of $h_l$ as the factor with highest absolute correlation with $c_l$. Let $q_1,...,q_M$ be the  eigen vectors of $X^\mathcal{S}(X^\mathcal{S})^T$ corresponding to eigenvalues $d_1^2 \geq d_2^2 \geq... \geq d_M^2$
		\begin{align}
		 	i^* &=  \argmax_{i \in \{1,...,M\}} \widehat{\Cor}(h_l, q_i) \\
			\hat{h_l} &= q_{i^*}
		\end{align}
\end{enumerate}	

Before detailing what it means for a variable to include the \emph{signature} of another variable and how this is done in the SVA methodology we remark on some advantages and disadvantages of the algorithm described above. We assume that the method for finding the variables $x_j$ that contain the signature of $c_l$ is adequate. 

\textbf{Advantages} 
\begin{itemize}
	\item A lot of the $f_{x_j}$ signal can be filtered out especially if $J>>L$ and the size of $\mathcal{S}$ is small (because a lot of the $\gamma_{lk}$ and $\beta_{kj}$ parameters are equal to zero), 
	\item If for all the $j \in \mathcal{S}$, there exists $k$ such that $\beta_{kj} \neq 0$ then the $x_j$ contain the signal of $f_{h_k}$.
\end{itemize}

\textbf{Disadvantages}
\begin{itemize}
	\item The $f_{x_j}$ signal for $j \in \mathcal{S}$ is not filtered out,
	\item By taking the  factor from the enriched matrix most correlated with $c_l$ we ensure that the estimated span of $h_k$ includes the signal from $c_l$ but not necessarily the signal from $f_{h_k}$ as that may be concentrated in the other factors,
	
	\item In general it is only clear that the estimated span of $h_k$ allows for the inclusion of signal from $y$ and that if $J$ is large and $|\mathcal{S}|$ small, it may filter out most of the $f_{x_j}(y)$ signal, however it isn't clear how succesful it is in including the $f_{h_k}(y)$ signal. 
\end{itemize}

\subsection{Finding the signature of $c_l$} \label{findSig}

To find the set $\mathcal{S}$ with the indices of the variables $x_j$ which have the signature of $c_l$ the following regression models are fitted:

\begin{align} \label{sig1}
	x_j = \beta_{j0} + \beta_{j1} c_l + \epsilon_j
\end{align}	

for $j \in \{1,...,J\}$. We say that $x_j$ contains the signature of $c_l$ if we can reject the null hypothesis $H_{j0}:\beta_{j1}=0$. This means that the set $\mathcal{S}$ is defined as:

\begin{align}
	\mathcal{S} := \{j: H_{j0} \textrm{ is rejected}\}
\end{align}	

Clearly $x_j$ also depend on $y$ so that the hypothesis tests and corresponding p-values are dependent and a conventional significance analysis will be invalid. It is not clear to the author why the SVA methodology does not use the following model:

\begin{align} \label{sig2}
	x_j = f_{j}(y) + \beta_{j1} c_l + N_j
\end{align}	

instead of model \ref{sig1}. However, even in this case the hypothesis tests would be dependent since as we saw in  \ref{errors} the error term $N_j$ is composed of error terms $N_{h_k}$ which are common accross different $j$:

\begin{align}
	N_j = \sum_{k=1}^K \beta_{kj}N_{h_k} + N_{x_j}
\end{align}	

There are two difficulties with this multiple hypotheses strategy, one of which we have touched upon (dependence of hypotheses). Before exploring these difficulties further, we classify the $m$ hypotheses of a multiple hypotheses problem according to the ground truth of the hypothesis and whether we reject them or not:

\begin{center}
\begin{tabular}{c|c|c|c}
\textbf{truth} & \textbf{rejected}  &  \textbf{not rejected} & \textbf{total} \\
\hline
\hline
 null        & $F$         & $m_0 -F $ & $m_0$ \\
 \hline
 alternative & $T$         &$m_1-T$    & $m_1$\\
 \hline
 \hline
 total       & $S$         & $m-S$     & $m$\\
\end{tabular}
\end{center}

\begin{enumerate}[1.]
	\item \textbf{Multiple hypotheses}: When we perform a single hypothesis $\phi$ and reject the null if the p-value is below a certain threshold we are controling the probability of making a Type I error:
	
	\begin{align}
			\mathbb{P}[\phi(X)=1 | H_0] \leq \alpha
	\end{align}	
	
	where $X$ is the sample upon which the hypothesis is based. If we perform multiple hypotheses $\phi_i$ and reject analogously this corresponds to controlling the \emph{false positive rate} (FPR):

	\begin{align}
			FPR = \frac{\mathbb{E}[F]}{m} \leq \alpha
	\end{align}
	
	However this does not provide effective control of the number of Type I errors we make. Typically, this is fixed by controlling the \emph{familiy wise error rate} (FWER) instead of the false positive rate:

	\begin{align}
			FWER = \mathbb{P}[F \geq 1] \leq \alpha
	\end{align}

	This can be achieved with the p-values  by using the \emph{Bonferroni correction}. Other methods exist which also control the FWER while increasing the power of the overall procedure. However, for many high dimensional applications this compound error measure is too strict because in these cases we often don't care that we make more than a few Type I errors so long as the proportion of these errors $F$, relative to the total number rejected hypotheses $S$ is small. This is the case when we try and find the variables $x_j$ that include the signature of a certain variable $c_l$. So long as the proportion of \emph{false discoveries} $F$, those variables $x_j$ which we declare contain the signature of $c_l$ but which in reality don't, are a small proportion of the total \emph{discoveries} $S$ then we are satisfied. This is why the SVA methodology, in performing the $m=J$ tests as described in Section \ref{sigMeth} seeks to control the \emph{false discovery rate} (FDR):
	
	\begin{align}
				FDR = \mathbb{E}\bigg[\frac{F}{S \vee 1}\bigg] =  \mathbb{E}\bigg[\frac{F}{S} \bigg|S>0\bigg]\mathbb{P}[S>0] \leq \alpha
	\end{align}
	
	Where the "$\vee$ 1" part takes care of the possibility that $S=0$ by setting the whole quotient to zero.
	
	\item \textbf{Dependent hypotheses tests}: We actually know that since the $x_j$ variables depend on $y$ for some $j \in \{1,...,J\}$ the hypotheses are dependent and so any joint significance analysis, which controlls either the FWER or the FDR, will be invalid in principle. However, if the signal $f_h(y)$ is weak compared to that of $c_l$ then the significance may still be valid.  To gauge whether this is the case we examine the p-value distribution. Qualitatively speaking it is simple to distinguish between valid and invalid p-value distributions. We may assume that each p-value has the following mixture distribution:
	
	\begin{align} \label{mixture}
		P_i \sim \pi_0 F_0 + \pi_1 F
	\end{align}	
	
	where
	\begin{itemize}
		\item $\pi_0$ is the proportion of null hypothesis tests $H_i$,
		\item $\pi_1 = 1- \pi_0$ is the proportion of alternative hypothesis tests $H_i$,
		\item $\mathcal{F}_0$ is the distribution of a uniform random variable with support on $[0,1]$ since assuming a one-sided hypothesis test we have that:
		
			\begin{align} \label{uniform}
				\mathbb{P}[P_i \leq p | H_0] = \mathbb{P}[\mathcal{F}_0(T) \leq p ] = \mathbb{P}[T \leq \mathcal{F}_0^{-1}(p) ] = \mathcal{F}_0(\mathcal{F}_0^{-1}(p))= p 
			\end{align}	
		
		and,
		\item $\mathcal{F}$ is a right-skewed probability distribution since it is the distribution of alternative p-values which have a greater probability of being on the low end of the $[0,1]$ interval. 
	\end{itemize}	
\end{enumerate}	

\begin{figure}[H]
  \begin{subfigure}{.4\textwidth}
    \centering
	\includegraphics[width=1\textwidth]{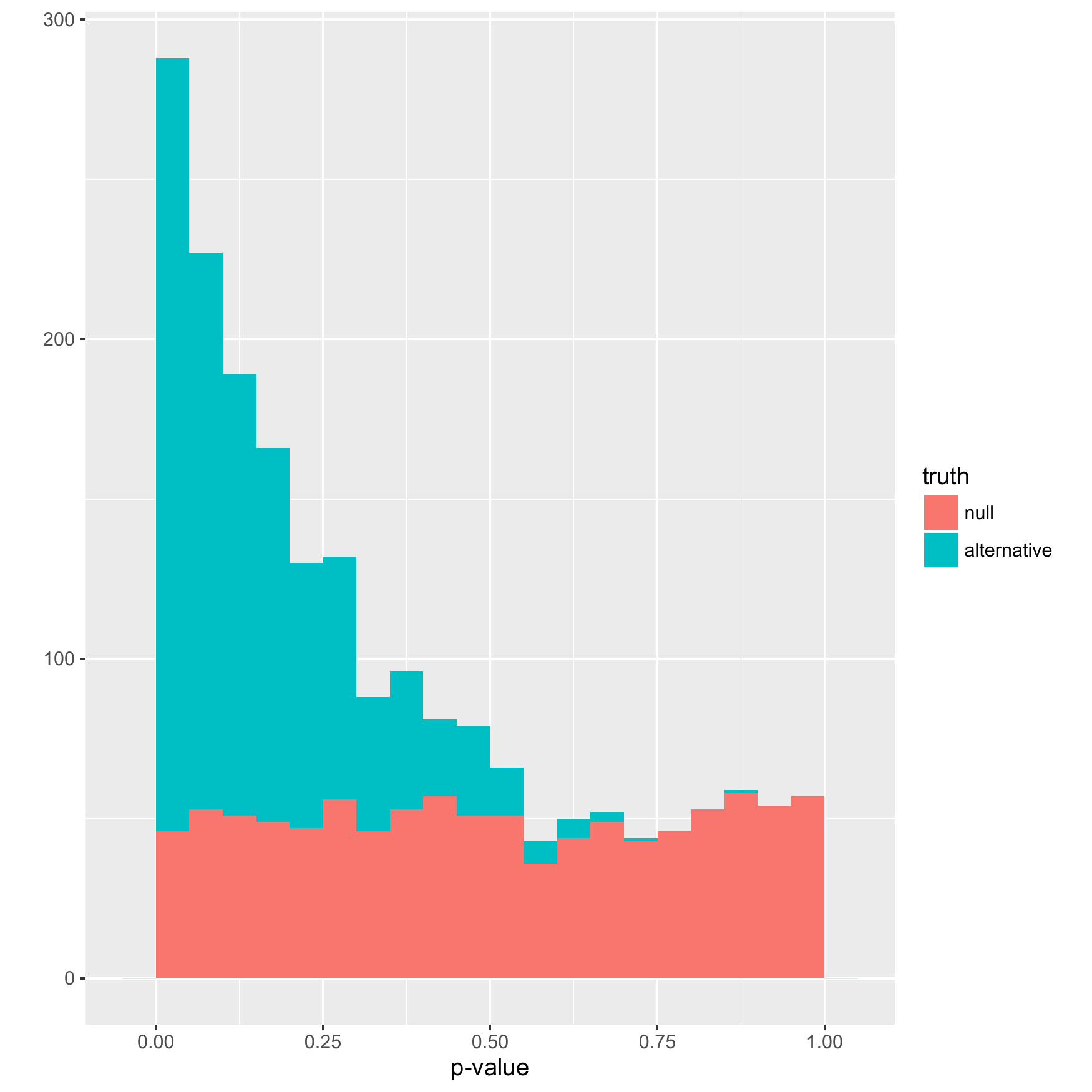} 
    \caption{valid p-value distribution}
    \label{fig:pvalue_val}
  \end{subfigure}%
  \begin{subfigure}{.4\textwidth}
    \centering
    \includegraphics[width=1\textwidth]{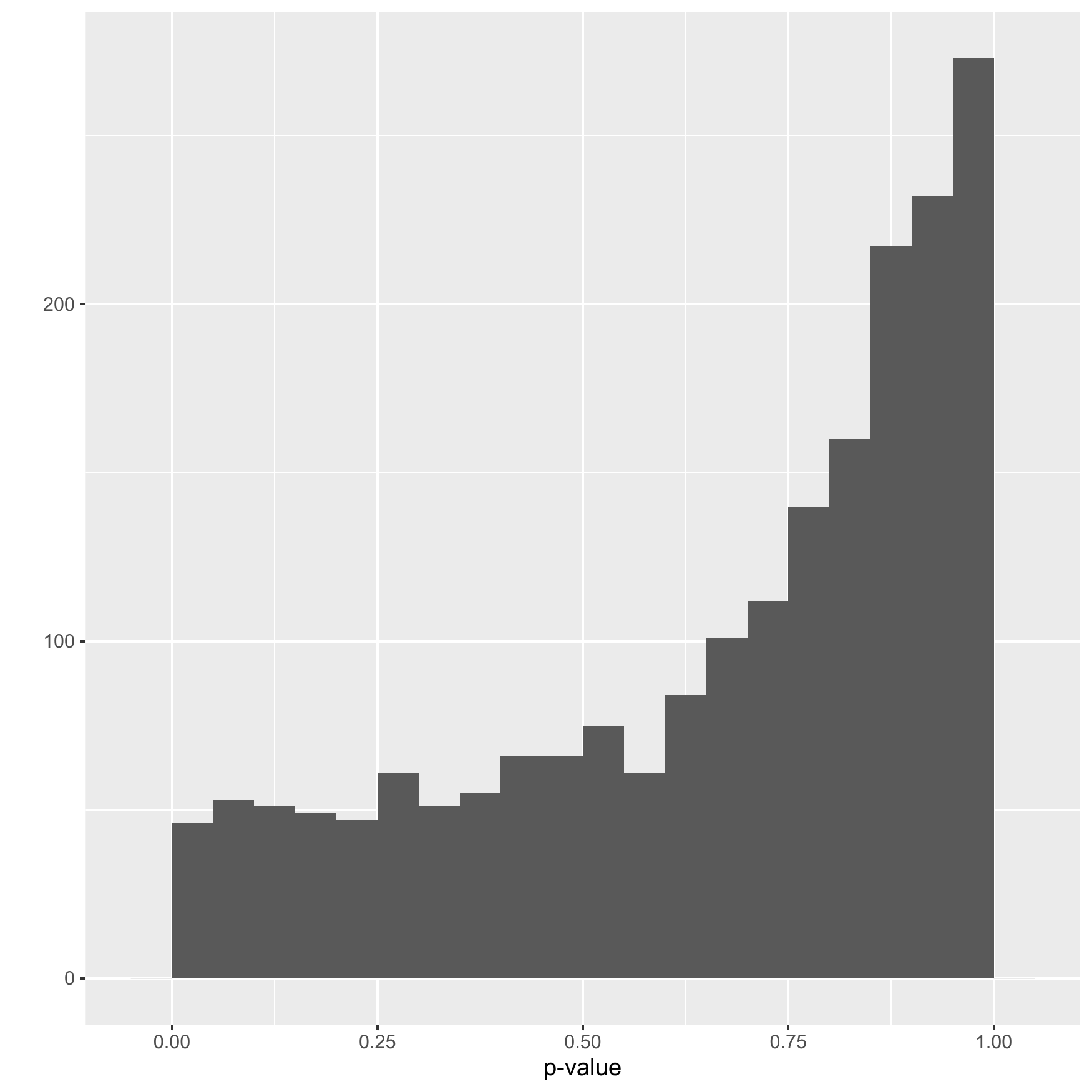}
    \caption{invalid p-value distribution}
    \label{fig:pvalue_inval}
  \end{subfigure}
  \caption[examples of p-value distribution]
  {examples of p-value distribution}
  \label{fig:pvalue}
\end{figure}

Graphically the a valid p-value distribution should be the superposition of a uniform distribution and a right-skewed distribution. Figure \ref{fig:pvalue} shows examples of valid and invalid p-distributions.

The SVA methodology uses a \emph{point-based} method - to be explored in Section \ref{FDR} - for controlling the FDR which:

\begin{enumerate}[I.]
	\item Is still valid under \emph{weak dependence} of the hypotheses tests (see remark D of \cite{statSig} or  \cite{weakDep} for a definition of weak dependence) and,
	\item In case of an invalid p-value distribution caused by dependence between the hypotheses tests automatically \emph{detects} this so that the number of variables $x_j$ estimated as having the signature of $c_l$ is set to zero. 
\end{enumerate}	

\subsection{FDR based significance analysis} \label{FDR}

Two related quantities to the FDR are the \emph{positive false discovery rate} (pFDR) 

	\begin{align}
				 pFDR= \mathbb{E}\bigg[\frac{F}{S} \bigg|S>0\bigg]    
	\end{align}

and the \emph{marginal false discovery rate} (mFDR) 

	\begin{align}
		mFDR= \frac{\mathbb{E}[F]}{\mathbb{E}[S]}
	\end{align}	
	
As \citeauthor{FDR} surmises, there is some disagreemnt as to which quantity is more adequate to control however, for settings when there is a large number of hypothesis and the probability that $S=0$ is negligible FDR, pFDR and mFDR are similar. 

There are two approaches to control FDR:

\begin{enumerate}[1.]
	\item \textbf{FDR Control}. Fix FDR at a certain level $\alpha$ and come up with a data-dependent thresholding rule such as the \emph{FDR controlling algorithm} proposed by \citeauthor{Simes} and proven to control FDR at level $\alpha$ by \citeauthor{BenjHoch}.  
	\item \textbf{Point estimate of FDR}. Fix a p-value threshold $t$ for rejecting hypothesis and estimate the corresponding $FDR(t)$ conservatively. \emph{Q-values} and \emph{local false discovery rates} (lFDR) represent two point estimate approaches for controling FDR. 
\end{enumerate}	

\citeauthor{weakDep} show that under a certain choice for the estimation of $m_0$ and threshold $t$ the FDR controlling algorithm and q-value approach are equivalent.  

We first focus on the q-value point estimation method and then go on to explore the local false discovery rate which is used in the  \Rp package \verb|sva| to control the FDR. Let:

\begin{align}
	F(t) &= \#\{\textrm{null } P_i \leq t: i=1,...,m\}\\
	S(t) &= \#\{ P_i \leq t: i=1,...,m\}
\end{align}	

Where $P_i$ are i.i.d. random variables. We want to estimate FDR in terms of the threshold $t$. For $m$ large we have that

\begin{align}
	FDR(t) \approx pFDR(t) \approx mFDR(t) = \frac{\mathbb{E}[F(t)]}{\mathbb{E}[S(t)]}
\end{align}	

We can estimate $\mathbb{E}[S(t)]$, the expected number of significant variables simply as the observed number of significant variables:

\begin{align}
	\hat{\mathbb{E}}[S(t)] = \#\{p_i \leq t, i=1,...,m \}
\end{align}	

Now

\begin{align}
	\mathbb{E}[F(t)] = \mathbb{E}[\#\{\textrm{null } P_i \leq t\}] = m_0t
\end{align}	

since $P_i|H_0 \sim U[0,1]$. We can estimate $m_0$, the total number of null hypotheses, in terms of the total number of hypotheses $m$ and the proportion of null hypotheses $\pi_0$:

\begin{align}
	\hat{m}_0 = \hat{\pi}_0m
\end{align}	

We estimate $\pi_0$ using the p-value histogram. P-values greater than a certain threshold, say $\lambda$, will be mostly null p-values. We can estimate the proportion of null p-values as the height of the histogram for values larger than $\lambda$ divided by the average height of the histogram. 

	 \begin{figure}[H]
	   \centering
	   \includegraphics[width=.7\textwidth]{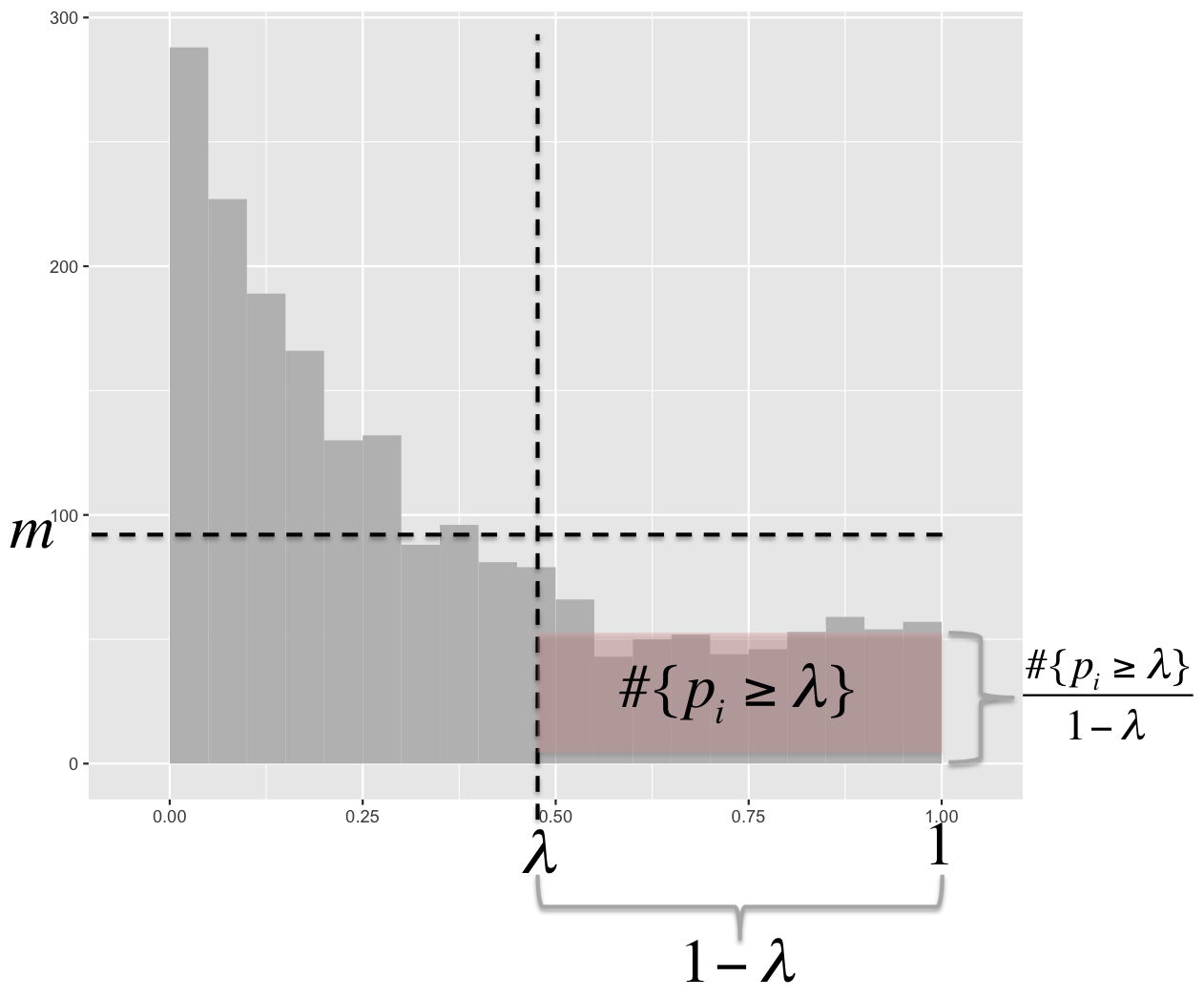} 
	   \caption[estimation of $\pi_0$]
	   {estimation of $\pi_0$}
	   \label{fig:estim}
	 \end{figure}

\begin{align}
	\hat{\pi}_0(\lambda) = \frac{\#\{p_i > \lambda\}}{m(1-\lambda)}
\end{align}	
	
	To the right of some point $\lambda$ close to 1 there are almost only null p-values so we could use this $\lambda$ to calculate $\pi_0$. However, the closer we get to $\lambda=1$ the less data points we use to compute $\hat{\pi}_0$ so the the higher the variance of the estimator: there is a bias-variance trade-off in choosing $\lambda$. For the histogram of figure \ref{fig:estim} we may plot the estimate $\hat{\pi}_0$:

	 \begin{figure}[H]
	   \centering
	   \includegraphics[width=.7\textwidth]{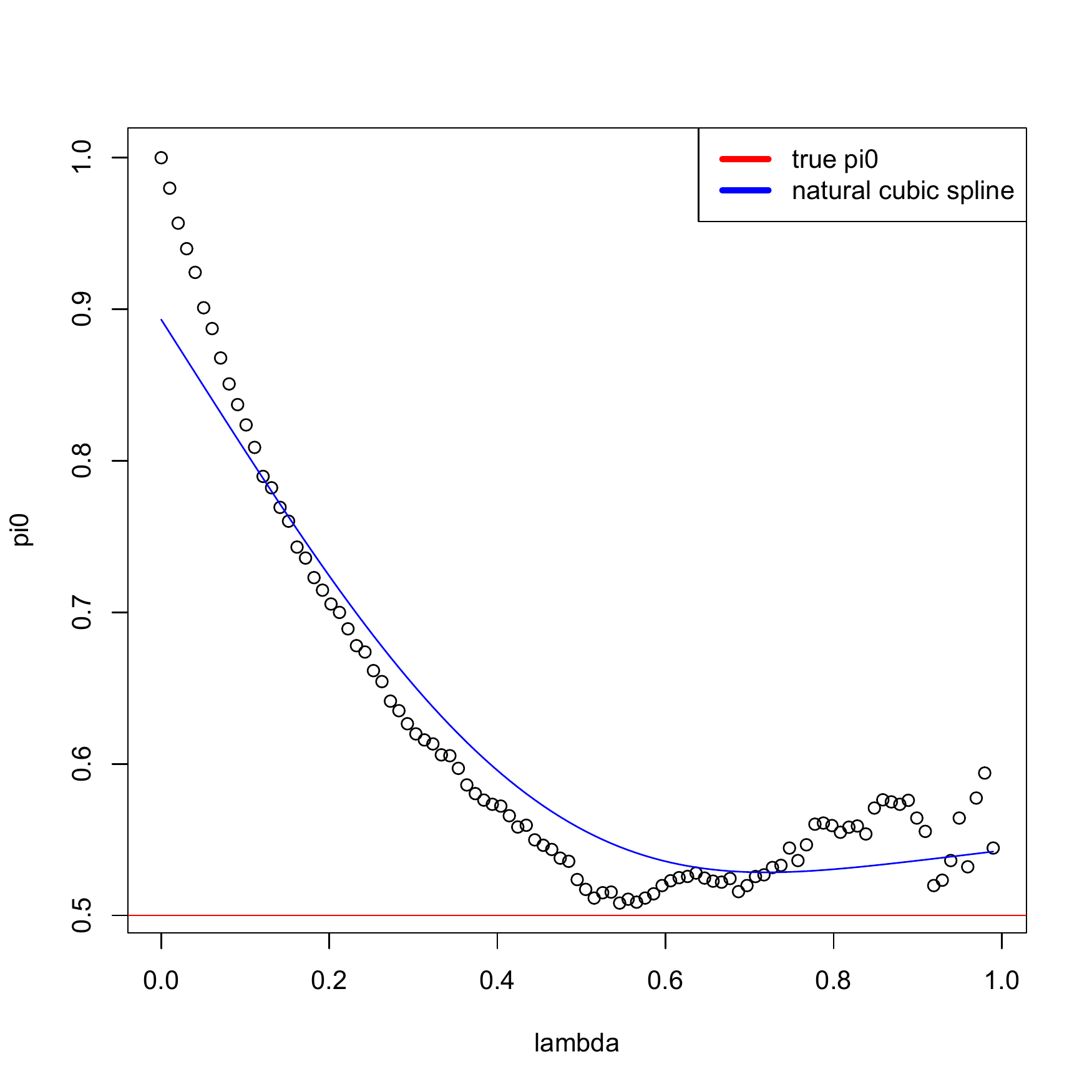} 
	   \caption[$\hat{\pi}_0$ as a function of $\lambda$]
	   {$\hat{\pi}_0$ as a function of $\lambda$}
	   \label{fig:estim2}
	 \end{figure}

As prescribed in \cite{statSig} a natural cubic spline $\hat{f}(\lambda)$  is fitted to $\hat{\pi}_0(\lambda)$ and $\pi_0$ is estimated as $\hat{f}(1)$. This method borrows strength accross $\hat{\pi}_0(\lambda)$ giving a balance between variance and bias as is explained in remark B of \cite{statSig}.

Notice that if we have an invalid p-value distribution such as that of figure \ref{fig:pvalue_inval} then $\hat{\pi}_0 >= 1$. If we set values larger than 1 to 1, then as mentioned at the end of Section \ref{findSig} we have a way of automatically detecting invalid p-value distributions and setting the number of discoveries to zero. 

Going back to our estimate of $FDR(t)$ we have:

\begin{align}
	\widehat{FDR}(t) = \frac{\hat{\mathbb{E}}[F(t)]}{\hat{\mathbb{E}}[S(t)]} = \frac{m \hat{\pi}_0 t}{\#\{p_i \leq t, i=1,...,m \}}
\end{align}	

Notice that if we threshold p-values with $t=1$ then $FDR(1)=\hat{\pi}_0$, in other words if we fail to reject all hypothesis our FDR will be $\hat{\pi}_0$. 

Since $FDR(t)$ is a non-increasing function, we can't simply choose a level $\alpha$ and then find a $t$ such that if we threshold the p-values with a value of $t$ or lower we guarantee that that $FDR(t) \leq \alpha$. This means the p-value loses its meaning of expressing the significance of each feature because a feature might have a lower p-value, but thresholding at the lower p-value might induce a higher FDR. For illustrative purposes suppose we have 6 hypothesis tests with the following p-values which we have ordered for convenience:

\begin{center}
\begin{tabular}{|c|c|}
\hline
\textbf{Hypothesis} & \textbf{p-value}  \\
\hline
\hline
 $p_0:=$        & 0\\
 \hline
 $p_1$ & 0.05 \\
 \hline
 $p_2$ & 0.1 \\
 \hline
 $p_3$ & 0.2 \\
 \hline
 $p_4$ & 0.35 \\
 \hline
 $p_5$ & 0.6 \\
 \hline
 $p_6$ & 0.85 \\
 \hline
 $p_7:=$ & 1 \\
 \hline
 \hline
\end{tabular}
\end{center}

	 \begin{figure}[H]
	   \centering
	   \includegraphics[width=.7\textwidth]{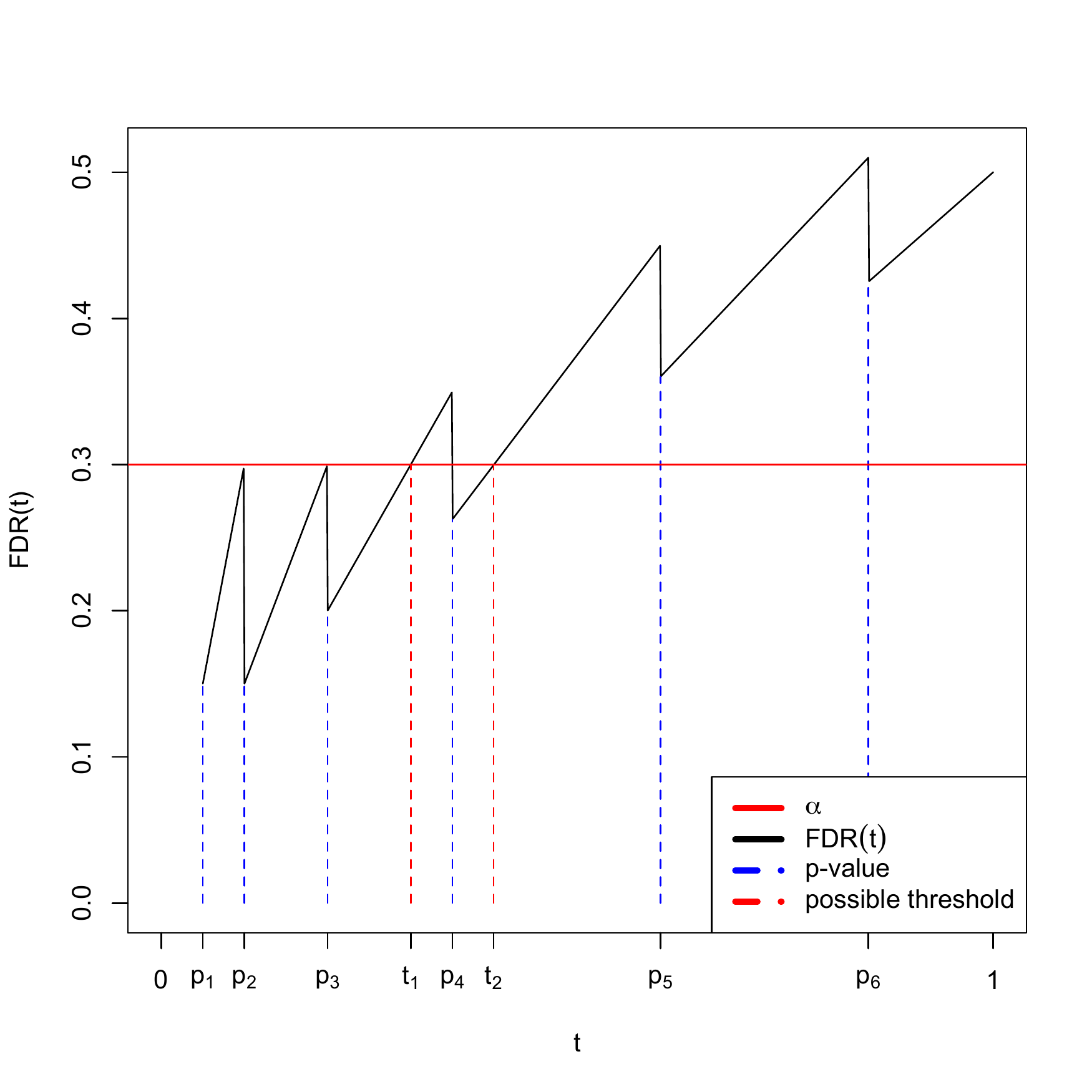} 
	   \caption[$FDR(t)$]
	   {$FDR(t)$}
	   \label{fig:FDRt}
	 \end{figure}

We can see that we could threshold p-values at $t_1$ or $t_2$ and obtain the same $FDR$. We need a new definition for a value which expresses the significance of each feature. This value is called the $\emph{q-value}$. The q-value of a feature is the minimum FDR that can be attained if we include a feature (call it significant):

\begin{align}
	q_i = q(p_i) := \min_{t \geq p_i} FDR(t) = \min_{t \geq p_i} pFDR(t)
\end{align}	

This definition ensures we can use q-values to control FDR since we choose the thresholding value that minimizes FDR and in doing so implicitly define a strictly increasing function $q^*$ which we can use to control FDR. Define the function $q^*$ as:

\begin{align}
	q^*(t) :=  \inf_{s \in [t,1]} \widehat{FDR}(s)
\end{align}	

Also note that

\begin{align}
\widehat{FDR}(t) = \frac{m \hat{\pi}_0 t}{\#\{p_i \leq t, i=1,...,m \}} = m \hat{\pi}_0 t \sum_{i=1}^m \frac{1}{i}\mathbbm{1}_{[p_{(i)},p_{(i+1)}]}(t)
\end{align}

for $t \in [p_{(1)},1]$ and

\begin{align}
	\frac{d\widehat{FDR}(t)}{dt} = m \hat{\pi}_0 \sum_{i=1}^m \frac{1}{i}\mathbbm{1}_{[p_{(i)},p_{(i+1)}]}(t)
\end{align}	

for $t \in [p_{(1)},1] \setminus \{p_1,...,p_m\}$

We can see how $q*(t)$ can be used to set the p-value threshold $t$ and to obtain the significance q-values for each feature graphically. 

	 \begin{figure}[H]
	   \centering
	   \includegraphics[width=.7\textwidth]{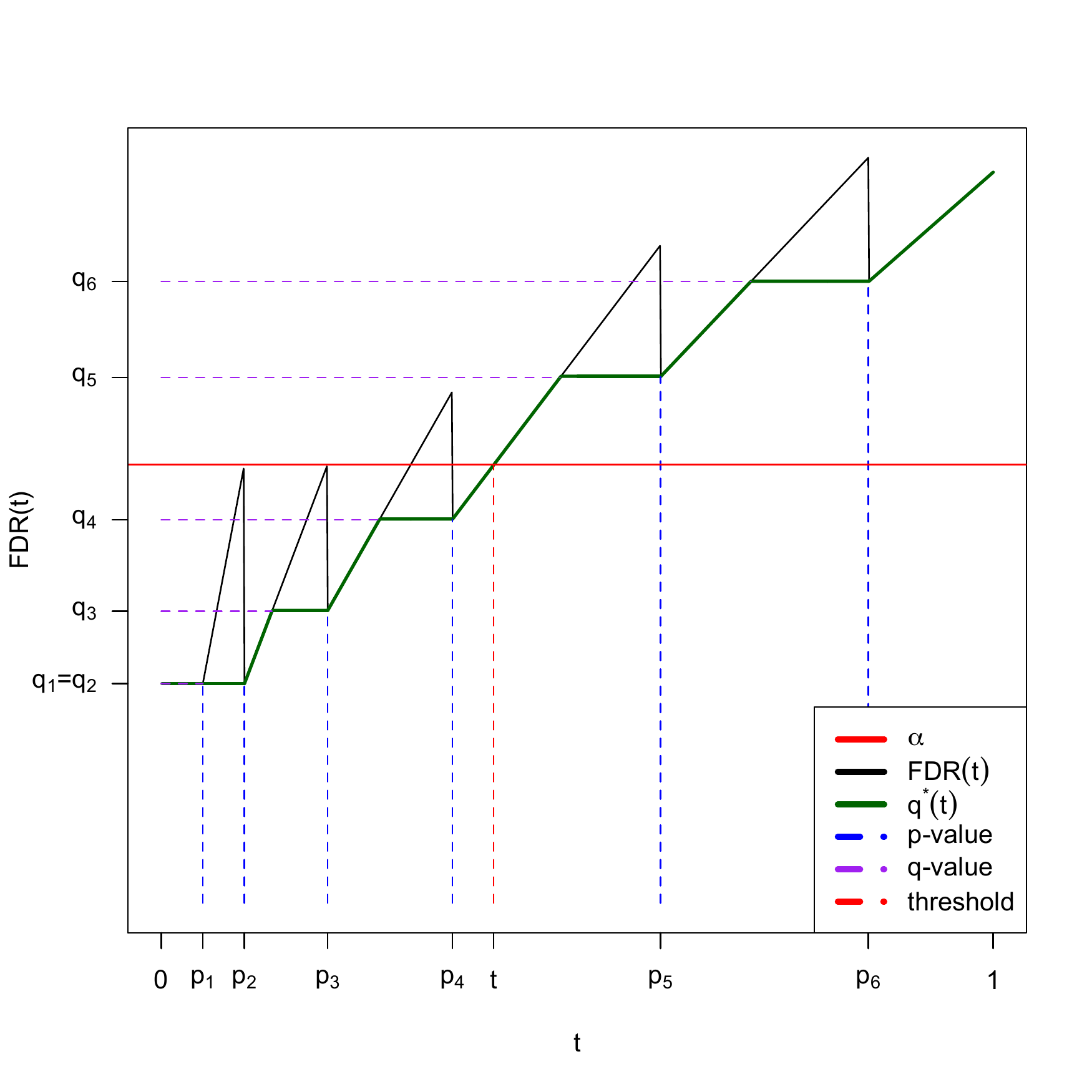} 
	   \caption[$q^*(t)$]
	   {$q^*(t)$}
	   \label{fig:qt}
	 \end{figure}

Note that since:

\begin{align}
	p_i \leq t \iff q_i = q^*(p_i) \leq q^*(t) \leq \widehat{FDR}(t) = \alpha \iff q_i \leq \alpha
\end{align}	

we can threshold the q-values themselves at $\alpha$ to control the $FDR$ at this level. 

To find the variables $x_j$ which contain the signature of $c_l$ then, one could either choose the $\hat{m}_0=\hat{\pi}_0*m$ variables $x_j$ with the lowest p-values for the hypothesis $H_{j0}:\beta_{j1}=0$ or equivalently calculate the q-values and threshold at a the desired level $\alpha$. For reasons that are not clear to the author, in the implementation of the \Rp package \verb|sva| (specifically function \verb|edge.lfdr|), another \emph{point estimate} approach is used based on \emph{local false discovery rates} (lFDR). This approach follows from a Bayesian interpretation of q-values.

In Theorem 1 of \cite{qValues}, \citeauthor{qValues} proves that the positive false discovery rate for a given threshold $t$ for the p-value, $pFDR(t$) is equal to the conditional probability that the corresponding hypothesis is null given that the p-value is below the threshold:

\begin{align}
	pFDR(t)= \mathbb{E}\bigg[\frac{F(t)}{S(t)} \bigg|S(t)>0\bigg] = \mathbb{P}[H_i=0|P_i \leq t]
\end{align}	

This allows us to interpret the $pFDR(t)$ as a posterior Bayesian Type I error where the prior probability of a hypotheses being null is $\pi_0$. Since

 \begin{align}
 	q_i = q(p_i)  = \min_{t \geq p_i} pFDR(t)
 \end{align}

we can interpret the q-value as a posterior Bayesian  p-value. We can then express $pFDR(t)$ as:

\begin{align}
	pFDR(t)=  \mathbb{P}[H=0|P \leq t] = \int \mathbb{P}[H = 0 | P=p]dG(p|p\leq t)
\end{align}	

Where $G$ is the distribution of $P_i$ from \ref{mixture}. We then define the local false discovery rate $lFDR(p)$ of a hypothesis as:

\begin{align}
  lFDR(p) := \mathbb{P}[H = 0 | P=p]
\end{align}	

Notice that this is not a function of the p-value thresholding level $t$ and so it gives a measure of significance of the p-value without taking into account the multiple hypotheses being carried out. Using Bayes' Theorem we have 

\begin{align}
	lFDR(p) = \mathbb{P}[H = 0 | P=p]  = \frac{f_{P_0}(p) \pi_0}{f_P(p)} = \frac{\pi_0}{f_P(p)}
\end{align}	

where
\begin{itemize}
	\item $f_P(p)$ is the density of the p-values and
	\item $f_{P_0}(p)=1$ is the density of the null p-values which we know is uniform by \ref{uniform}. 
\end{itemize}	

To estimate $lFDR(p)$ we estimate $\pi_0$ as was illustrated in figures \ref{fig:estim} and \ref{fig:estim2}. To estimate $f_P(p)$  standard kernel-density estimators can be used however these tend to not perform very well for random variables with bounded support, and so in the \Rp package \verb|sva| (specifically function \verb|edge.lfdr|), the p-values are first transformed so that the underlying random variable has unbounded support. A standard change of variable theorem (see for example pg 153 of \cite{DeGroot}) is then used to calculate the density of the P-values in terms of the density of the transformed P-values which we an estimate effectively using kernel-density estimation. Let $S = \Phi^{-1}(P)$ where $\Phi$ is the standard normal cumulutive distribution function, with associated $\phi$ probability density function, then:

\begin{align}
	f_P(p) = \frac{f_S(s)}{\phi(s)} = \frac{f_S(\Phi^{-1}(p))}{\phi(\Phi^{-1}(p))} 
\end{align}	

The estimate for $lFDR(p)$ is then:

\begin{align}
	\widehat{lfDR}(p) = \frac{\hat{\pi}_0}{\hat{f}_P(p)} = \frac{\hat{\pi}_0 \phi(\Phi^{-1}(p))}{\hat{f}_s(\Phi^{-1}(p))}
\end{align}

\chapter{Simulation Expermiments} \label{ch:sims}

\section{Methods to be compared} \label{sec:methods}

Given $n$ i.i.d. observations of $y$ and $x_j$ for $j \in \{1,...,J\}$ simulated from model \ref{eq:model}, following \citeauthor{SVA}, we will fit 4 different models to the data:

\begin{align}
	 x_j &= f(y) + \epsilon_j\\
	 x_j &= f_{x_j}(y) + \sum_{k=1}^{K-1} \beta_{kj}\hat{h}_k^{svdx} + \epsilon_j\\
	 x_j &= f_{x_j}(y) + \sum_{k=1}^M \beta_{kj}\hat{h}_k^{svdr} + \epsilon_j\\
	 x_j &= f_{x_j}(y) + \sum_{k=1}^N \beta_{kj}\hat{h}_k^{sva} + \epsilon_j\\
\end{align}	

for $j \in \{1,...,J\}$ where

\begin{itemize}
	\item $f(y)$ and $f_{x_j}(y)$ are polynomials of the form $a_1 y + a_2 y^2 + ... + a_n y^n$, 
	\item $\hat{h}_k^{svdx}$ is the estimate of $h_k$ obtained by performing SVD on the observation matrix $X$ (instead of on residual matrix $R$) to obtain $K$ factors, as per sections \ref{subsec:SVD} and \ref{subsec:PA}. The factor $h_k$ with largest absolute correlation with $y$ is removed and the remaining $K-1$ factors are included as covariates. The idea is that if we exclude the strongest part of the $y$ signal we will only be left with the unobserved $h_k$ signal, however the other factors may still include undersirable $f_{x_j}(y)$ signal and the removed factor could also include desirable $f_{h_k}(y)$ signal,
	\item $\hat{h}_k^{svdr}$ is the estimate of $h_k$ obtained by performing SVD on the observation matrix $R$ as in section \ref{subsec:SVD} to obtain $M$ estimates for the $c_l$ factors. The estimates for $c_l$ are plugged in as the $h_k$ estimates. In this case we completely filter out all $y$ signal including $f_{h_k}(y)$ signal meaning that $\hat{h}_k^{svdr}$ are uncorrelated to $y$ which may lead to bias in the estimate of $h_k$, and
	\item $\hat{h}_k^{sva}$ is the estimate of $h_k$ obtained by performing SVA on the observations of $y$ and $x_j$.
	
\end{itemize}	

The four models are forthwith refered to as \emph{vanilla} (or \emph{van}), SVDX, SVDR and SVA.  The main objective is to see if the SVA method has significant advantages in estimating $f_{x_j}(y)$ for $j \in \{1,...,J\}$.  This also entails that valid significance analysis is produced for the joint hypotheses tests $H_{j0}: f_{x_j}(y) = 0$.

\section{Low-dimensional experiments} \label{sec:lowdim}

\subsection{Design} \label{lowDimDesign}

As a first approximation we simulated $n=100$ observations of the following low-dimensional SEM so as to follow the performance of each estimate $\hat{f}_{x_j}$:

\begin{itemize}
	\item $J = K = L = 4$
	\item $y = N_y \sim N(0,1)$
	\item $c_l = N_{c_l} \sim N(0,1)$ for $l \in \{1,...,4\}$
	\item $h_k = f_{h_k}(y) + c_l$ for $k \in \{1,...,4\}$
	\item $f_{h_1}(y)=f_{h_2}(y)=0$
	\item $f_{h_3}(y) = -0.28y + 1.29y^2$
	\item $f_{h_4}(y) = 1.54y + 0.59y^2$
	\item $x_j = f_{x_j}(y) + \sum_{k=1}^K \beta_{kj}h_k + N_{x_j}$ for $j \in \{1,...,4\}$
	\item Non-zero $\beta_{kj}$ values were generated randomly with standard normal generator:
	\item $f_{x_1}(y)=f_{x_2}(y)=0$
	\item $f_{x_3}(y) = -0.92y$
	\item $f_{x_4}(y) = 1.24y - 1.48y^2$, and
	\item $N_{x_j} \sim N(0,1)$ for $j \in \{1,...,4\}$
\end{itemize}	

The DAG for this SEM is the following:

	 \begin{figure}[H]
	   \centering
	   \includegraphics[width=.7\textwidth]{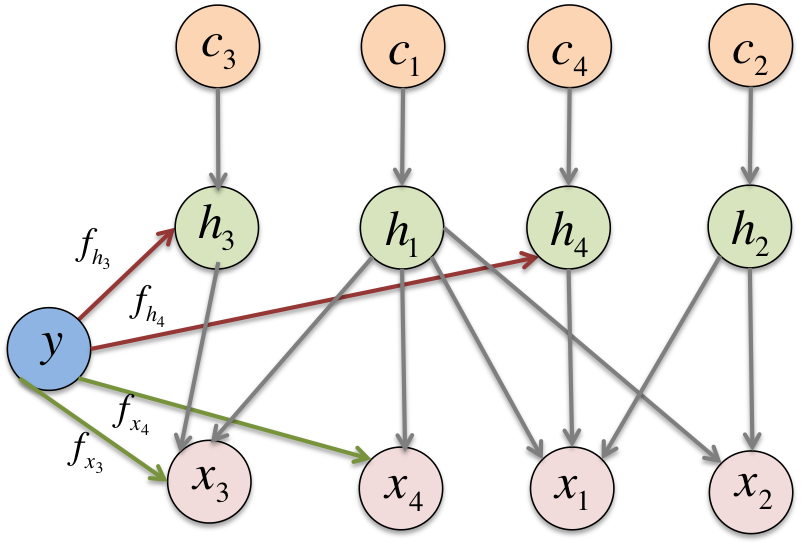} 
	   \caption[DAG of simulated low dimensional SEM]
	   {DAG of simulated low dimensional SEM}
	   \label{fig:lowDimSEM}
	 \end{figure}

We chose a SEM with this DAG so that we get variety in the way $x_j$ depend on $y$:

\begin{enumerate}[1.]
	\item $x_1$: depends on $y$ only through $f_{h_4}$,
	\item $x_2$: doesn't depend on $y$,
	\item $x_3$: depends on $y$ through $f_{h_3}$ and $f_{x_3}$, and
	\item $x_4$: depends on $y$ only through $f_{x_4}$.
\end{enumerate}	

Moreover for the estimation of $f_j(y)$ and $f_{x_j}(y)$ we chose polynomials of the form:

\begin{align}
	g(y) = a_1y + a_2y^2
\end{align}	

\subsection{Result for one repetition} \label{lowDim1Rep}

For $x_j$ with $j \in \{1,2,3,4\}$ we display the results of the simulation with the following graphs:

\begin{itemize}
	\item A comparison of $\hat{f}_{x_j}(y)$ using SVA to the real $f_{x_j}$,
	\item A comparison of $\hat{\mathbb{E}}[x_j|y,\hat{h}_k^{sva}]$ for the vanilla and SVA methods,
	\item A comparison of the residuals produced with the four methods, van, SVDX, SVDR and SVA, and
	\item A Tukey-Anscombe residual plot, with $\sum_{k=1}^K \beta_{kj}h_k$ on the x-axis.
\end{itemize}

\subsubsection{$\hat{f}_{x_j}(y)$ using SVA}

\begin{figure}[H]
  \begin{subfigure}{.5\textwidth}
    \centering
	\includegraphics[width=1\textwidth]{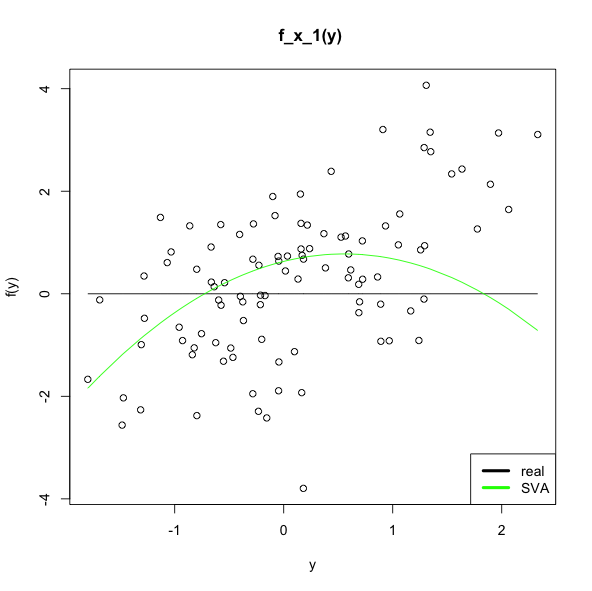} 
    \caption{$x_1$}
    \label{fig:fX1}
  \end{subfigure}%
  \begin{subfigure}{.5\textwidth}
    \centering
    \includegraphics[width=1\textwidth]{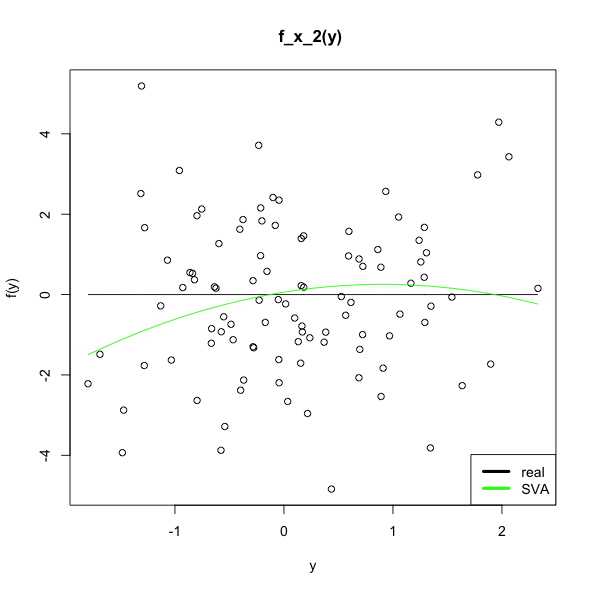}
    \caption{$x_2$}
    \label{fig:fX2}
  \end{subfigure}
  \begin{subfigure}{.5\textwidth}
    \centering
    \includegraphics[width=1\textwidth]{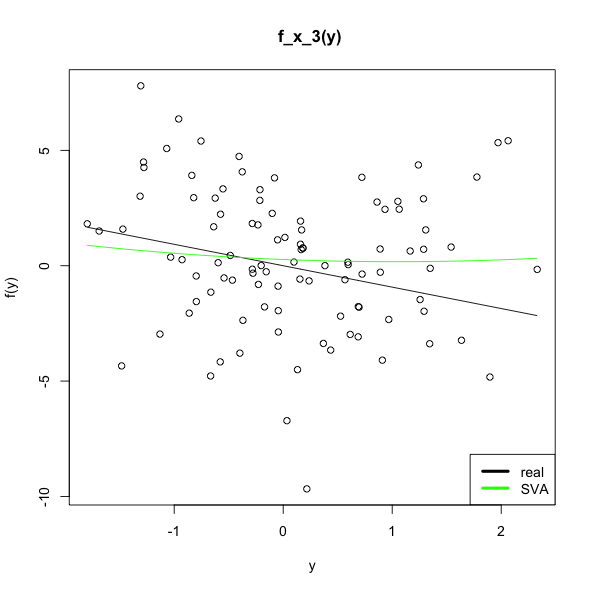}
    \caption{$x_3$}
    \label{fig:fX3}
  \end{subfigure}
  \begin{subfigure}{.5\textwidth}
    \centering
    \includegraphics[width=1\textwidth]{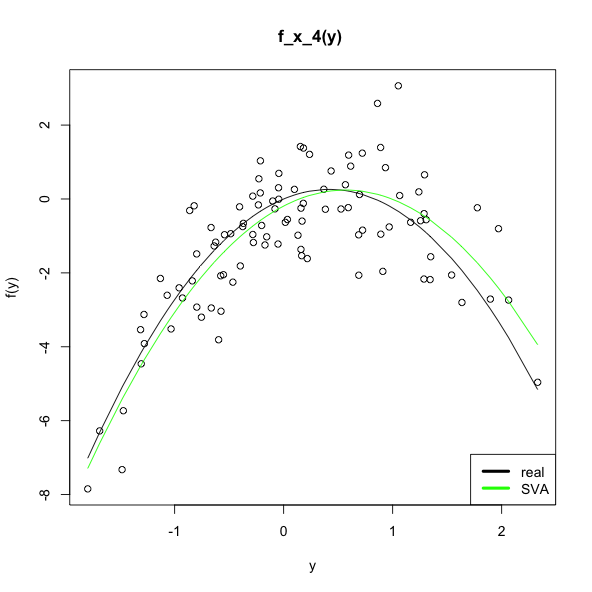}
    \caption{$x_4$}
    \label{fig:fX4}
  \end{subfigure}
  \caption[Comparison of $\hat{f}_{x_j}(y)$ using SVA to the real $f_{x_j}$]
  {Comparison of $\hat{f}_{x_j}(y)$ using SVA to the real $f_{x_j}$}
  \label{fig:fXj}
\end{figure}

We can see that the SVA method does fairly well in estimating $f_{x_j}$ when it has similar complexity to the polynomial being used to estimate it (in this case a quadratic polynomial). This is because the SVA method does not apply regularization since this would not be feasible for high dimensional cases.  Since we are using quadratic polynomials to estimate first, $f_{j}$ and subsequently, $f_{x_j}$, we overfit the data. 

\subsubsection{$\hat{\mathbb{E}}[x_j|y,\hat{h}_k^{sva}]$}

\begin{figure}[H]
  \begin{subfigure}{.5\textwidth}
    \centering
	\includegraphics[width=1\textwidth]{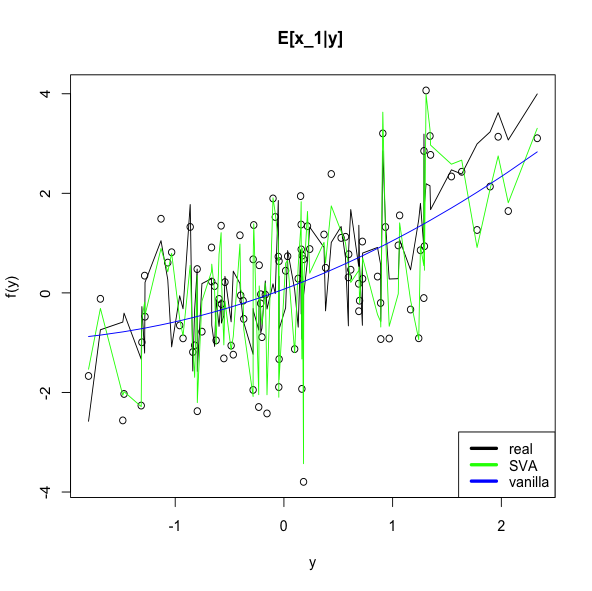} 
    \caption{$x_1$}
    \label{fig:ExpX1}
  \end{subfigure}%
  \begin{subfigure}{.5\textwidth}
    \centering
    \includegraphics[width=1\textwidth]{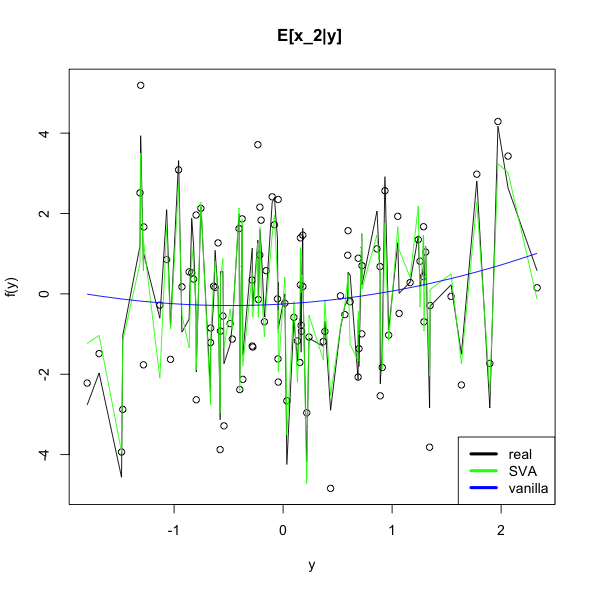}
    \caption{$x_2$}
    \label{fig:ExpX2}
  \end{subfigure}
  \begin{subfigure}{.5\textwidth}
    \centering
    \includegraphics[width=1\textwidth]{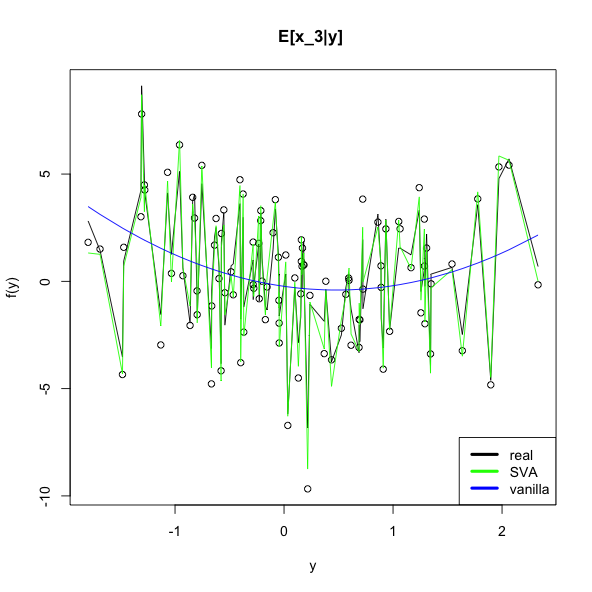}
    \caption{$x_3$}
    \label{fig:ExpX3}
  \end{subfigure}
  \begin{subfigure}{.5\textwidth}
    \centering
    \includegraphics[width=1\textwidth]{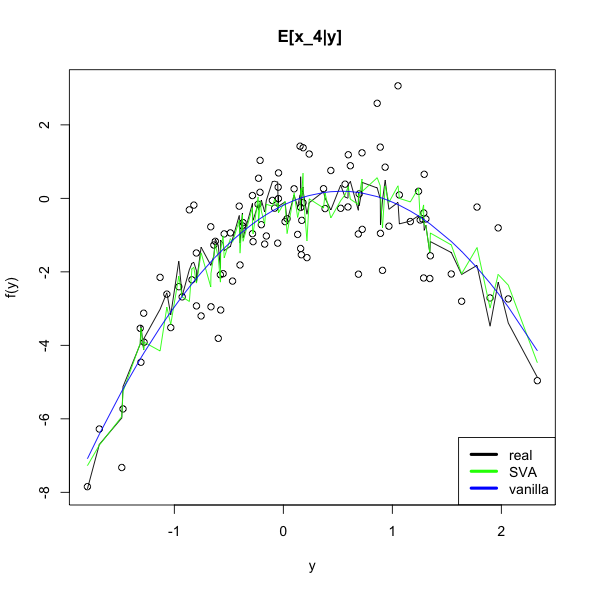}
    \caption{$x_4$}
    \label{fig:ExpX4}
  \end{subfigure}
  \caption[Comparison of $\hat{\mathbb{E}} \mbox{[} x_j|y,\hat{h}_k^{sva} \mbox{]}$ for the vanilla and SVA methods]
  {Comparison of $\hat{\mathbb{E}} \mbox{[} x_j|y,\hat{h}_k^{sva} \mbox{]}$ for the vanilla and SVA methods}
  \label{fig:ExpXj}
\end{figure}

The SVA estimation seems to have very poor performance for $\hat{\mathbb{E}}[x_1|y,\hat{h}_k^{sva}]$ where overfitting was worse. This could be because the overfitting of $f_1(y)$  leads to bad estimation of the $\beta_{k1}$ coefficients to compensate. This shows up in the graph with a green $SVA$ line that is \emph{noisier} than the underlying true  $\mathbb{E}[x_1|y,h_k]$.

\subsubsection{Comparison of residuals for 4 methods}

\begin{figure}[H]
  \begin{subfigure}{.5\textwidth}
    \centering
	\includegraphics[width=1\textwidth]{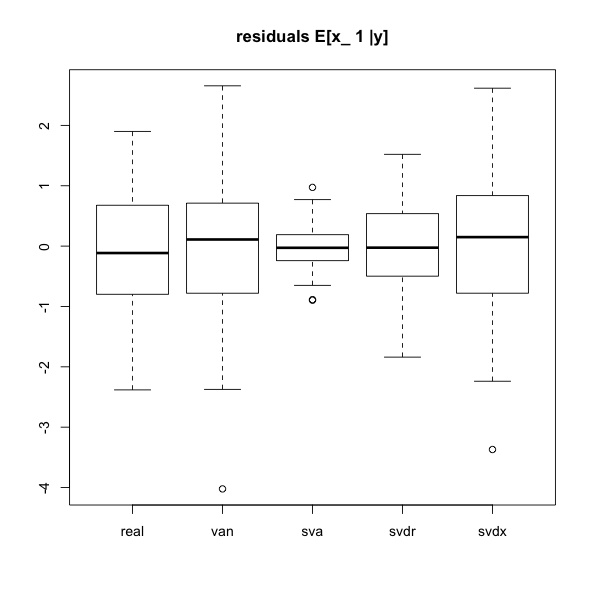} 
    \caption{$x_1$}
    \label{fig:res1}
  \end{subfigure}%
  \begin{subfigure}{.5\textwidth}
    \centering
    \includegraphics[width=1\textwidth]{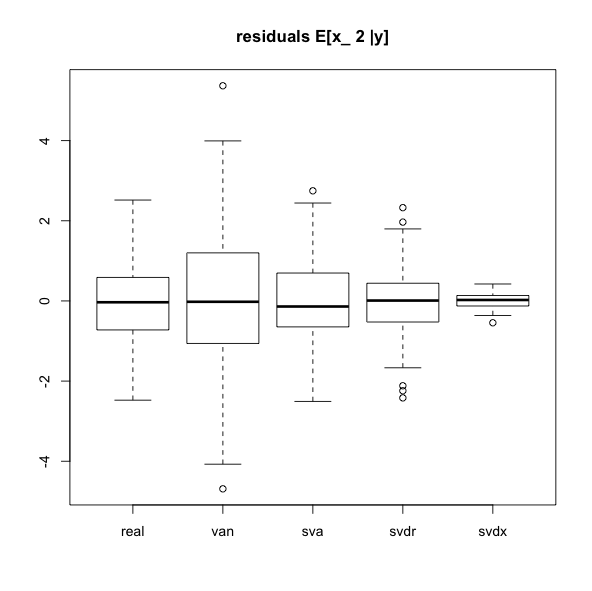}
    \caption{$x_2$}
    \label{fig:res2}
  \end{subfigure}
  \begin{subfigure}{.5\textwidth}
    \centering
    \includegraphics[width=1\textwidth]{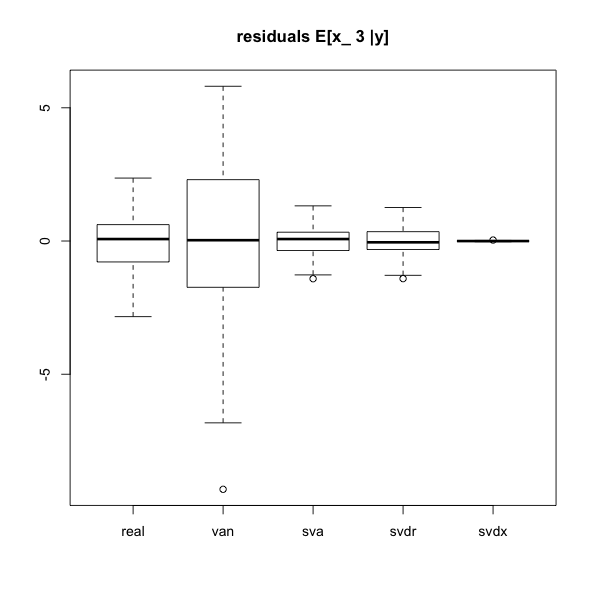}
    \caption{$x_3$}
    \label{fig:res3}
  \end{subfigure}
  \begin{subfigure}{.5\textwidth}
    \centering
    \includegraphics[width=1\textwidth]{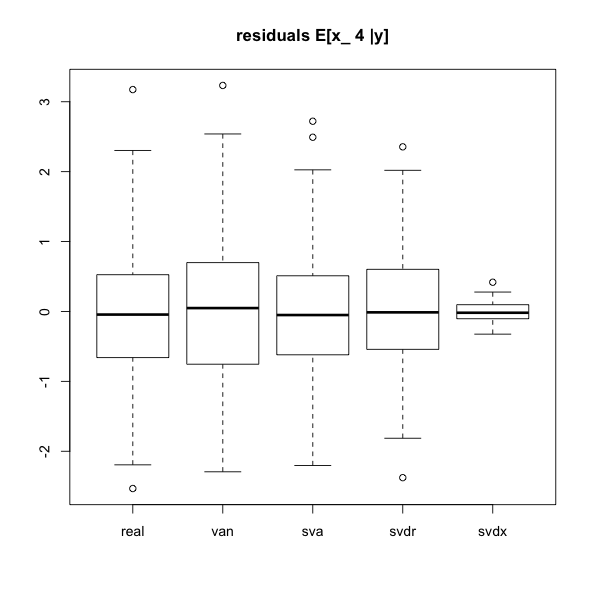}
    \caption{$x_4$}
    \label{fig:res4}
  \end{subfigure}
  \caption[A comparison of the residuals produced with the four methods, van, SVDX, SVDR and SVA]
  {A comparison of the residuals produced with the four methods, van, SVDX, SVDR and SVA}
  \label{fig:res}
\end{figure}

A \emph{good} estimation method would show residuals similar to those of the real errors $N_{x_j}$. None of the methods show consistency in this respect accross the four $r_j$. Again for $r_4$ where the true model complexity is similar to the basis function model fitted SVA appears to do well. 

\subsubsection{Tukey-Anscombe plots}

\begin{figure}[H]
  \begin{subfigure}{.5\textwidth}
    \centering
	\includegraphics[width=1\textwidth]{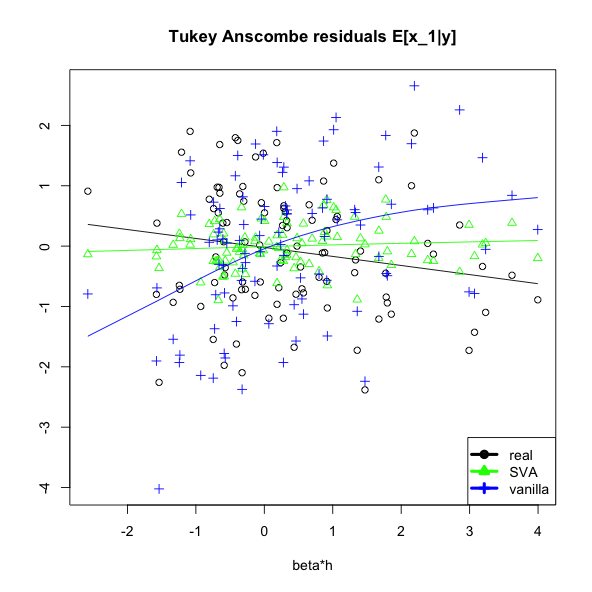} 
    \caption{$x_1$}
    \label{fig:tukey1}
  \end{subfigure}%
  \begin{subfigure}{.5\textwidth}
    \centering
    \includegraphics[width=1\textwidth]{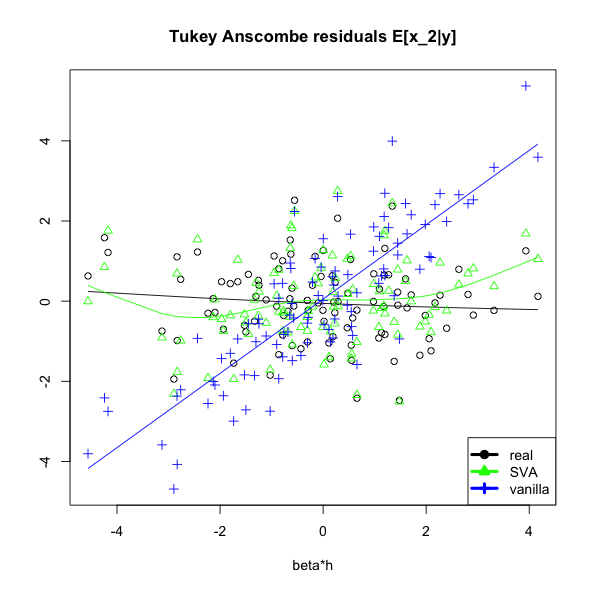}
    \caption{$x_2$}
    \label{fig:tukey2}
  \end{subfigure}
  \begin{subfigure}{.5\textwidth}
    \centering
    \includegraphics[width=1\textwidth]{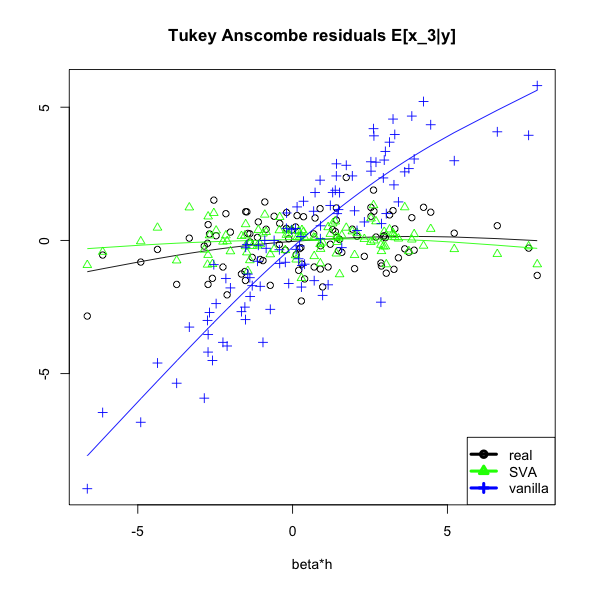}
    \caption{$x_3$}
    \label{fig:tukey3}
  \end{subfigure}
  \begin{subfigure}{.5\textwidth}
    \centering
    \includegraphics[width=1\textwidth]{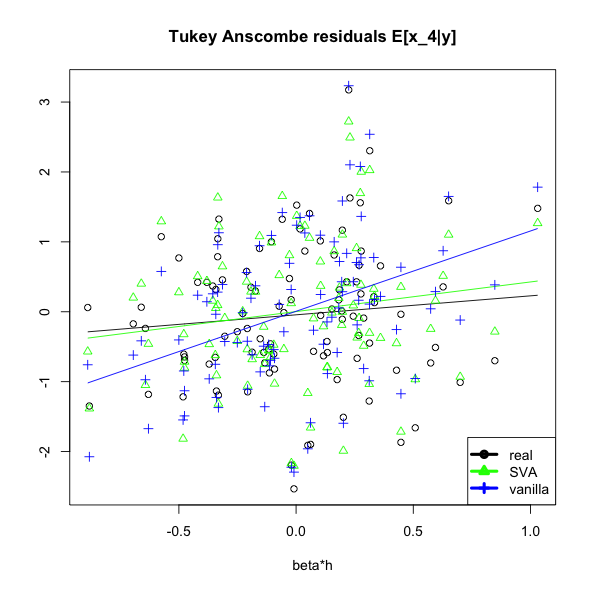}
    \caption{$x_4$}
    \label{fig:tukey4}
  \end{subfigure}
  \caption[Tukey-Anscombe residual plot, with $\sum_{k=1}^K \beta_{kj}h_k$ on the x-axis]
  {Tukey-Anscombe residual plot, with $\sum_{k=1}^K \beta_{kj}h_k$ on the x-axis}
  \label{fig:tukey}
\end{figure}

In general we see that SVA outperforms the vanilla method, and especially for $r_3$ and $r_4$ where the true model complexity is similar to the basis function model being fitted. 

If the significance analysis is done properly, if it is valid, then under the null hypotheses $H_{j0}: f_{x_j}(y)=0$ p-values should be distributed uniformly. In this case, with only one repetition of the experiment, there are only 4 p-values per method (one for each $j$) so that confirming uniformity is not possible. However in some cases it is clear that we can rule out the possibility of the p-values being uniformly distributed. The following table shows the p-values for each $j$ and method. These were obtained by simulating under the null. 

\begin{center}
\begin{tabular}{|c|c|c|c|c|}
\hline
\textbf{$j$} & \textbf{sva} & \textbf{svdx} & \textbf{svdr} & \textbf{van}  \\
\hline
\hline
 1 & 0.820 & 0.000 & 0.000 & 0.000 \\
 \hline
 2 & 0.230 & 0.000 & 0.016 & 0.390 \\
 \hline
 3 & 0.235 & 0.000 & 0.000 & 0.014 \\
 \hline
 4 & 0.804 & 0.000 & 0.016 & 0.055 \\
 \hline
 \hline
\end{tabular}
\end{center}

It seems that only the SVA method produces null p-values which could potentially be uniformly distributed.

\subsection{Results for 1000 repetitions} \label{lowDim1000reps}

We repeated the above simulation experiments this time for $M=1000$ repetitions and tracked the following variables to measure the performance of the four methods:

\begin{enumerate}[1.]
	\item \textbf{$c_l$-node span estimation.} \% overlap between estimated $c_l$-node span and real $c_l$-node span. Using canonical correlation analysis (CCA) we measured what \% of the linear span of $c_l$ and $\hat{c}_l$ is shared. This measures our ability to accurately estimate the $c_l$-node span. Since SVA is the only method that estimates the span of $c_l$ differently to the span of $h_k$ we only measure this for the SVA method. 
	\item \textbf{$h_k$-node span estimation.} \% overlap between estimated $h_k$-node span and real $h_k$-node span (using CCA). In this case we are able to compare the SVA, SVDR and SVDX methods since each has an estimate for $h_k$ ($\hat{h}_k^{sva}$, $\hat{h}_k^{svdr}$ and $\hat{h}_k^{svdx}$ as described in Section \ref{sec:methods}). This measures our ability to accurately estimate the $h_k$-node span. 
	\item \textbf{Dependence between $h_k$ nodes and $y$.} R2 between real $y$ and $h_k$ nodes minus R2 between real $y$ and estimated $h_k$ nodes. The R2 corresponds to a simple linear regression with $y$ as the independent variable and $h_k$ or $\hat{h}_k$ as the dependent variables. This measures our ability to accurately model the dependence between $h$ and $y$ (to linear approximation).
	\item \textbf{$f_{x_j}$ estimation.} Mean absolute error in estimation of $f_{x_j}$ calculated as:
	
	\begin{align}
		\frac{\sum_{i=1}^n|f_{x_j}(y_i)-\hat{f}_{x_j}(y_i)|}{n}
	\end{align}
	
	The ultimate goal of all four methods is to estimate $f_{x_j}(y)$ accurately and this measure helps us evaluate this. 
		
	\item \textbf{Valid significance analysis.} We perform a \emph{nested} Kolgomorov-Smirnov hypothesis test (nested KS test) to see if, under the null hypotheses $H_{j0}: f_{x_j}(y)=0$ for $j \in \{1,...,J\}$, the p-values are distributed for this hypothesis test are distributed uniformly. This measures whether we are performing valid significance analysis.
	
\end{enumerate}	

	\textbf{Kolmogorov-Smirnov test}
	Suppose we have a sample of $n$ realizations from a random variable $X$. We want to perform the hypothesis test $H_0:X_i \sim F$ vs. $H_a:X_i \sim F' \neq F$. We may use the Kolmogorov-Smirnov statistic $D_n$ for $F$.  We know $\sqrt{n} D_n$ converges in distribution to a Kolmogorov random variable $K$ and so can perform an asymptotic test with this statistic:
	
	\begin{align}
		D_n &= \sup_{x}|F_n(x) - F_(x)| 
	\end{align}	
	
	\begin{align}
		 \sqrt{n} D_n  & \overset{d}{\rightarrow}  K	
	\end{align}
	
	where 
	\begin{itemize}
		\item $F_n$ is the empirical distribution function: $F_n(x) = \frac{1}{n} \sum_{i=1}^n \mathbbm{1}_{[-\infty,x](x_i)}$ and
		\item $K$ is a Kolmogorov random variable
	\end{itemize}	
	
	For a large enough $n$ if $\sqrt{n}D_n > K_\alpha$, where $\mathbb{P}[K \leq K_\alpha] = 1-\alpha$, then $H_0$ is rejected. 

	\textbf{Nested KS test}
	
	The idea of the nested KS test is that for a given repetition of the experiment $m$ we have $J$ p-values (call these outer p-values) and use the test statistic $D_n$ for $F \sim U[0,1]$ to test for uniformity. Under the null $H_{m0}^{outer}: \{J \mbox{ outer p-values uniform}\}$ the p-value (call this a nested p-value) for this test is again uniform. If we collect the $M$ nested p-values from $M$ KS tests, one for each repetition of the $J$ hypothesis tests $H_{mj0}: f_{x_j}(y)=0$, then under the null they should be distributed uniformly. We perform another \emph{nested} KS test $H_0^{nested}: \{ M  \mbox{nested p-values uniform}\}$  to conclude on the uniformity of the nested p-value distribution. As is mentioned in the section \emph{definition of a correct procedure} of \cite{SVA} the nested KS test is more robust to chance fluctuations as a set of individual outer p-values corresponding to a repetition $m$, $H_{mj0}$, $j \in \{1,...,J\}$ may not be uniform due to chance fluctuations, but this will only contribute one of $M$ nested p-values. 

The following figure shows the first four indicators for the results of the 1000 repetitions of our low-dimensional simulation experiment.

\begin{figure}[H]
  \begin{subfigure}{.5\textwidth}
    \centering
	\includegraphics[width=1\textwidth]{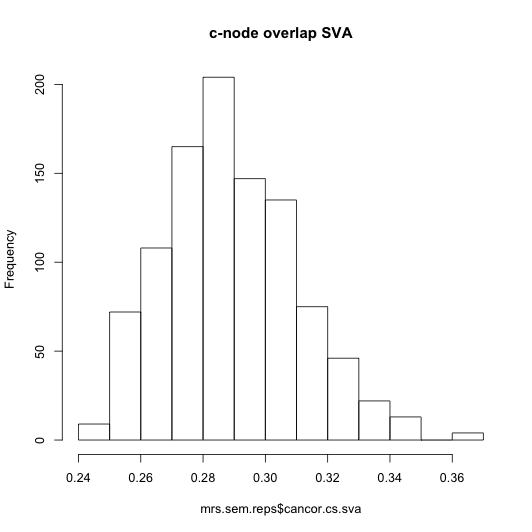} 
    \caption{$c_l$ node span SVA}
    \label{fig:lowDim_cnode}
  \end{subfigure}%
  \begin{subfigure}{.5\textwidth}
    \centering
    \includegraphics[width=1\textwidth]{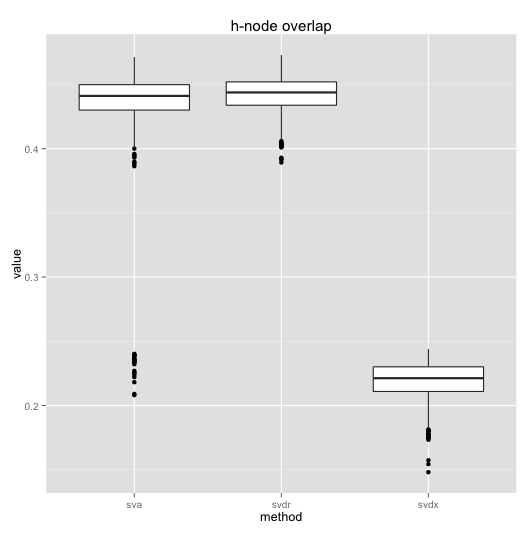}
    \caption{$h_k$ node span SVA, SVDR and SVDX}
    \label{fig:lowDim_hnode}
  \end{subfigure}
  \begin{subfigure}{.5\textwidth}
    \centering
    \includegraphics[width=1\textwidth]{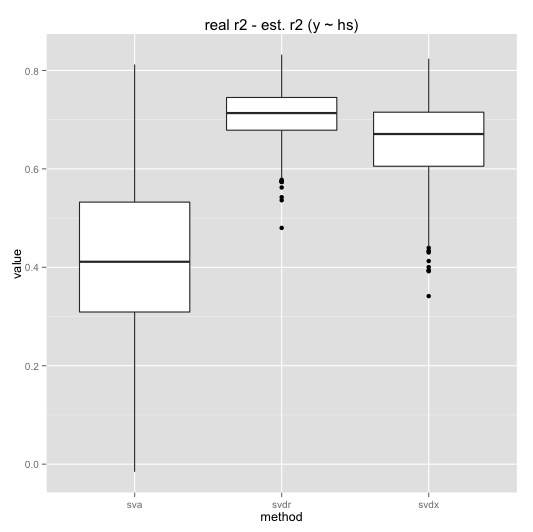}
    \caption{$y$-$h_k$ dependence SVA, SVDR and SVDX}
    \label{fig:lowDim_R2}
  \end{subfigure}
  \begin{subfigure}{.5\textwidth}
    \centering
    \includegraphics[width=1\textwidth]{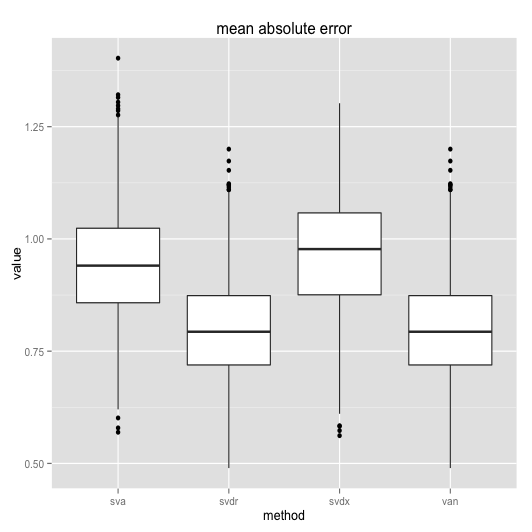}
    \caption{$f_{x_j}$ estimation SVA, SVDR, SVDX and van}
    \label{fig:lowDim_fxj}
  \end{subfigure}
  \caption[Evaluation of performance for low-dim sem with 1000 repetitions: $c_l$ and $h_k$ node span estimation, dependence between $h_k$ nodes and $y$ and $f_{x_j}$ estimation ]
  {Evaluation of performance for low-dim sem with 1000 repetitions: $c_l$ and $h_k$ node span estimation, dependence between $h_k$ nodes and $y$ and $f_{x_j}$ estimation}
  \label{fig:lowDim1000reps}
\end{figure}

The graphs above show that not only does SVA produce estimates of the span of $h_k$ with a large overlap with the real span of $h_k$ but it also manages to capture the relationship between $h_k$ and $y$ better than the other methods. However, this did not translate into more accurate estimates for $f_{x_j}$. Lets see how valid the significance analysis for each method is by looking at the nested KS test statistic and p-values.

\begin{center}
\begin{tabular}{|c|c|c|c|c|}
\hline
 & \textbf{sva} & \textbf{svdx} & \textbf{svdr} & \textbf{van}  \\
\hline
\hline
 KS statistic & 0.308 & 0.312 & 0.527 & 0.265 \\
 \hline
 p-value & 0.000 & 0.000 & 0.000 & 0.000 \\
 \hline
 \hline
\end{tabular}
\end{center}

Although in all cases we reject the hypothesis that p-values under null $H_{j0}:f_{x_j}(y)$ are distributed uniformly, and so in all cases significance analyses are invalid, we are closer to not rejecting for vanilla method. Since the vanilla method also produced the most accurate estimates of $f_{x_j}$ it seems that the SVA methodology does not perform well in this low-dimensional setting. 

\section{High-dimensional experiments} \label{sec:highdim}

\subsection{Design} \label{highDimDesign}

We now explore a high-dimensional setting more akin to the gene expression setting for which SVA was designed. We simulated $M=100$ repetitions of experiments with $n=100$ realizations. The simulated SEM had the following characteristics:

\begin{itemize}
	\item $J = 1000$, $K = L = 10$
	\item $y = N_y \sim N(0,1)$
	\item $c_l = N_{c_l} \sim N(0,1)$ for $l \in \{1,...,10\}$
	\item $h_k = f_{h_k}(y) + c_l$ for $k \in \{1,...,10\}$
	\item $f_{h_k}(y)= b_k y$
	\item $x_j = f_{x_j}(y) + \sum_{k=1}^K \beta_{kj}h_k + N_{x_j}$ for $j \in \{1,...,1000\}$
	\item Non-zero $\beta_{kj}$ values were generated randomly with standard normal generator.
	\item $f_{x_j}(y)=a_j y$
	\item $N_{x_j} \sim N(0,1)$ for $j \in \{1,...,1000\}$
	\item Non-zero $a_j$ and $b_k$ values were generated randomly with standard normal generator.
	\item We control the \emph{sparsity} of the SEM, the number of edges it has, with four parameters:
		\begin{enumerate}[i.]
			\item $p_{0j}$: the proportion of $j \in \{1,...,J\}$ such that $f_{x_j}(y)=0$. We set this to 0.5. The $j$ are chosen uniformly at random from $\{1,...,J\}$. Denote this randomly selected set as $\mathcal{J}_0$
			\item $p_{0k}$: the proportion of $k \in \{1,...,K\}$ such that $f_{h_k}(y)=0$. We set this to 0.5. The $k$ are chosen uniformly at random from $\{1,...,K\}$. Denote this randomly selected set as $\mathcal{K}_0$
			\item $p_{0\beta}$: the \textbf{minimum} proportion of $(k,j) \in \{1,...,K\} \times \{1,...,J\}$ such that $\beta_{kj}=0$. We set this to 0.5. The pairs $(k,j)$ are chosen uniformly at random from $\{1,...,K\} \times \{1,...,J\}$. Denote this randomly selected set as $\mathcal{B}_0$
			\item $p_{\dse}$: the proportion of $j \in \{1,...,J\}$ such that $\dsep{x_j}{y}{\emptyset}$. We set this to 0.25. For this to be possible we need $p_{\dse} \leq p_{0j}$. We sample $\lceil J*p_{\dse} \rceil$ uniformly at random from $\mathcal{J}_0$. Denote this randomly selected set as $\mathcal{J}_{\dse} \subset \mathcal{J}_0$. Then for all $(k,j)$ such that $k \notin \mathcal{K}_0$  and $j \in \mathcal{J}_{\dse}$ we set  $\beta_{kj}=0$.  Notice that the actual proportion of $(k,j) \in \{1,...,K\} \times \{1,...,J\}$ such that $\beta_{kj}=0$ is  $\frac{|\mathcal{J}_0 \cup \mathcal{J}_{\dse} |}{J}$, which we do not control and can be larger than $p_{0\beta}$ for a given random selection of $\mathcal{J}_0$ and $\mathcal{B}_0$.   
		\end{enumerate}	
\end{itemize}	

\subsection{Results for 100 repetitions} \label{highDim100reps}

The following figure shows the first four indicators for the results of the 100 repetitions of our high-dimensional simulation experiment.

\begin{figure}[H]
  \begin{subfigure}{.5\textwidth}
    \centering
	\includegraphics[width=1\textwidth]{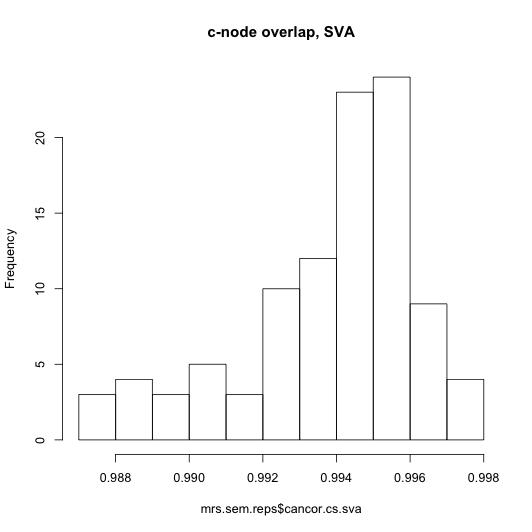} 
    \caption{$c_l$ node span SVA}
    \label{fig:highDim_cnode}
  \end{subfigure}%
  \begin{subfigure}{.5\textwidth}
    \centering
    \includegraphics[width=1\textwidth]{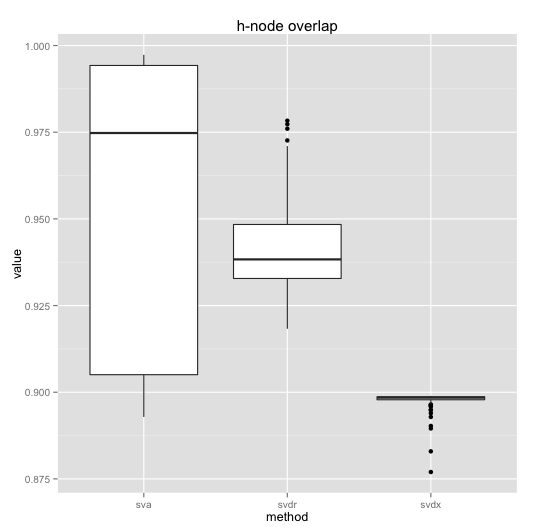}
    \caption{$h_k$ node span SVA, SVDR and SVDX}
    \label{fig:highDim_hnode}
  \end{subfigure}
  \begin{subfigure}{.5\textwidth}
    \centering
    \includegraphics[width=1\textwidth]{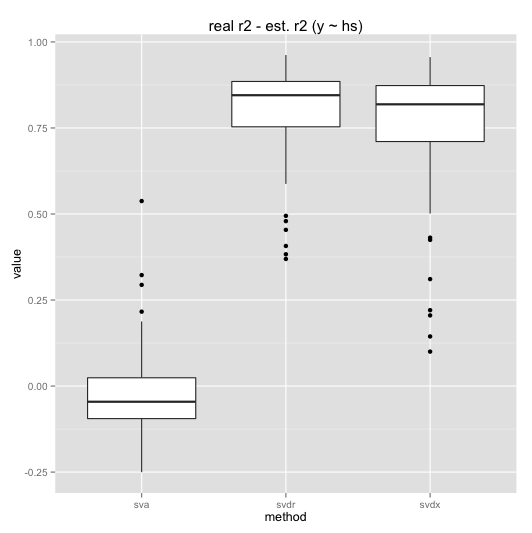}
    \caption{$y$-$h_k$ dependence SVA, SVDR and SVDX}
    \label{fig:highDim_R2}
  \end{subfigure}
  \begin{subfigure}{.5\textwidth}
    \centering
    \includegraphics[width=1\textwidth]{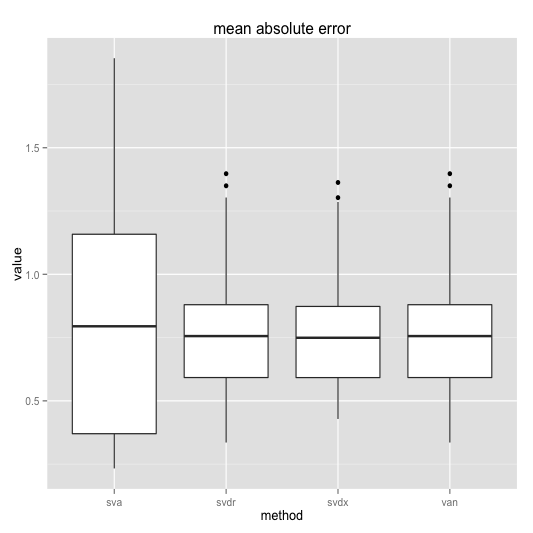}
    \caption{$f_{x_j}$ estimation SVA, SVDR, SVDX and van}
    \label{fig:highDim_fxj}
  \end{subfigure}
  \caption[Evaluation of performance for high-dim sem with 100 repetitions: $c_l$ and $h_k$ node span estimation, dependence between $h_k$ nodes and $y$ and $f_{x_j}$ estimation ]
  {Evaluation of performance for high-dim sem with 100 repetitions: $c_l$ and $h_k$ node span estimation, dependence between $h_k$ nodes and $y$ and $f_{x_j}$ estimation}
  \label{fig:highDim100reps}
\end{figure}

In this case the estimation of the span of the $c_l$ nodes with SVA is between 98.8\% and 99.9\% a lot higher than for the low dimensional experiment. The estimation of the span of $h_k$ nodes is clearly better with SVA and SVDR than with SVDX. In the median SVA is superior to SVDR in this respect although there is much more variance accross the different $M$ repetitions. For modeling of the dependence between $y$ and $h_k$  in the high dimensional setting, SVA really stands out as for more than 95\% of repetitions the R2 difference between the real $y$ and $h_k$ and the real $y$ and estimated $h_k$ was of 0.25 or less while for SVDR and SVDX the difference was of 0.5 or more for 90\% of repetitions. The result of this in terms of the accuracy of $f_{x_j}$ estimation is not clear-cut as for the median all methods perform similarly but SVA displays much greater variance accross repetitions having much better accuracy for some and much worse for others. Lets see how valid the significance analysis for each method is by looking at the nested KS test statistic and p-values.

\begin{center}
\begin{tabular}{|c|c|c|c|c|}
\hline
 & \textbf{sva} & \textbf{svdx} & \textbf{svdr} & \textbf{van}  \\
\hline
\hline
 KS statistic & 0.099 & 0.637 & 0.622 & 0.457 \\
 \hline
 p-value & 0.000 & 0.000 & 0.000 & 0.000 \\
 \hline
 \hline
\end{tabular}
\end{center}

Although in all cases we reject the hypothesis that p-values under null $H_{j0}:f_{x_j}(y)$ are distributed uniformly, and so in all cases significance analyses are invalid, we are closer to not rejecting for SVA method. The modeling of the relationship between the $y$ and $h_k$ nodes and the validity of the significance analysis show some evidence that SVA is superior to the other three methods considered. 

\textbf{Additional remark}: We observed that in general more variables $x_j$ for $j \in \{1,...,J\}$ are found to be contain the signature of $c_l$, for $l$ small than for $l$ large. This seems to indicate that the first few factors (standardized principal components) are composed of \emph{more} residuals $r_j$ variables than the latter factors (or rather the corresponding weights are more even).



The experiments carried out here were somewhat different to those carried out in \cite{SVA} especially in terms of:

\begin{itemize}
	\item $f_{x_j}$ and $f_{h_k}$ complexity: in \cite{SVA} only simple step functions where used while in this case we used linear functions , and 
	\item sparsity of the sems involved: the sems implicitly simulated in \cite{SVA} are somewhat sparser than those simulated here.
\end{itemize}	

In the following sections we explore how the performance of SVA and other methods changes as we change certain parameters of the simulated SEM, including the complexity of the functions $f_{x_j}$ and $f_{h_k}$ and the sparsity. 

\subsection{Sensitivity analysis} \label{sensib}

We performed univariate sensitivity analysis on the following parameters of the additive gene expression SEMs:

\begin{itemize}
\item Dimension of the additive gene expression SEM, specifically parameters $K$ and $J$,
\item Sparsity of the additive gene expression SEM, specifically parameters $p_{0k}$, $p_{0j}$, $p_{0\beta}$ and $p_{\dse}$, 
\item Variance of noise variables, specifically parameters $\sigma_{c_l}$ and $\sigma_{N_{x_j}}$,
\item Complexity of $f_{x_j}$ and $f_{h_k}$, specifically the maximum degree of the polynomials, and
\item Number of observations. 
\end{itemize}

In the following sections all parameters are as described in Section \ref{highDimDesign} unless explicitly stated otherwise. 

\subsubsection{Dimension of additive gene expression SEM} \label{sensDim}


We let the number of gene expression level variables take the following values $J \in \{100, 200,...,1000\}$.

\begin{figure}[H]
  \begin{subfigure}{.5\textwidth}
    \centering
	\includegraphics[scale=0.6]{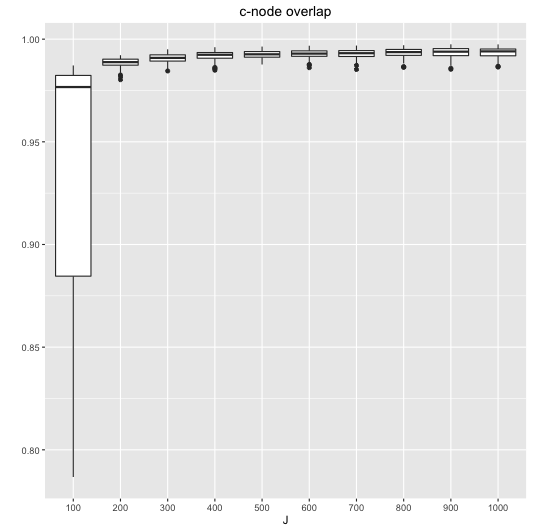} 
    \caption{$c_l$ node span SVA}
    \label{fig:Jsens_cnode}
  \end{subfigure}%
  \begin{subfigure}{.5\textwidth}
    \centering
    \includegraphics[scale=0.6]{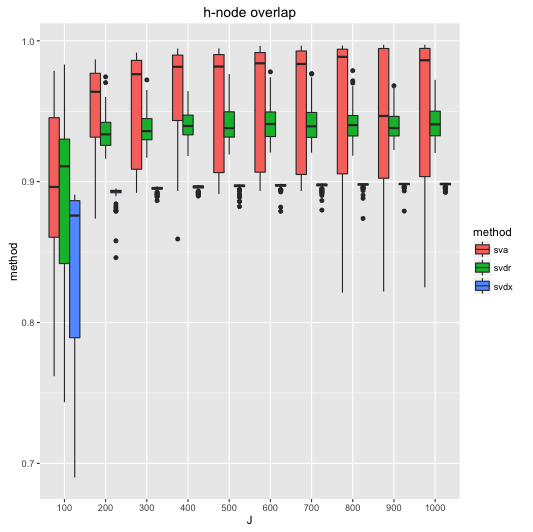}
    \caption{$h_k$ node span SVA, SVDR and SVDX}
    \label{fig:Jsens_hnode}
  \end{subfigure}
  \begin{subfigure}{.5\textwidth}
    \centering
    \includegraphics[scale=0.6]{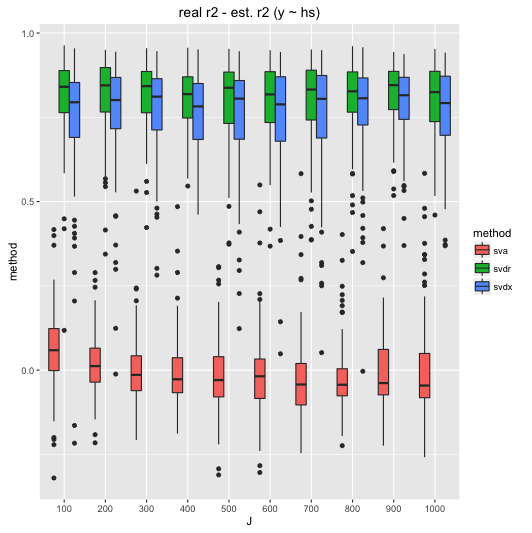}
    \caption{$y$-$h_k$ dependence SVA, SVDR and SVDX}
    \label{fig:Jsens_R2}
  \end{subfigure}
  \begin{subfigure}{.5\textwidth}
    \centering
    \includegraphics[scale=0.6]{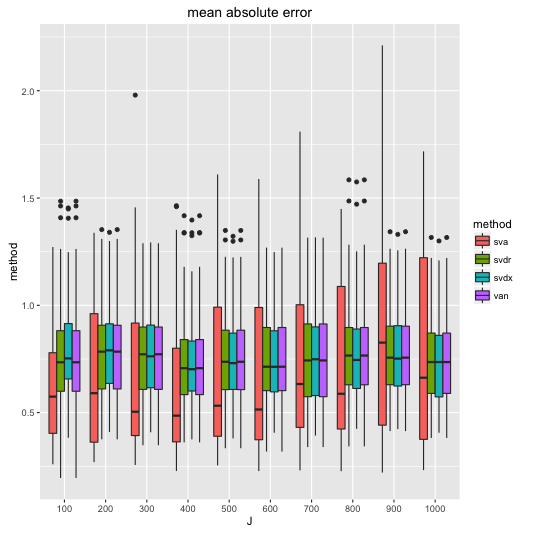}
    \caption{$f_{x_j}$ estimation SVA, SVDR, SVDX and van}
    \label{fig:Jsens_fxj}
  \end{subfigure}
  \begin{subfigure}{.5\textwidth}
    \centering
    \includegraphics[scale=0.6]{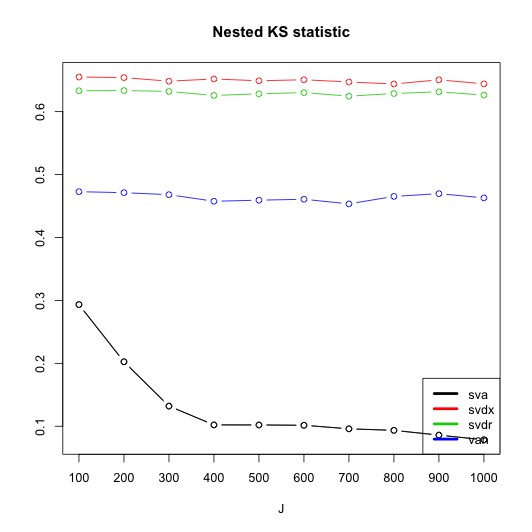}
    \caption{Validity of significance analysis: Nested KS for SVA, SVDR, SVDX and van}
    \label{fig:Jsens_KS}
  \end{subfigure}
  \caption[Sensitivity of parameter $J$]
  {Sensitivity of parameter $J$}
  \label{fig:sensJ}
\end{figure}

Increasing $J$ from 100 to 200 seems to improve all 5 measures studied for SVA both in absolute terms and in comparison with the other methods. Between $J=200$ and $J=500$ the estimation of the $h_k$-node span for SVA is high and has little variance. For higher values of $J$ it is still as high as for SVDR however there is much greater variance. This seems to be reflected in the accuracy of $f_{x_j}$ estimation since for values less than $J=500$ SVA is the most accurate method but for higher values the median accuracy is for all methods is similar but SVA has much more variance. For the range of $J$ values analyzed the nested KS statistic shows that significance analysis for SVA is much more valid than for the other three methods.  

\newpage


We let the number of gene expression level variables and the number of unobserved factors take the following values $(K,J) \in \{(4,40), (37,370),(70,700),...,(300,3000)\}$.

\begin{figure}[H]
  \begin{subfigure}{.5\textwidth}
    \centering
	\includegraphics[scale=0.6]{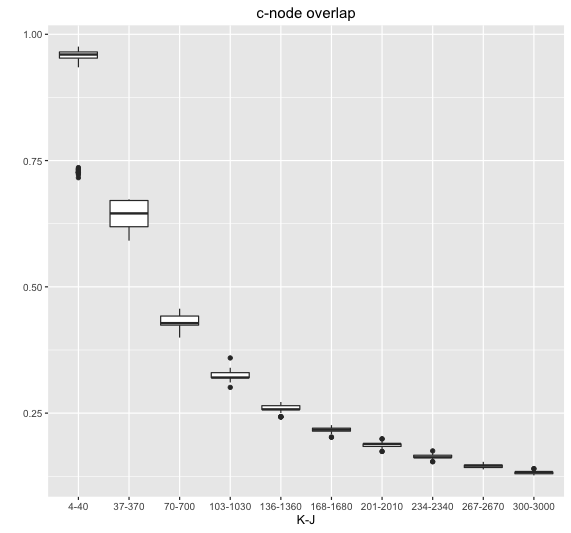} 
    \caption{$c_l$ node span SVA}
    \label{fig:KJsens_cnode}
  \end{subfigure}%
  \begin{subfigure}{.5\textwidth}
    \centering
    \includegraphics[scale=0.6]{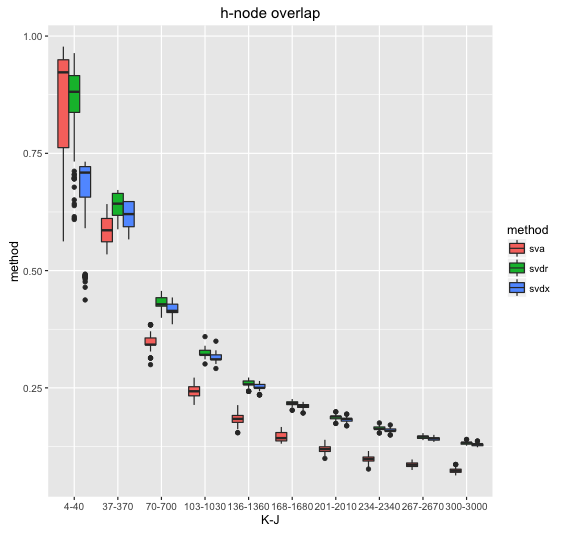}
    \caption{$h_k$ node span SVA, SVDR and SVDX}
    \label{fig:KJsens_hnode}
  \end{subfigure}
  \begin{subfigure}{.5\textwidth}
    \centering
    \includegraphics[scale=0.6]{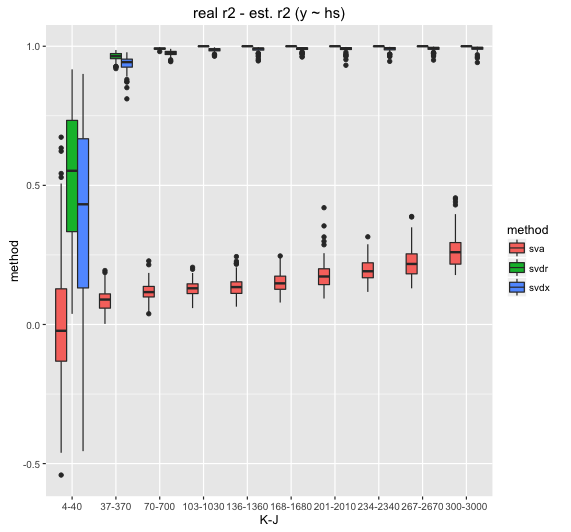}
    \caption{$y$-$h_k$ dependence SVA, SVDR and SVDX}
    \label{fig:KJsens_R2}
  \end{subfigure}
  \begin{subfigure}{.5\textwidth}
    \centering
    \includegraphics[scale=0.6]{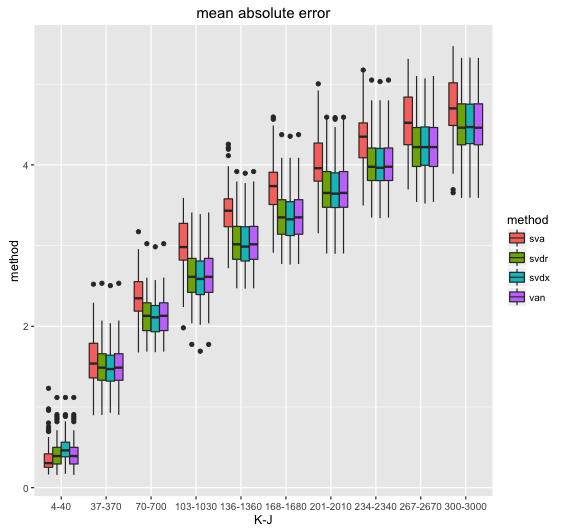}
    \caption{$f_{x_j}$ estimation SVA, SVDR, SVDX and van}
    \label{fig:KJsens_fxj}
  \end{subfigure}
  \begin{subfigure}{.5\textwidth}
    \centering
    \includegraphics[scale=0.6]{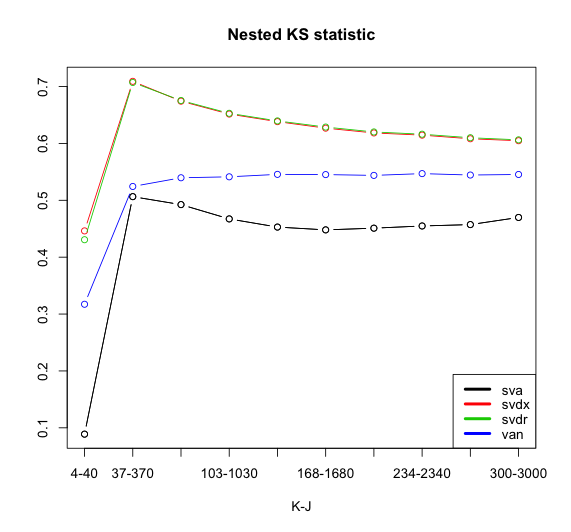}
    \caption{Validity of significance analysis: Nested KS for SVA, SVDR, SVDX and van}
    \label{fig:KJsens_KS}
  \end{subfigure}
  \caption[Sensitivity of parameters $(K,J)$]
  {Sensitivity of parameters $(K,J)$}
  \label{fig:sensKJ}
\end{figure}

Increasing the number of factors $K$ while keeping the $K:J$ ratio equal worsens all five measures for SVA both in absolute terms and in comparison with the other methods. For $K$ up to a value of 40 SVA has better or similar accuracy in the estimation of $f_{x_j}$ to the other methods but beyond this value it performs consistently worse. 

\newpage

\subsubsection{Sparsity of additive gene expression SEM} \label{sensSparse}


We let the proportion of values $k \in \{1,...,K\}$ such that $f_{h_k}(y)=0$ take the following values $p_{0k} \in \{0, 0.11, 0.22,...,0.88,1\}$.

\begin{figure}[H]
  \begin{subfigure}{.5\textwidth}
    \centering
	\includegraphics[scale=0.6]{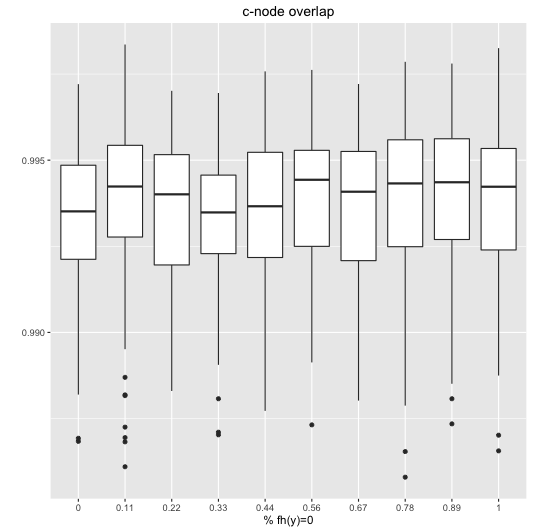} 
    \caption{$c_l$ node span SVA}
    \label{fig:p0ksens_cnode}
  \end{subfigure}%
  \begin{subfigure}{.5\textwidth}
    \centering
    \includegraphics[scale=0.6]{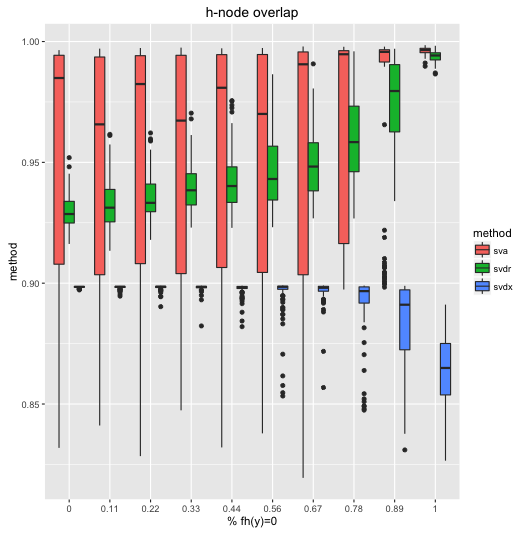}
    \caption{$h_k$ node span SVA, SVDR and SVDX}
    \label{fig:p0ksens_hnode}
  \end{subfigure}
  \begin{subfigure}{.5\textwidth}
    \centering
    \includegraphics[scale=0.6]{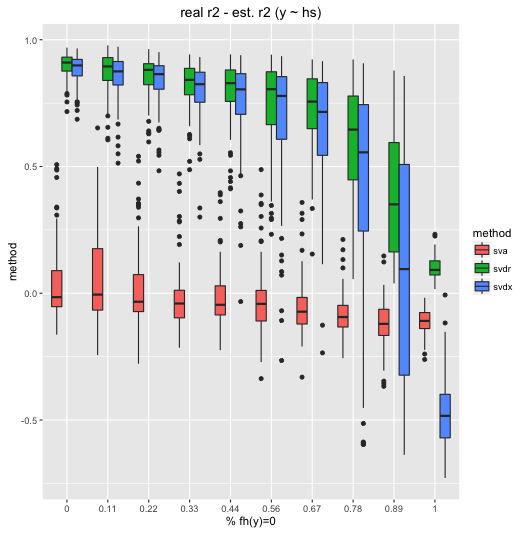}
    \caption{$y$-$h_k$ dependence SVA, SVDR and SVDX}
    \label{fig:p0ksens_R2}
  \end{subfigure}
  \begin{subfigure}{.5\textwidth}
    \centering
    \includegraphics[scale=0.6]{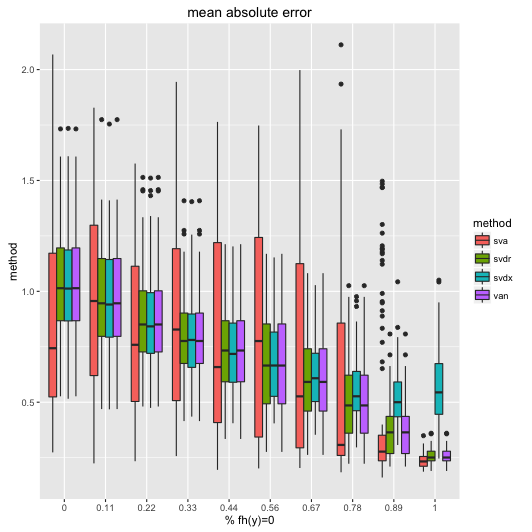}
    \caption{$f_{x_j}$ estimation SVA, SVDR, SVDX and van}
    \label{fig:p0ksens_fxj}
  \end{subfigure}
  \begin{subfigure}{.5\textwidth}
    \centering
    \includegraphics[scale=0.6]{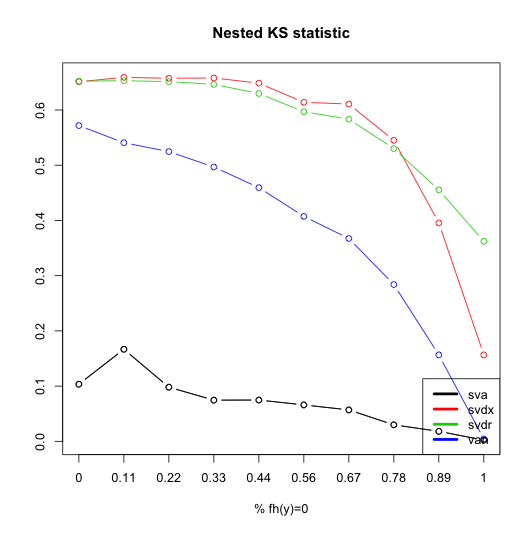}
    \caption{Validity of significance analysis: Nested KS for SVA, SVDR, SVDX and van}
    \label{fig:p0ksens_KS}
  \end{subfigure}
  \caption[Sensitivity of parameter $p_{0k}$]
  {Sensitivity of parameter $p_{0k}$}
  \label{fig:sensp0k}
\end{figure}

For values of $p_{0k}$ of 0.78 or higher $h_k$-node span estimation, $h_k-y$ dependence modeling, $f_{x_j}$ estimation and validity of significance analysis all improve for SVA both in absolute terms and relative to the other methods. This seems to indicate that for SVA to work better than the other methods the dependence between the $h_k$-node span and the primary variable $y$ must not be too strong. 

\newpage


We let the proportion of values $j \in \{1,...,J\}$ such that $f_{h_j}(y)=0$ take the following values $p_{0j} \in \{0.25, 0.33, 0.42, 0.5, 0.58, 0.67, 0.75, 0.83, 0.92, 1\}$.

\begin{figure}[H]
  \begin{subfigure}{.5\textwidth}
    \centering
	\includegraphics[scale=0.6]{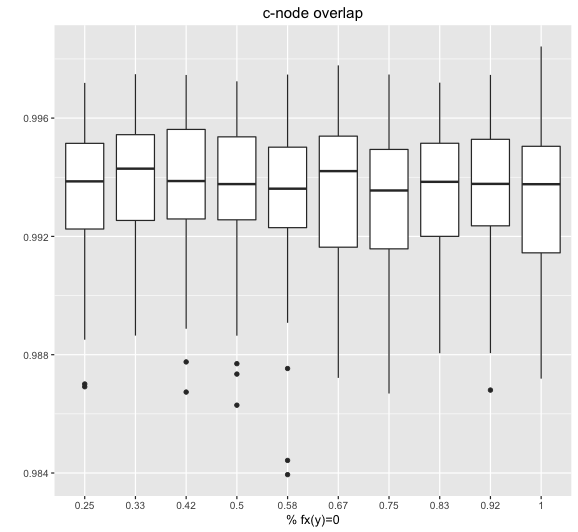} 
    \caption{$c_l$ node span SVA}
    \label{fig:p0jsens_cnode}
  \end{subfigure}%
  \begin{subfigure}{.5\textwidth}
    \centering
    \includegraphics[scale=0.6]{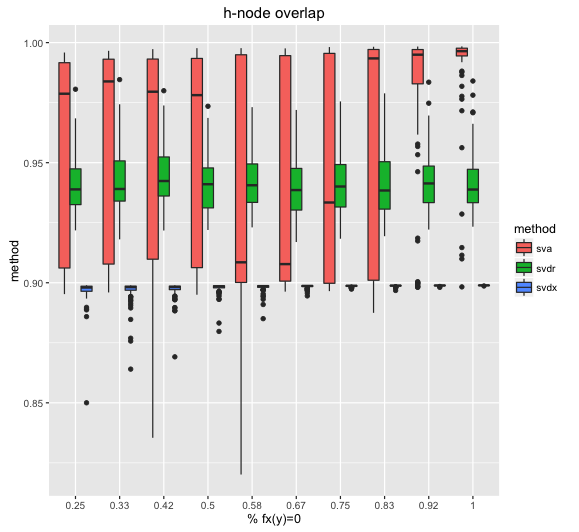}
    \caption{$h_k$ node span SVA, SVDR and SVDX}
    \label{fig:p0jsens_hnode}
  \end{subfigure}
  \begin{subfigure}{.5\textwidth}
    \centering
    \includegraphics[scale=0.6]{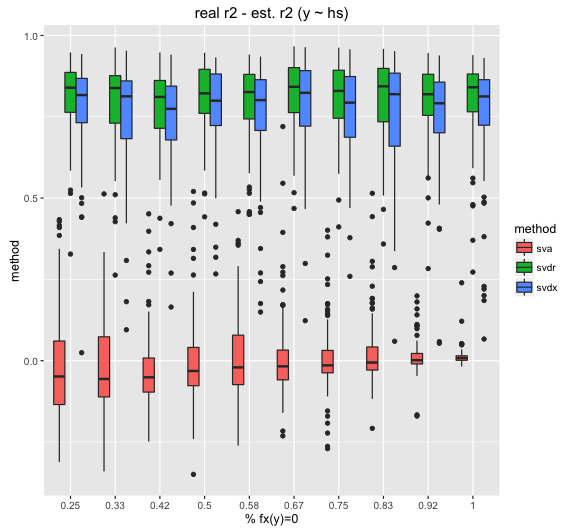}
    \caption{$y$-$h_k$ dependence SVA, SVDR and SVDX}
    \label{fig:p0jsens_R2}
  \end{subfigure}
  \begin{subfigure}{.5\textwidth}
    \centering
    \includegraphics[scale=0.6]{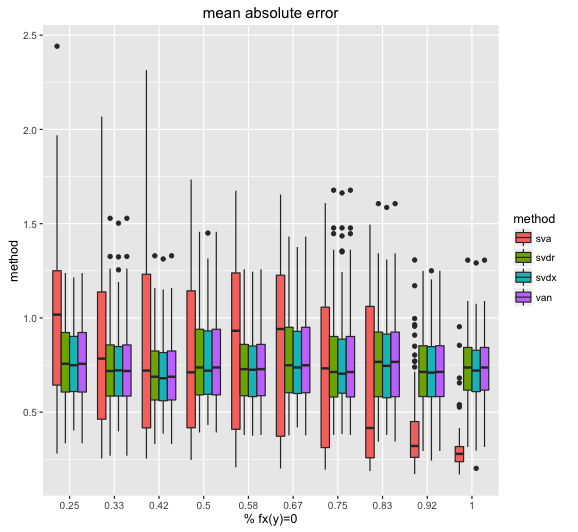}
    \caption{$f_{x_j}$ estimation SVA, SVDR, SVDX and van}
    \label{fig:p0jsens_fxj}
  \end{subfigure}
  \begin{subfigure}{.5\textwidth}
    \centering
    \includegraphics[scale=0.6]{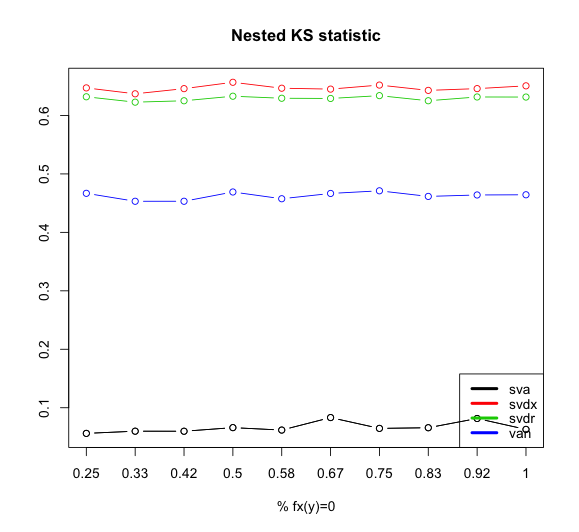}
    \caption{Validity of significance analysis: Nested KS for SVA, SVDR, SVDX and van}
    \label{fig:p0jsens_KS}
  \end{subfigure}
  \caption[Sensitivity of parameter $p_{0j}$]
  {Sensitivity of parameter $p_{0j}$}
  \label{fig:sensp0j}
\end{figure}

For very high values of $p_{0.j}$, when the primary variable only affects 8\% or less of the gene expression levels, the $h_k$-node span estimation and the modeling of the $h_k-y$ dependency for SVA improves significantly resulting in better accuracy in the estimation of $f_{x_j}$ for this method than the other methods. 

\newpage

We let the proportion of values $(k,j) \in \{1,...,K\} \times \{1,...,J\}$ such that $\beta_{kj}=0$ take the following values $p_{\beta} \in \{0, 0.11, 0.22,...,0.99\}$.

\begin{figure}[H]
  \begin{subfigure}{.5\textwidth}
    \centering
	\includegraphics[scale=0.6]{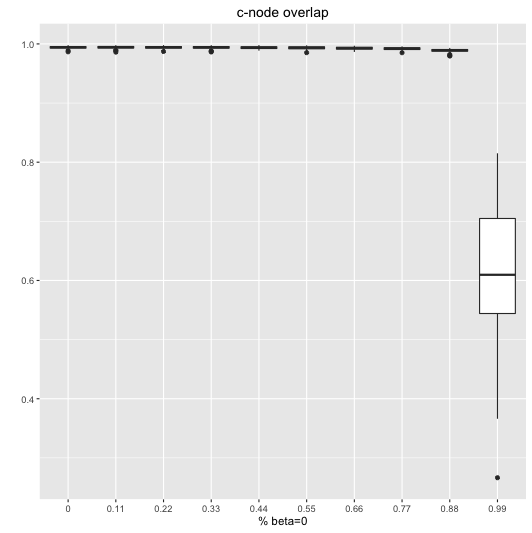} 
    \caption{$c_l$ node span SVA}
    \label{fig:p0betasens_cnode}
  \end{subfigure}%
  \begin{subfigure}{.5\textwidth}
    \centering
    \includegraphics[scale=0.6]{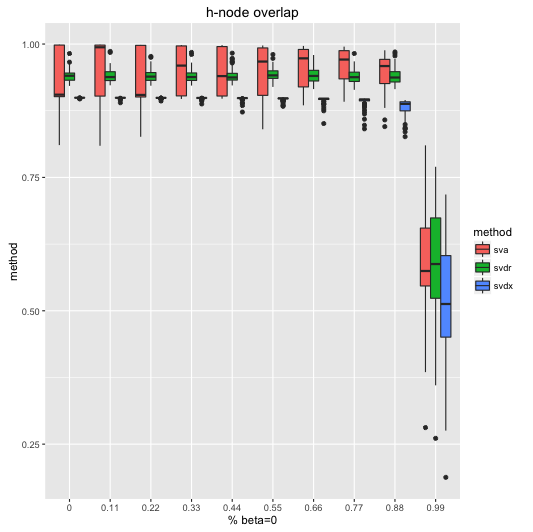}
    \caption{$h_k$ node span SVA, SVDR and SVDX}
    \label{fig:p0betasens_hnode}
  \end{subfigure}
  \begin{subfigure}{.5\textwidth}
    \centering
    \includegraphics[scale=0.6]{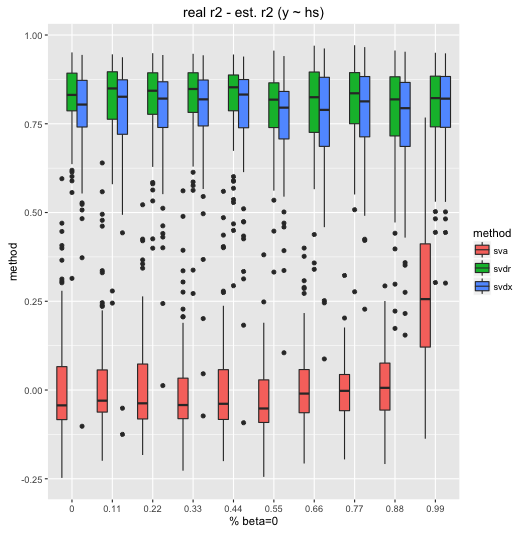}
    \caption{$y$-$h_k$ dependence SVA, SVDR and SVDX}
    \label{fig:p0betasens_R2}
  \end{subfigure}
  \begin{subfigure}{.5\textwidth}
    \centering
    \includegraphics[scale=0.6]{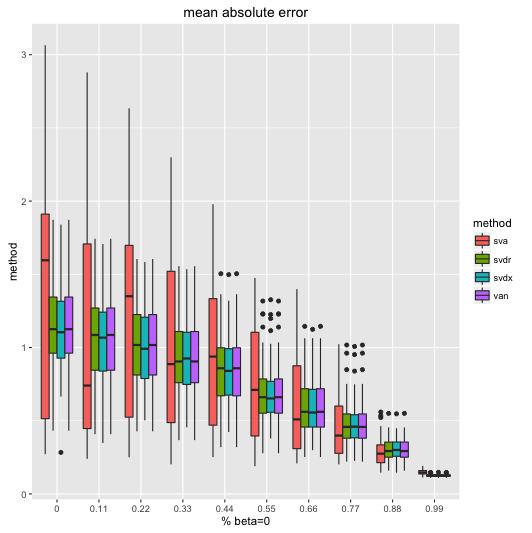}
    \caption{$f_{x_j}$ estimation SVA, SVDR, SVDX and van}
    \label{fig:p0betasens_fxj}
  \end{subfigure}
  \begin{subfigure}{.5\textwidth}
    \centering
    \includegraphics[scale=0.6]{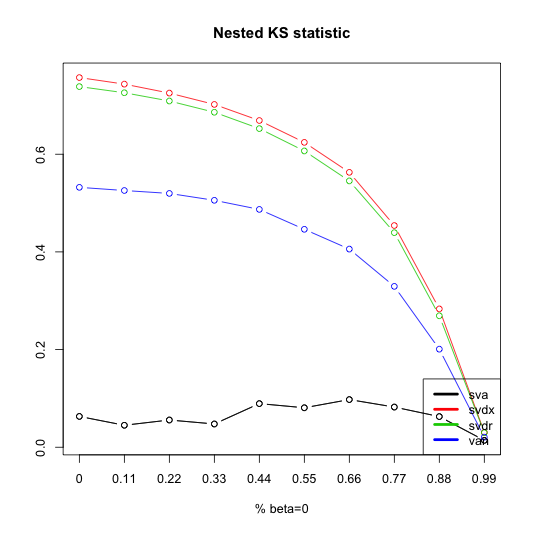}
    \caption{Validity of significance analysis: Nested KS for SVA, SVDR, SVDX and van}
    \label{fig:p0betasens_KS}
  \end{subfigure}
  \caption[Sensitivity of parameter $p_{0\beta}$]
  {Sensitivity of parameter $p_{0\beta}$}
  \label{fig:sensp0beta}
\end{figure}

Sparsity in the \emph{edges} that link $h_k$ nodes and $x_j$ nodes, represtented by the $\beta_{kj}$ coefficients improves the accuracy of $f_{x_j}$ estimation for all four methods. This is due to the fact that the $h_k$ estimations become less important since, as $p_{0\beta}$
increases, the $x_j$ gene expression levels depend less on these variables. For low values of $p_{0\beta}$ the the estimation accuracy of $f_{x_j}$ is similar for all four methods however the variance is much higher for SVA. For high values of $p_{0\beta}$ we start to see higher accuracy and similar variance for SVA compared to the other methods. 

\newpage

We let the proportion of $j \in \{1,...,J\}$ such that $f_{h_j}(y)=0$ take the following values $p_{0j} \in \{0, 0.11, 0.22,...,0.99\}$
and we let the proportion of $j \in \{1,...,J\}$ such that $\dsep{x_j}{y}{\emptyset}$ be the square of the corresponding $p_{0j}$ value, ie we let $(p_{0j},p_{\dse}) \in \{(0,0),(0.11,0.0121),(0.22,0.0484),...,(0.99,0.9801)\}$.

\begin{figure}[H]
  \begin{subfigure}{.5\textwidth}
    \centering
	\includegraphics[scale=0.6]{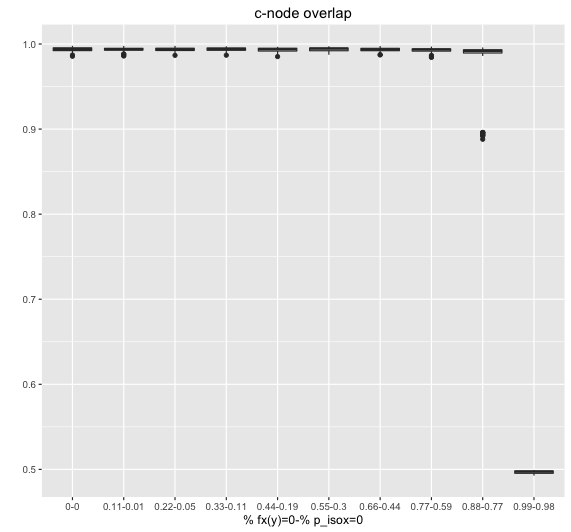} 
    \caption{$c_l$ node span SVA}
    \label{fig:p0dsesens_cnode}
  \end{subfigure}%
  \begin{subfigure}{.5\textwidth}
    \centering
    \includegraphics[scale=0.6]{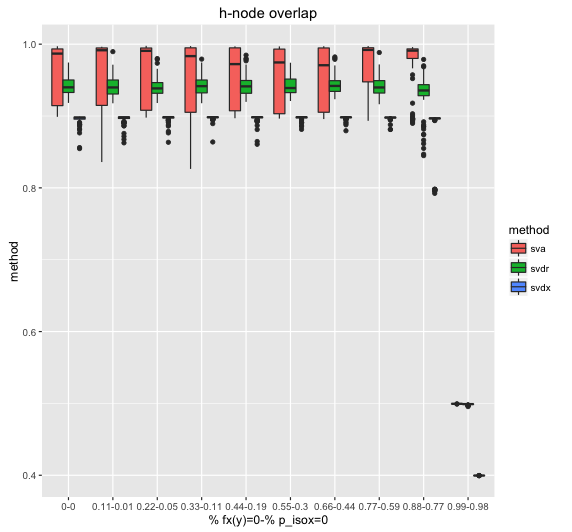}
    \caption{$h_k$ node span SVA, SVDR and SVDX}
    \label{fig:p0dsesens_hnode}
  \end{subfigure}
  \begin{subfigure}{.5\textwidth}
    \centering
    \includegraphics[scale=0.6]{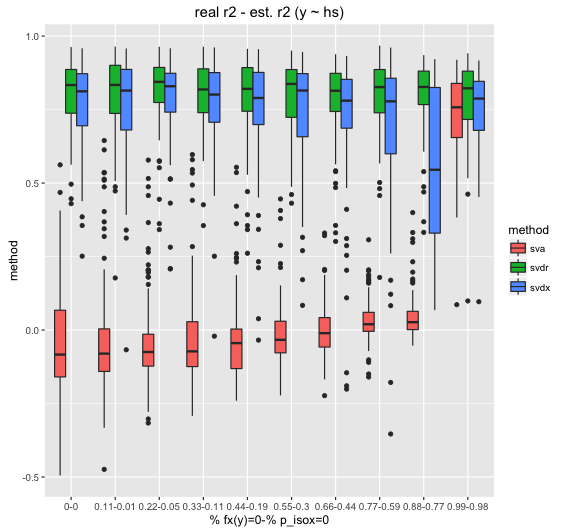}
    \caption{$y$-$h_k$ dependence SVA, SVDR and SVDX}
    \label{fig:p0dsesens_R2}
  \end{subfigure}
  \begin{subfigure}{.5\textwidth}
    \centering
    \includegraphics[scale=0.6]{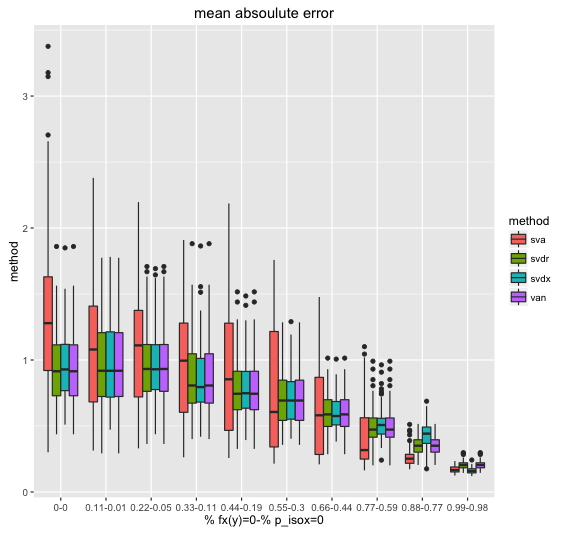}
    \caption{$f_{x_j}$ estimation SVA, SVDR, SVDX and van}
    \label{fig:p0dsesens_fxj}
  \end{subfigure}
  \begin{subfigure}{.5\textwidth}
    \centering
    \includegraphics[scale=0.6]{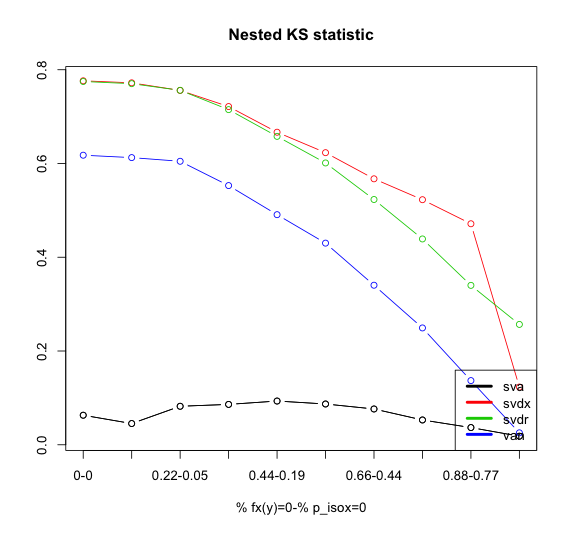}
    \caption{Validity of significance analysis: Nested KS for SVA, SVDR, SVDX and van}
    \label{fig:p0dsesens_KS}
  \end{subfigure}
  \caption[Sensitivity of parameters $(p_{0j},p_{\dse})$]
  {Sensitivity of parameters $(p_{0j},p_{\dse})$}
  \label{fig:sensp0dse}
\end{figure}

The sensitivity of all 5 measures to changes in $p_{\dse}$ behaves similarly to that for changes to $p_{0\beta}$: for increases in both parameters (individually)  estimation of the span of $h_k$ nodes and of the $h_k-y$ dependence gets worse, but the accuracy of $f_{x_j}$ estimation gets better. As $p_{\dse}$ increases, the gene expression levels depend less on the primary variable $y$ and more on the unobserved factors $h_k$, the opposite of what happens as $p_{0\beta}$ increases.  In terms of the accuracy of $f_{x_j}$ estimation, it would seems SVA performs better when either the dependence on $y$ or the dependence on the span of $h_k$ is weak. However, this is hard to explain in this case since for high values of $p_{\dse}$, the $x_j$ variables depend predominantly on the unobserved factors $h_k$ but these are now poorly estimated. The poor estimation of the span of $h_k$ nodes is due to a poor estimaton of the span of $c_l$ nodes. One reason for this could be that there is overfitting of the $f_j(y)$ functions in the initial step of SVA due to the fact that many $x_j$ variables don't depend on $y$. This would cause a bad estimation of residuals and therefore of the span of $c_l$ nodes. The fact that estimation $f_{x_j}$ estimation for SVA still improves despite the poor estimation of the span of $h_k$ nodes could be explained by the fact that most functions $f_{x_j}=0$. 

\newpage
\subsubsection{Variance of noise variables} \label{sensNoise}


We let the standard deviation of the $c_l$ noise variables take the following values $\sigma_{N_{c_l}} \in \{\sfrac{1}{10}, \sfrac{1}{5}, \sfrac{1}{3}, \sfrac{1}{2}, 1, 2\}$

\begin{figure}[H]
  \begin{subfigure}{.5\textwidth}
    \centering
	\includegraphics[scale=0.6]{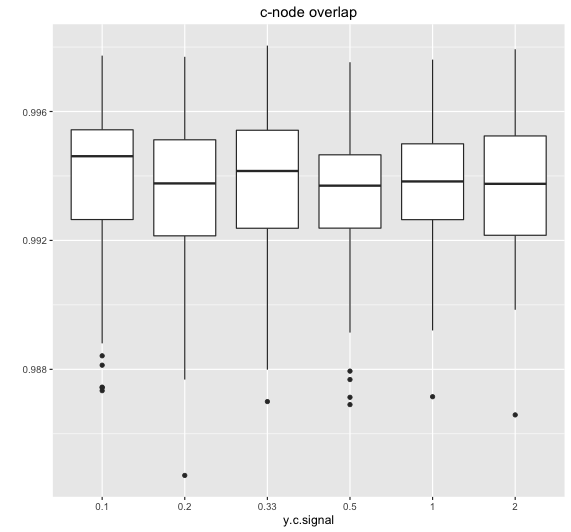} 
    \caption{$c_l$ node span SVA}
    \label{fig:sigmaclsens_cnode}
  \end{subfigure}%
  \begin{subfigure}{.5\textwidth}
    \centering
    \includegraphics[scale=0.6]{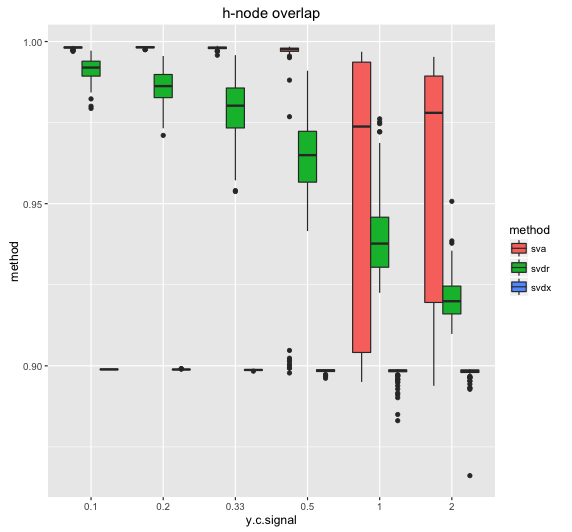}
    \caption{$h_k$ node span SVA, SVDR and SVDX}
    \label{fig:sigmaclsens_hnode}
  \end{subfigure}
  \begin{subfigure}{.5\textwidth}
    \centering
    \includegraphics[scale=0.6]{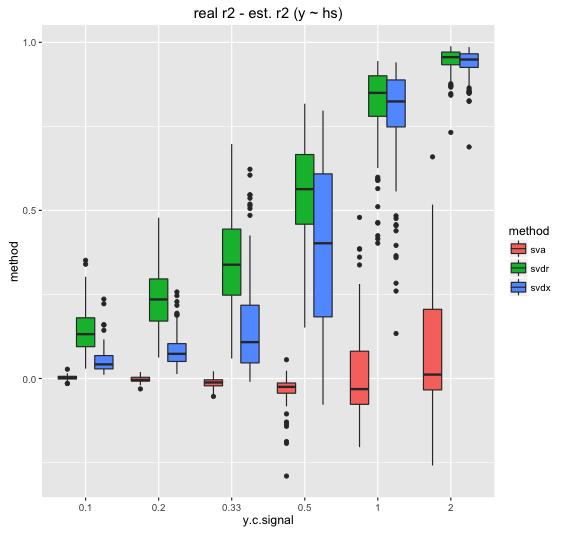}
    \caption{$y$-$h_k$ dependence SVA, SVDR and SVDX}
    \label{fig:sigmaclsens_R2}
  \end{subfigure}
  \begin{subfigure}{.5\textwidth}
    \centering
    \includegraphics[scale=0.6]{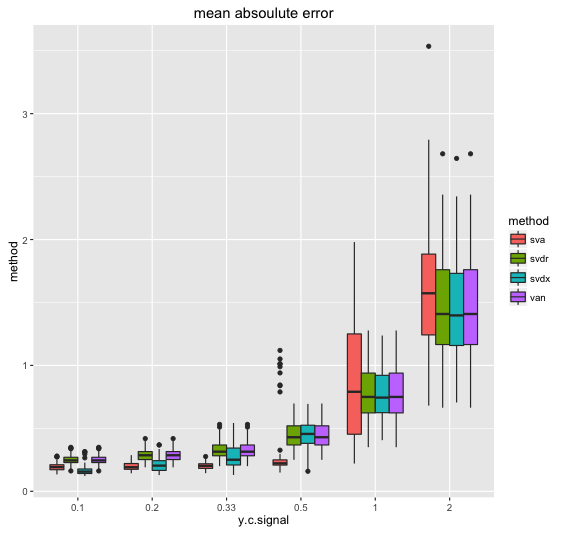}
    \caption{$f_{x_j}$ estimation SVA, SVDR, SVDX and van}
    \label{fig:sigmaclsens_fxj}
  \end{subfigure}
  \begin{subfigure}{.5\textwidth}
    \centering
    \includegraphics[scale=0.6]{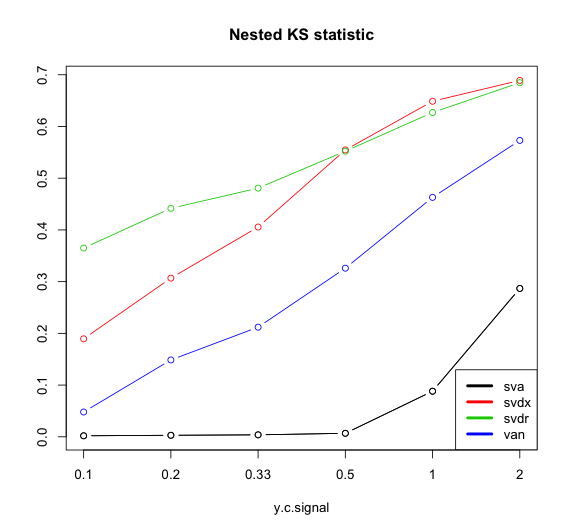}
    \caption{Validity of significance analysis: Nested KS for SVA, SVDR, SVDX and van}
    \label{fig:sigmaclsens_KS}
  \end{subfigure}
  \caption[Sensitivity of parameter $\sigma_{c_l}$]
  {Sensitivity of parameter $\sigma_{c_l}$}
  \label{fig:senssigmacl}
\end{figure}

Greater variance in the noise of $c_l$ nodes results in poorer estimation of span of $h_k$ nodes. This leads to more variance in the estimation of the $h_k-y$ dependency. For low values (0.1-0.5) of $\sigma_{c_l}$ SVA is the most accurate method for estimating $f_{x_j}$ however for higher values (>0.5) it is worse both in terms of the median and the variance. Although SVA is still the best method for producing valid significance analysis even for high levels of $\sigma_{c_l}$ the quality of the significance analyis drops dramatically as this parameter increases. 

\newpage

We let the standard deviation of the $x_j$ noise variables take the following values $\sigma_{N_{x_j}} \in \{\sfrac{1}{10}, \sfrac{1}{5}, \sfrac{1}{3}, \sfrac{1}{2}, 1, 2, 5, 10\}$

\begin{figure}[H]
  \begin{subfigure}{.5\textwidth}
    \centering
	\includegraphics[scale=0.6]{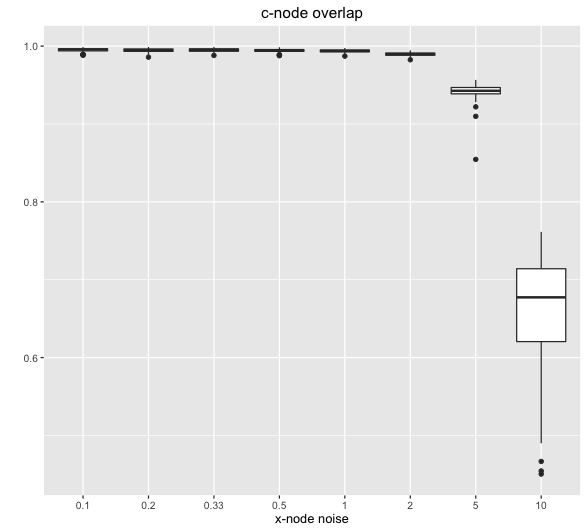} 
    \caption{$c_l$ node span SVA}
    \label{fig:sigmaxjsens_cnode}
  \end{subfigure}%
  \begin{subfigure}{.5\textwidth}
    \centering
    \includegraphics[scale=0.6]{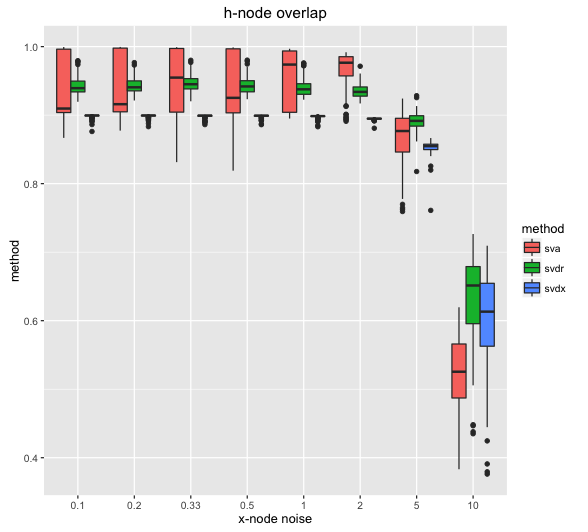}
    \caption{$h_k$ node span SVA, SVDR and SVDX}
    \label{fig:sigmaxjsens_hnode}
  \end{subfigure}
  \begin{subfigure}{.5\textwidth}
    \centering
    \includegraphics[scale=0.6]{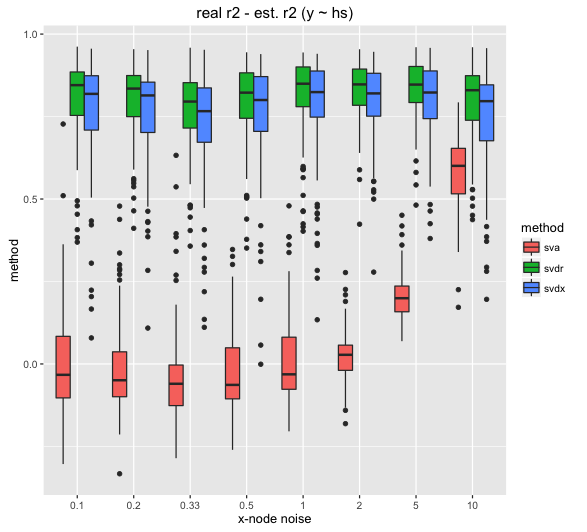}
    \caption{$y$-$h_k$ dependence SVA, SVDR and SVDX}
    \label{fig:sigmaxjsens_R2}
  \end{subfigure}
  \begin{subfigure}{.5\textwidth}
    \centering
    \includegraphics[scale=0.6]{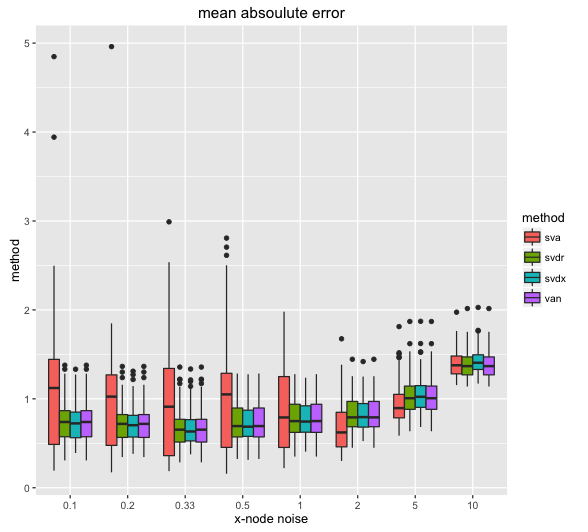}
    \caption{$f_{x_j}$ estimation SVA, SVDR, SVDX and van}
    \label{fig:sigmaxjsens_fxj}
  \end{subfigure}
  \begin{subfigure}{.5\textwidth}
    \centering
    \includegraphics[scale=0.6]{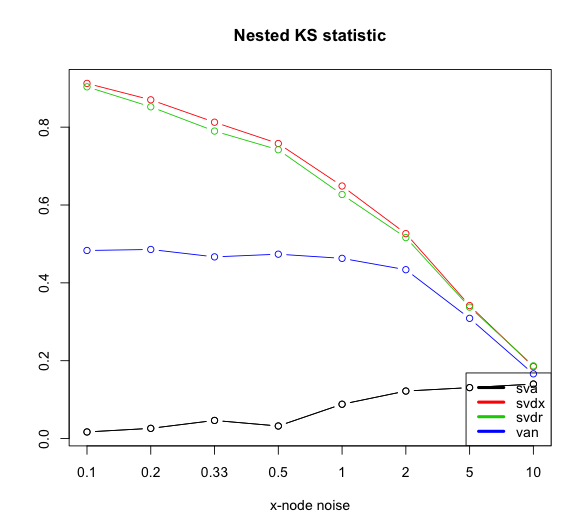}
    \caption{Validity of significance analysis: Nested KS for SVA, SVDR, SVDX and van}
    \label{fig:sigmaxjsens_KS}
  \end{subfigure}
  \caption[Sensitivity of parameter $\sigma_{N_{x_j}}$]
  {Sensitivity of parameter $\sigma_{N_{x_j}}$}
  \label{fig:senssigmaxj}
\end{figure}

As the $x_j$ node noise variance, $\sigma_{x_j}$   gets larger it is harder to recover the contribution of the $h_k$ and $c_l$ nodes so the estimation of their span worsens, and consequently that of the $h_k-y$ dependency. However, since the $h_k$ nodes constitute an ever decreasing component of the gene expression levels $x_j$ it is possible to model them with or without an estimate for $h_k$ and so all methods perform equally well for high $\sigma_{x_j}$ values. This also means that the significance analysis for the multiple hypothesis $H_{j0}:f_{x_j}(y)=0$ become less dependent and so significance analysis for all 4 methods becomes more valid. 

\newpage
\subsubsection{Complexity of $f_{x_j}$ and $f_{h_k}$} \label{sensComplexity}


We let the maximum degree  for any $f_{x_j}(y)$ or $f_{h_k}(y)$ polynomial function take the following values $d^{\max} \in \{1,2,3,4,5\}$. This is also the order of the polynomial basis function model used in estimation. \textbf{Remark}: the actual degree of any polynomial $f_{x_j}(y)$ for $j \in \{1,...,J\}$ and $f_{h_k}(y)$ for $k \in \{1,...,K\}$ is sampled uniformly from the set $\{1,...,d^{\max}\}$.

\begin{figure}[H]
  \begin{subfigure}{.5\textwidth}
    \centering
	\includegraphics[scale=0.6]{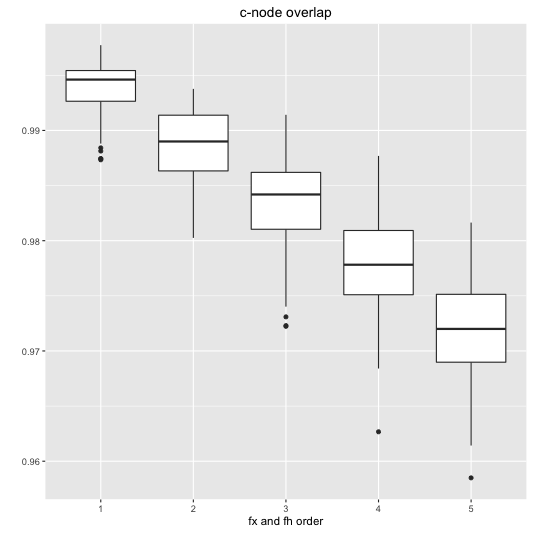} 
    \caption{$c_l$ node span SVA}
    \label{fig:dmaxsens_cnode}
  \end{subfigure}%
  \begin{subfigure}{.5\textwidth}
    \centering
    \includegraphics[scale=0.6]{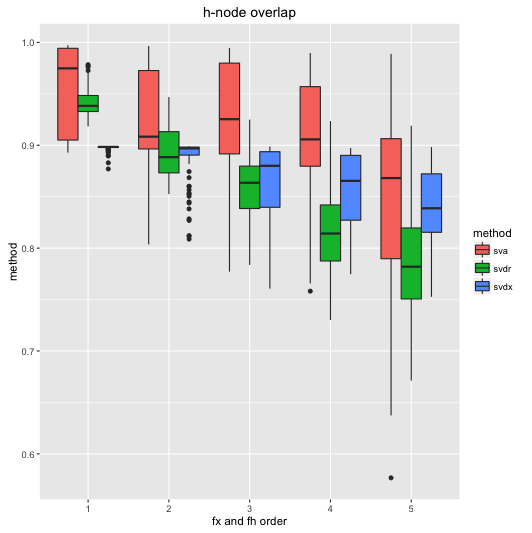}
    \caption{$h_k$ node span SVA, SVDR and SVDX}
    \label{fig:dmaxsens_hnode}
  \end{subfigure}
  \begin{subfigure}{.5\textwidth}
    \centering
    \includegraphics[scale=0.6]{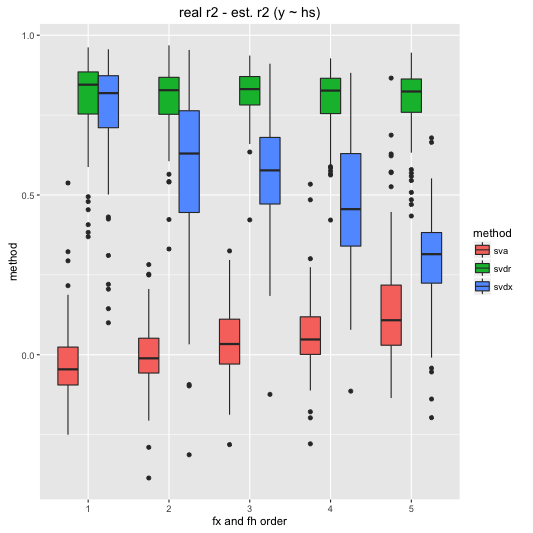}
    \caption{$y$-$h_k$ dependence SVA, SVDR and SVDX}
    \label{fig:dmaxsens_R2}
  \end{subfigure}
  \begin{subfigure}{.5\textwidth}
    \centering
    \includegraphics[scale=0.6]{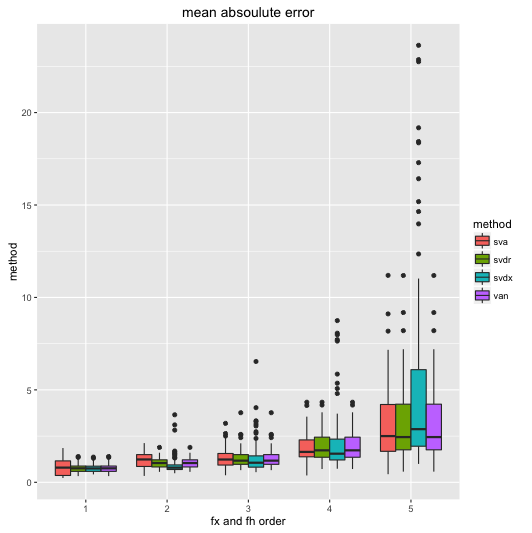}
    \caption{$f_{x_j}$ estimation SVA, SVDR, SVDX and van}
    \label{fig:dmaxsens_fxj}
  \end{subfigure}
  \begin{subfigure}{.5\textwidth}
    \centering
    \includegraphics[scale=0.6]{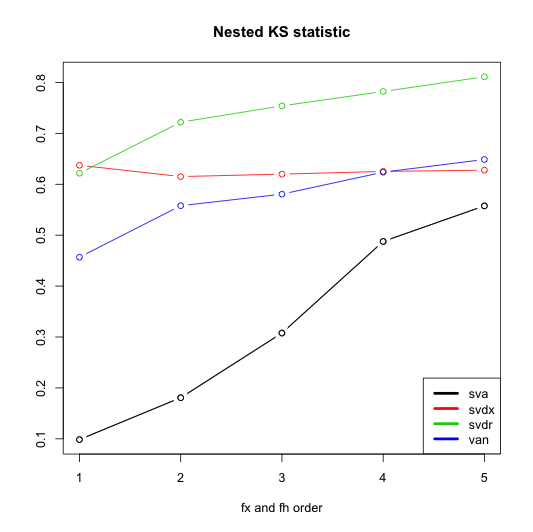}
    \caption{Validity of significance analysis: Nested KS for SVA, SVDR, SVDX and van}
    \label{fig:dmaxsens_KS}
  \end{subfigure}
  \caption[Sensitivity of parameter $d^{\max}$]
  {Sensitivity of parameter $d^{\max}$}
  \label{fig:sensdmax}
\end{figure}

As the maximum order of the polynomials $f_{x_j}$ and $f_{h_k}$ increases the estimation of the $c_l$-node span for SVA deteriorates probably because the variance of the signal from $y$ to $x_j$ dominates the signal coming from the $c_l$ nodes to $x_j$ making it hard to recover the latter. This in turn means that as complexity of $f_{x_j}$ and $f_{h_k}$ increases the $h_k$-node span and $h_k-y$ dependency estimation deteriorates leading to ever poorer $f_{x_j}$ estimation. However, greater $f_{x_j}$ and $f_{h_k}$ complexity lead to poorer $f_{x_j}$ estimation for all four methods. The validity of significance analysis also deteriorated for all four methods as the complexity of $f_{x_j}$ and $f_{h_k}$ increased. This is probably because the dependence of the gene expression levels $x_j$ on $h_k$ is stronger.

\newpage
\subsubsection{Number of observations} \label{sensNumObs}


We let the number of observations $n$ simulated for each repetition $m \in \{1,2,...,100\}$ take the following values $n \in \{25, 133, 242,...,1000\}$.

\begin{figure}[H]
  \begin{subfigure}{.5\textwidth}
    \centering
	\includegraphics[scale=0.6]{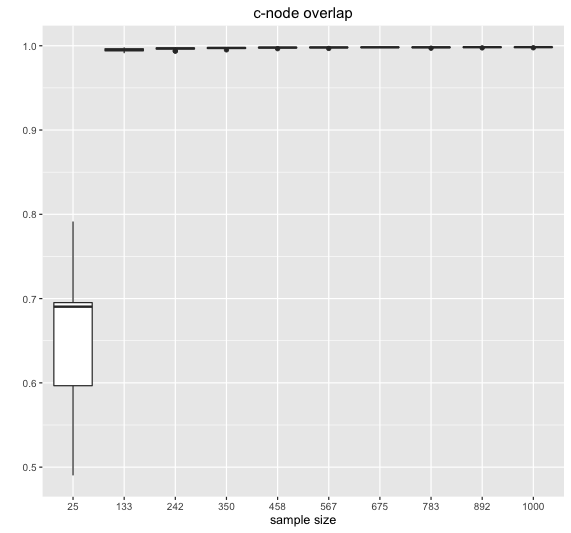} 
    \caption{$c_l$ node span SVA}
    \label{fig:nsens_cnode}
  \end{subfigure}%
  \begin{subfigure}{.5\textwidth}
    \centering
    \includegraphics[scale=0.6]{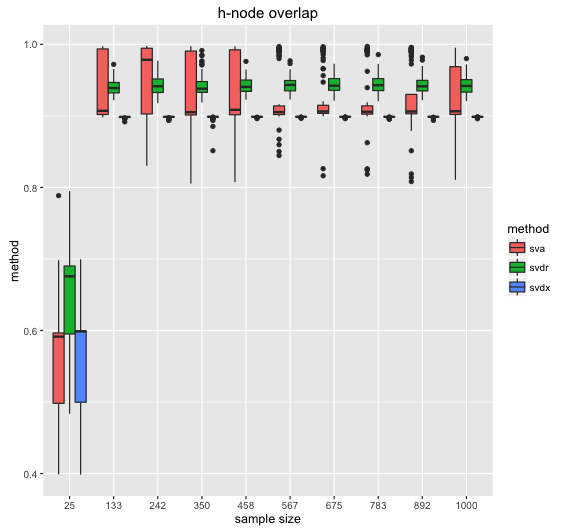}
    \caption{$h_k$ node span SVA, SVDR and SVDX}
    \label{fig:nsens_hnode}
  \end{subfigure}
  \begin{subfigure}{.5\textwidth}
    \centering
    \includegraphics[scale=0.6]{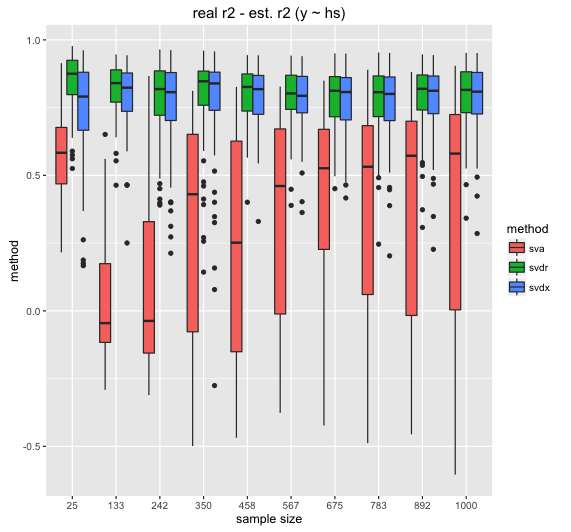}
    \caption{$y$-$h_k$ dependence SVA, SVDR and SVDX}
    \label{fig:nsens_R2}
  \end{subfigure}
  \begin{subfigure}{.5\textwidth}
    \centering
    \includegraphics[scale=0.6]{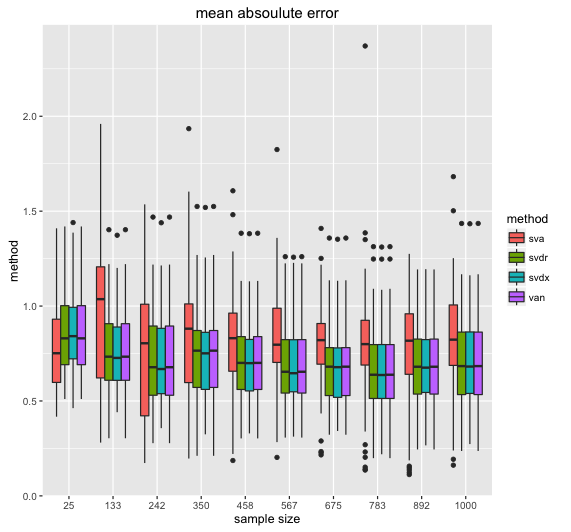}
    \caption{$f_{x_j}$ estimation SVA, SVDR, SVDX and van}
    \label{fig:nsens_fxj}
  \end{subfigure}
  \begin{subfigure}{.5\textwidth}
    \centering
    \includegraphics[scale=0.6]{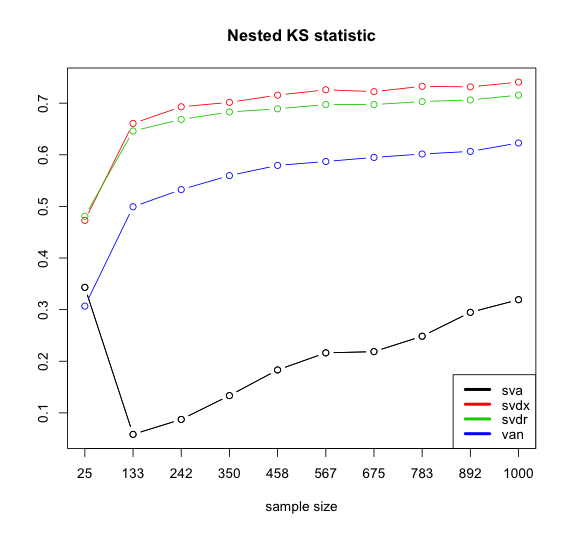}
    \caption{Validity of significance analysis: Nested KS for SVA, SVDR, SVDX and van}
    \label{fig:nsens_KS}
  \end{subfigure}
  \caption[Sensitivity of parameter $n$]
  {Sensitivity of parameter $n$}
  \label{fig:sensn}
\end{figure}

As $n$ increases above 500 observations the performance measures for the SVA estimation do not seem to improve, if anything they get worse. This may relate to what \citeauthor{multTest} refer to as a \emph{reversal of the curse of dimensionality} where a \emph{low dimensional kernel}, in this case the $K$ $h_k$ unobserved variables, \emph{fully captures the dependence structure in an observed high-dimensional data set}, in this case the $x_j$ variables. As $n$ increases the the data set of $x_j$ variables approaches a non high-dimensional setting so it may become harder to capture the dependence structure with a low dimensional set of variables. 

The sensibility analysis carried out is univariate meaning we can't choose the best parameter values form each sensitvity analysis to obtain a set of parameters where SVA performs ideally or better than the other methods. However assuming that performance measures vary with respect to the different sensitivity parameters somewhat independently this analyis suggests the following parameters for ideal SVA performance:

\begin{itemize}
	\item \textbf{Dimension}: Low number of unobserved factors $h_k$ and high number of gene expression levels $x_j$ ($J$ 20 to 50 times bigger than $K$ based on sensitivity analysis),
	\item \textbf{Sparsity}: High level of sparsity in gene expression SEM favours a high level of accuracy of $f_{x_j}$ estimation for SVA both in absolut terms and in comparison with other methods ($p_{0k} \geq 0.88$, $p_{0j} \geq 0.92$, $p_{0\beta} \geq 0.88$ and $p_{\dse} \geq 0.77$ as roughly indicated by sensitivity analysis),
	\item \textbf{Variance of noise}: Sensitivity analysis suggests that lower variance for $c_l$ variables than for $y$ variable and a slightly higher variance for $x_j$ variables than for $y$ variable are ideal conditions for SVA both in absolute terms and relative to other methods (sensitivity analysis suggests $\sigma_{c_l} \leq 0.5$, $\sigma_y=1$ and $\sigma_{N_{x_j}}=2$ are good values for example )
	\item \textbf{Complexity of $f_{x_j}$ and $f_{h_k}$}: simpler functions lead to better estimation accross all four methods, 
	\item \textbf{Number of observations}: There is some evidence that SVA performs better relative to other methods when $n$ is smaller, probably in relation to $K$ and $J$.
\end{itemize}


\chapter{Summary}
\label{s:Summary}

In Chapter \ref{ch:intro} we introduced the SVA methodology in the context of the modeling of a large number of gene expression levels, $x_j$ with $j \in \{1,...,J\}$, in terms of an observed primary variable (or vector) $y$ and unobserved factors $g_l$. We justified the use of an additive model of the form in \ref{eq:geneModel} and argued that we must estimate the variables $g_l$ in order to obtain an unbiased estimate of $f_{x_j}(y)$ which is the ultimate goal. We concluded that it is not necessary to obtain an explicit estimate of $g_l$ for $l \in \{1,...,L\}$ rather we can estimate variables $h_k$ for $k \in \{1,...,K\}$ such that they generate the same linear space as the $g_l$ variables. 

In Chapter \ref{ch:causality} we framed the SVA modeling problem as a SEM, graphical model and causality estimation problem. In Section \ref{DAGsSEMs} we defined DAGs and SEMs, established their relationship and the properties, such as the Markov property, faithufulness and causal minimality, that determine whether a given DAG is a sound and complete map for the independencies implicit in a corresponding SEM. We defined a class of \emph{gene expression } SEMs (\ref{GESEM}) that we assume can be used to model gene expression data but concluded that it is too large since it includes SEMs with different \emph{unidentifiable} DAG structures. In Section \ref{sec:causality} we defined the interventional distribution of a SEM and used it to define causal effects. We explored causality as a tool for identifying the DAG structure of a SEM and gave some results that help establish the causal relationships between the variables in a SEM with a given DAG. In Section \ref{sec:addGESEM} we defined a subset class of SEMs denoted \emph{additive gene expression} SEMs (\ref{adGESEM}) and justified the use of this smaller class to model gene expression data whose behavior, we assumed previously, could be that of any SEM from the broader class. All the DAGs of this class are equivalent, up to deletion of edges, and so identifiability is not a problem when working with this class. Using the theory developed through out the chapter we derive some of the independence and causality porperties of the additive gene expression SEMs. 

Chapter \ref{ch:SVA} gives a detailed description of the SVA estimation methodology including its implementation in the \Rp package \verb|sva|. Section \ref{sec:overview} gives a general overview of the three main steps involved and how they relate to the class of additive gene expression SEMs defined in Section \ref{sec:addGESEM}. Sections \ref{sec:cl} and \ref{sec:hk} give a detailed description of the first two steps. The third step simply involves fitting a standard basis function model so a further description is not given. Section \ref{sec:cl} shows how the span of $c_l$ for $l \in \{1,...,L\}$, from the additive gene expression SEM, is estimated. In Section \ref{SVAest} we showed how the first step of SVA, involving fitting a standard basis function model of the form $x_j=f_j(y)+\epsilon$ and factorizing the residuals, corresponds to estimating the span of the $c_l$ variables from the additive gene expression SEM. In Section \ref{subsec:basisModels} we detailed the type of standard basis function model fitted in the \Rp package \verb|sva|. Section \ref{subsec:SVD} describes the SVD factorization of residuals to estimate the span of the $c_l$ variables and its relationship to principal component analysis (PCA), while in Section \ref{subsec:PA} we described the parallel analysis method used to select the number of factors to be used and detailed how this method is implemented in the \Rp package \verb|sva|. Section \ref{sec:hk} shows how the span of $h_k$ for $k \in \{1,...,K\}$ from the additive gene expression SEM is estimated. Section \ref{sigMeth} describes the filtering procedure in general and remarks on its limitations. Part of this procedure involves finding the \emph{signature} of $c_l$ variables in the $x_j$ gene expression level variables. Section \ref{findSig} describes the $J$ regressions, and subsequent FDR based significance analysis of the corresponding hypotheses, that is carried out for this purpose. In Section \ref{FDR} we develop various concepts related to FDR such as positive FDR (pFDR), marginal FDR (mFDR), local FDR (lFDR) and q-values, and show how they can be used to control the FDR at a certain level $\alpha$. We show how q-values control FDR and describe a method for estimating them conservatively. We also show how lFDR is estimated and used to decide which hypotheses tests are significant (discoveries) in the \Rp package \verb|sva|, as part of the procedure for finding the signature of a variable $c_l$ in $x_j$ for $j \in \{1,...,J\}$. 

Chapter \ref{ch:sims} includes the simulations carried out to evaluate the performance of the SVA methodology in different, relevant data environments. Section \ref{sec:methods}  includes the description of three alternate methods for estimating the effect of $y$ on the $x_j$ variables which serve as benchmarks for SVA. In Section \ref{sec:lowdim} we assess the performance of SVA for a low dimensional additive gene expression SEM, specifically when $J=K=L=4$. Section \ref{lowDimDesign} includes the details of the simulation design. Section \ref{lowDim1Rep} includes the simulation results for $M=1$ repetition of a simulation with $n=100$ observations. The results seem to show that SVA has good performance when the complexity of $f_{x_j}$ and $f_{h_k}$ is similar to the complexity of the standard basis model chosen. By the same token there is a clear potental for overfitting for cases when, for example, the complexity of $f_{x_j}$ for at least one $j \in \{1,...,J\}$, is less than the standard basis function model chosen. This is a weakness of the methodology borne of the fact that regularization for large $J$ is costly. In Section \ref{lowDim1000reps} we perform $M=1000$ repetitions of the simulation experiment reporting five metrics which evaluate different aspects of the SVA methodology: 
\begin{enumerate}[1.]
\item $c_l$-node span estimation,
\item $h_k$-node span estimation,
\item dependence between span of $h_k$ nodes and $y$,
\item $f_{x_j}$ estimation accuracy, and
\item validity of significance analysis. 
\end{enumerate}

We conclude that in this low dimensional setting, SVA had good performance for metrics 1-3 and 5, but this did not translate into a more accurate estimation of the $f_{x_j}$ effects. 

In Section \ref{sec:highdim} we assess the performance of SVA in a high dimensional setting, specifically when $J=1000$ and $K=L=10$. Section \ref{highDimDesign} includes the details of a \emph{base} simulation design. Section \ref{highDim100reps} includes the results of performing $M=100$ repetititons of the simulation experiment each with $n=100$ observations. The results of this base scenario show that SVA outperforms the SVDR and SVDX methods in terms of $h_k$-node span estimantion and modeling of the dependence between the span of $h_k$ and $y$ and also in terms of the validity of the significance analysis but this did not translate into consistently better results in terms of the accuracy of the $f_{x_j}$ estimation. In Section \ref{sensib} we perform univariate sensitivity analysis with the scenario described in Section \ref{highDimDesign} as the base scenario to explore for what type of additive gene expression SEMs and for what parameter settings SVA performs better. We performed sensitivity analysis on the dimension of the additive gene expression SEM ($K$ and $J$ parameters), the sparsity of the additive gene expression SEM ($p_{0k}$, $p_{0j}$, $p_{0\beta}$ and $p_{\dse}$ parameters), variance of noise variables ($\sigma_{c_l}$ and $\sigma_{N_{x_j}}$ parameters), complexity of $f_{x_j}$ and $f_{h_k}$ (maximum degree of polynomials) and number of observations. We found that, assuming that performance measures vary independently with respect to the different parameters,  for gene expression SEMS with low number of unobserved factors $h_k$, high level of gene expression variables $x_j$ (around 20 to 50 times), high sparsity, a low variance ratio $\frac{\sigma_{c_l}}{\sigma_y}$ and low complexity for the signals $f_{x_j}$ and $f_{h_k}$ SVA has superior performance to the other methods considered.

\section{Future Work}
\label{ss:FutureWork}

The following is a list of possible ways to extend or improve the work presented here:

\begin{itemize}
	\item Define the class of additive gene expression SEMs in terms of the more general case where $x_j$ can depend on $x_k$ for $j \neq k$. Also add this feature to simulation experiments,
	\item Explore the possibility of using Independent Component Analysis (ICA) to factorize and obtain independent and not only uncorrelated $c_l$ variables,
	\item Understand why the model $x_j = \beta_0 + \beta_1 c_l + \epsilon$ instead of the model $x_j = f_j(y) + \beta_1 c_l + N_j$ is used to find the variables $x_j$ for $j \in \{1,...,J\}$ with the signature of $c_l$,
	\item Understand why local false discovery rates instead of q-values are used to find the variables $x_j$ for $j \in \{1,...,J\}$ with the signature of $c_l$,
	\item Perform simulations from non additive genetic expression sems and see how good the approximation is to test the approximation of real life gene expression data generating mechanisms with additive gene expression SEMs,
	\item In simulation experiments include noise for $h_k$ equations and allow $h_k$ nodes to be affected by more than one $c_l$ variable,
	\item Investigate if the error shown in low-dim experiment is mainly due to overfitting of $f_j(y)$, 
	\item Perform simulations with default conditions closer to those of the simulation of experiments in \cite{SVA} where the functions $f_{x_j}(y) and f_{h_k}(y)$ are simple step functions of the form $g(y)=\mathbbm{1}_{y \leq a} + \mathbbm{1}_{y > a}$ and the simulated SEMs are sparser. Perform sensitivity analysis with these default conditions as a starting point,
	\item Provide performance measures for different categories of $x_j$ variables separately: those that depend on $y$ only, on the span of $h_k$ only or on both, 
	\item Separate sensibility analysis of $f_{x_j}$ complexity from $f_{h_k}$ complexity, and
	\item Perform a multivariate sensitivity analysis based upon different scenarios for the additive gene expression SEM parameters. 
\end{itemize}


\addtocontents{toc}{\vspace{.5\baselineskip}}
\cleardoublepage
\phantomsection
\addcontentsline{toc}{chapter}{\protect\numberline{}{Bibliography}}
\bibliography{myReferences}


\begin{appendices}
\chapter{Proofs}
\label{app:proofs}

\section{Proof of Proposition \ref{propertiesAddSEM}} \label{proofPropertiesAddSEM}

\begin{proof}
	\hspace{1mm}
	\begin{enumerate}[1.]
		\item All paths from $y$ to $c_l$ or from $c_l$ to $c_m$ must go through at least one of four types of v-structures: 
			\begin{itemize}
				\item $y \rightarrow h_k \leftarrow c_s$  
				\item $y \rightarrow x_j \leftarrow h_k$,
				\item $h_k \rightarrow x_j \leftarrow h_r$ or 
				\item $c_s \rightarrow h_k \leftarrow c_t$.
			\end{itemize}
				Where  $j \in \{1,...,J\}$, $k,r \in \{1,...,K\}$ and $s,t \in \{1,...,L\}$. This means that $y$ is d-separated from $c_l$ by the empty set and $c_l$ is d-separated from $c_m$ by the empty set. Since the DAG is induced by a SEM, by Proposition \ref{SEMarkov} $\pr^{\mathbb{X} \cup C}$ is Markov with respect to $\mathcal{G}$ which means that $y \indep c_l$ and $c_l \indep c_m$. 
 
		\item Suppose $\pr^{\mathbb{X} \cup C}$ does not satisfy causal minimiality with respect to $\mathcal{G}$. Then by Proposition \ref{minimCond} we have that there must exist a variable $w \in \{x_1,...,x_J,h_1,...,h_K\}$ with a parent $z \in PA_w^\mathcal{G} \subset \{y,h_1,...,h_K,c_1,...,c_L\}$ such that $w \indep z | PA_w^\mathcal{G} \setminus z$. 
		\begin{enumerate}[a.]
			\item Suppose $w=h_k$. This means that $z \in PA_w^\mathcal{G} \subseteq \{y,c_1,...,c_L\}$.
				\begin{enumerate}[i.]
					\item Suppose $z=y$. This means that $h_k \indep y | PA_{h_k}^\mathcal{G} \setminus y$, but given $PA_{h_k}^\mathcal{G} \setminus y \subseteq \{c_1,...,c_L\}$ we have that $h_k = f_{h_k}(y) + \sum_{l=1}^L \gamma_{lk}c_l = f_{h_k}(y) + b $ where $b$ is a constant. Since $y \in PA_{h_k}^\mathcal{G}$ we have that $f_{h_k}(y)$ is not constant. This leads to the contradictory conclusion that  $h_k \nindep y | PA_{h_k}^\mathcal{G} \setminus y$.
					\item Suppose $z=c_l$. This means that $h_k \indep c_l | PA_{h_k}^\mathcal{G} \setminus c_l$, but given $PA_{h_k}^\mathcal{G} \setminus c_l \subseteq \{y,c_1,...,c_{l-1},c_{l+1},...,c_L\}$ we have that $h_k = f_{h_k}(y) + \sum_{l=1}^L \gamma_{lk}c_l = \gamma_{lk}c_l + b $ where $b$ is a constant. Since $c_l \in PA_{h_k}^\mathcal{G} $ we have that $ \gamma_{lk} \neq 0$. This leads to the contradictory conclusion that  $h_k \nindep c_l | PA_{h_k}^\mathcal{G} \setminus c_l$.
				\end{enumerate}	
			\item Suppose $w=x_j$. This means that $z \in PA_w^\mathcal{G} \subseteq \{y,h_1,...,h_K\}$.
			\begin{enumerate}[i.]
				\item Suppose $z=y$. This means that $x_j \indep y | PA_{x_j}^\mathcal{G} \setminus y$, but given $PA_{x_j}^\mathcal{G} \setminus y \subseteq \{h_1,...,h_K\}$ we have that $x_j = f_{x_j}(y) + \sum_{k=1}^K \beta_{kj}h_k = f_{x_j}(y) + b $ where $b$ is a constant. Since $y \in PA_{x_j}^\mathcal{G}$ we have that $f_{x_j}(y)$ is not constant. This leads to the contradictory conclusion that  $x_j \nindep y | PA_{x_j}^\mathcal{G} \setminus y$.
				\item Suppose $z=h_k$. This means that $x_j \indep h_k | PA_{x_j}^\mathcal{G} \setminus h_k$, but given $PA_{x_j}^\mathcal{G} \setminus h_k \subseteq \{y,h_1,...,h_{k-1},h_{k+1},...,h_K\}$ we have that $x_j = f_{x_j}(y) + \sum_{k=1}^K \beta_{kj}h_k = \beta_{kj}h_k + b $ where $b$ is a constant. Since $h_k \in PA_{x_j}^\mathcal{G}$ we have that $\beta_{kj} \neq 0$. This leads to the contradictory conclusion that  $x_j \nindep y | PA_{x_j}^\mathcal{G} \setminus h_k$.
			\end{enumerate}
			
			We conclude that $\pr^{\mathbb{X} \cup C}$ satisfies causal minimality. 
		\end{enumerate}	
		
		\item We look at the SEM $\widetilde{\mathcal{S}}$ that results from replacing the equations $S_{h_k}$ of Definition \ref{adGESEM} with $\widetilde{S}_{h_k}=\widetilde{N}_{h_k}$ where $\widetilde{N}_{h_k} \sim \mathcal{N}(0,1)$, say. This means that for $j$ such that $\beta_{kj} \neq 0$ we have:
		\begin{align}
			x_j = f_{x_j}(y) + \sum_{r=1}^K \beta_{rj} h_r + N_{x_j} = f_{x_j}(N_y) + \sum_{r \neq k} \beta_{rj} N_{h_r} + N_{x_j} + \beta_{kj}\widetilde{N}_{h_k}
		\end{align}	
		
		Since $\beta_{kj} \neq 0$ and the random noise variables are mutually independent it is clear that $x_j \nindep h_k$ in $\pr_{\widetilde{\mathcal{S}}}^{\mathbb{X} \cup C}$.
		 
		\item Since $y$ has no parents, intervening on $y$ corresponds to changing $N_y$ with some $\widetilde{N}_y$ and we have:
		
		\begin{align}
			x_j &= f_{x_j}(y) + \sum_{k=1}^K \beta_{kj} h_k + N_{x_j} \\
			&= f_{x_j}(y) + \sum_{k=1}^K \beta_{kj}(f_{h_k}(y)+ \sum_{l=1}^L \gamma_{lk}c_l+N_{h_k}) + N_{x_j}\\
			&= \bigg( f_{x_j}(y) + \sum_{k=1}^K \beta_{kj}f_{h_k}(y) \bigg) + \bigg( \sum_{k=1}^K \sum_{l=1}^L \beta_{kj}\gamma_{lk}c_l+ \sum_{k=1}^K \beta_{kj} N_{h_k} \bigg) + N_{x_j} \\
			&= f_j(y) + \bigg( \sum_{k=1}^K \sum_{l=1}^L \beta_{kj}\gamma_{lk}c_l+ \sum_{k=1}^K \beta_{kj} N_{h_k} \bigg) + N_{x_j} \\
			&= f_j(\widetilde{N}_y) + \bigg( \sum_{k=1}^K \sum_{l=1}^L \beta_{kj}\gamma_{lk}N_{c_l}+ \sum_{k=1}^K \beta_{kj} N_{h_k} \bigg) + N_{x_j} 
		\end{align}
		
		Where $f_j(y):= f_{x_j}(y) + \sum_{k=1}^K \beta_{kj}f_{h_k}(y)$. Since $f_j(y)$ is assumed not to be constant and the random noise variables are mutually independent it is clear that $x_j \nindep y$ in $\pr_{\widetilde{\mathcal{S}}}^{\mathbb{X} \cup C}$.
		
		
		
		
		
	\end{enumerate}	
	
	
\end{proof}

\chapter{\Rp Code}
\label{app:rcode}

We include the code in \Rp that was used to produce all results in this work.  

\section{Functions}
\lstinputlisting{functions.R}
\section{Script}
\begin{lstlisting}
########################################################################
# SCRIPT
########################################################################

library(reshape)
library(ggplot2)
library(sva)
library(abind) #bind arrays

source("./functions.R")

########################################################################
# LOW DIMENSIONAL SEM
########################################################################

# Simulate the SEM from class of SEMS
seed <- 4
K <- 4
J <- 4
deg.max <- 2
p0_fh <- 0.5
p0_fx <- 0.5
p_isox <- 0
p0_beta <- 0
set.seed(seed)
beta <- matrix(rnorm(k*m),k,m)
mat.indx <- matrix(c(2,3,3,3,4,4,4,1,2,4,2,4),6,2)
beta[mat.indx] <- 0
sem.lin <- sim.sem.sparse(K,J, deg.max,p0_fh, p0_fx, p_isox, p0_beta, beta)

# Simulate the repetitions of the hole SEM which is now fixed
n <- 100
sigma <- 1
y.c.signal <- 1
set.seed(seed)
sem.sim <- sim.lin(sem=sem.lin, n, sigma, y.c.signal)

# estimate surrogate variables
ord <- deg.max
n.sv.r <- max(2,num.sv(dat=t(sem.sim$xs), mod))
n.sv.x <- min(min(n,m),2*k)  #max(3, fa.parallel(t(sem.sim$xs), fa="pc", n.iter=20)$ncomp)
beta <- NULL
mod <- model.matrix(~poly(sem.sim$y, ord, raw=TRUE))
sva.lin   <- sva.mod(dat=t(sem.sim$xs), y=sem.sim$y, mod, n.sv.r=n.sv.r, n.sv.x=n.sv.x, beta=beta)
sva.lin0 <- sva.mod(dat=t(sem.sim$xs0), y=sem.sim$y, mod, n.sv.r=n.sv.r, n.sv.x=n.sv.x, beta=beta)

# Visualize different models
models <- get.Models(sem.sim, sva.lin, ord)
graph.Models(sem.sim, models, select=c("real","sva","van"), 1)
graph.Models(sem.sim, models, select=c("real","sva","van"), 2)
graph.Models(sem.sim, models, select=c("real","sva","van"), 3)
graph.Models(sem.sim, models, select=c("real","sva","van"), 4)


# Calculate some performance measures for our simulated sem
(msrs.sem <- measures.sem(sem.sim, sva.lin=sva.lin, sva.lin0=sva.lin0, mod, ord))

# Perform some repetitions

reps <- 1000
K <- 4
J <- 4
deg.max <- 1
p0_fh <- 0.5
p0_fx <- 0.5
p_isox <- 0.25
p0_beta <- 0.5
n <- 100
sigma <- 1
y.c.signal <- 1
ord <- deg.max


set.seed(1)
pm <- proc.time()		
mrs.sem.reps <- do.n.reps(reps, K,J, deg.max, n, sigma, y.c.signal,  ord, sem=sem.lin ,cheat=FALSE, sparse=TRUE, p0_fh=p0_fh, p0_fx=p0_fx, p_isox=p_isox, p0_beta=p0_beta, trace=TRUE)
proc.time() - pm

#217 seconds for reps = 10,000
#414 seconds for reps = 20,000

mrs.sem.reps$count.no.svs
hist(mrs.sem.reps$ext.rate, main="c-node extraction rate")
hist(mrs.sem.reps$cancor.cs.sva, main="c-node overlap SVA")
boxPlot(mrs.sem.reps$cancor.hs, "h-node overlap")
boxPlot(mrs.sem.reps$r2yh, "real r2 - est. r2 (y ~ hs)")
boxPlot(mrs.sem.reps$err.fx, "mean absolute error")
mrs.sem.reps$ks.nested

########################################################################
# HIGH DIMENSIONAL SEM
########################################################################

# Simulate the SEM from class of SEMS
seed <- 4
K <- 10
J <- 1000
deg.max <- 2
p0_fh <- 0.5
p0_fx <- 0.5
p_isox <- 0.25
p0_beta <- 0.5
set.seed(seed)
sem.lin <- sim.sem.sparse(K,J, deg.max,p0_fh, p0_fx, p_isox, p0_beta)

# Simulate the repetitions of the hole SEM which is now fixed
n <- 100
sigma <- 1
y.c.signal <- 1
set.seed(seed)
sem.sim <- sim.lin(sem=sem.lin, n, sigma, y.c.signal)

# estimate surrogate variables
ord <- deg.max
mod <- model.matrix(~poly(sem.sim$y, ord, raw=TRUE))

n.sv.r <- max(2,num.sv(dat=t(sem.sim$xs), mod))
n.sv.x <- n.sv.r  #max(3, fa.parallel(t(sem.sim$xs), fa="pc", n.iter=20)$ncomp)
beta <- NULL

sva.lin   <- sva.mod(dat=t(sem.sim$xs), y=sem.sim$y, mod, n.sv.r=n.sv.r, n.sv.x=n.sv.x, beta=beta)
sva.lin0 <- sva.mod(dat=t(sem.sim$xs0), y=sem.sim$y, mod, n.sv.r=n.sv.r, n.sv.x=n.sv.x, beta=beta)

# Visualize different models
models <- get.Models(sem.sim, sva.lin, ord)
graph.Models(sem.sim, models, select=c("real","sva","van"), 1)
graph.Models(sem.sim, models, select=c("real","sva","van"), 2544)
graph.Models(sem.sim, models, select=c("real","sva","van"), 6664)
graph.Models(sem.sim, models, select=c("real","sva","van"), 8532)


# Calculate some performance measures for our simulated sem
(msrs.sem <- measures.sem(sem.sim, sva.lin=sva.lin, sva.lin0=sva.lin0, mod, ord))
hist(msrs.sem$pval0[,"sva"])
hist(msrs.sem$pval0[,"svdr"])
hist(msrs.sem$pval0[,"van"])
ks.test(msrs.sem$pval0[,"sva"], "punif",alternative = "two.sided")
ks.test(msrs.sem$pval0[,"svdr"], "punif",alternative = "two.sided")
ks.test(msrs.sem$pval0[,"van"], "punif",alternative = "two.sided")

# Perform 100 repetitions

reps <- 100
K <- 10
J <- 1000
deg.max <- 1
p0_fh <- 0.5
p0_fx <- 0.5
p_isox <- 0.25
p0_beta <- 0.5
n <- 100
sigma <- 1
y.c.signal <- 1
ord <- deg.max


set.seed(2)
pm <- proc.time()		
mrs.sem.reps <- do.n.reps(reps, K,J, deg.max, n, sigma, y.c.signal,  ord, sem=NULL ,cheat=FALSE, sparse=TRUE, p0_fh=p0_fh, p0_fx=p0_fx, p_isox=p_isox, p0_beta=p0_beta, trace=TRUE)
proc.time() - pm

#40 minutes for reps = 20, K=100, J=10,000
#6.7 minutes for reps = 20, K=50, J=2,500


mrs.sem.reps$count.no.svs
hist(mrs.sem.reps$ext.rate, main="c-node extraction rate")
hist(mrs.sem.reps$cancor.cs.sva, main="c-node overlap, SVA")
boxPlot(mrs.sem.reps$cancor.hs, "h-node overlap")
boxPlot(mrs.sem.reps$r2yh, "real r2 - est. r2 (y ~ hs)")
boxPlot(mrs.sem.reps$err.fx, "mean absolute error")
mrs.sem.reps$ks.nested


########################################################################
# SENSIBILITY
########################################################################

#############################
# size of model (just J)
#############################


reps <- 100
K <- 10
J <- round(seq(100, 1000, length.out=10))
deg.max <- 1
n <- 100
sigma <- 1
y.c.signal <- 1
ord <- deg.max
p0_fh <- 0.5
p0_fx <- 0.5
p_isox <- 0.25
p0_beta <- 0.5
params <- expand.grid(K=K,J=J, deg.max=deg.max, n=n, sigma=sigma, y.c.signal=y.c.signal, ord=ord, p0_fh=p0_fh, p0_fx=p0_fx, p_isox=p_isox, p0_beta=p0_beta)


set.seed(2)
pm <- proc.time()		
sens.sem.km <- sens.model(reps, params,cheat=FALSE, sparse=TRUE, trace=FALSE)
proc.time() - pm
# 47 mins

labs <- as.character(J)
xaxis <- "J"
yaxis <- ""
boxPlot2(sens.sem.km$ext.rate, labs, xaxis, "",title="c-node extraction rate")
boxPlot2(sens.sem.km$cancor.cs.sva, labs,xaxis,"", title="c-node overlap")
boxPlot3(sens.sem.km$cancor.hs, labs, xaxis, "method" , title="h-node overlap")
boxPlot3(sens.sem.km$r2yh, labs, xaxis, "method" , title="real r2 - est. r2 (y ~ hs)")
boxPlot3(sens.sem.km$err.fx, labs, xaxis, "method" , title="mean absolute error")
ks.plot(sens.sem.km$ks.nested, labs, xaxis, yaxis)

#############################
# size of model (K and J)
#############################

reps <- 100
K <- round(seq(4,300, length.out=10))
J <- round(K/0.1)
deg.max <- 1
n <- 100
sigma <- 1
y.c.signal <- 1
ord <- deg.max
p0_fh <- 0.5
p0_fx <- 0.5
p_isox <- 0.25
p0_beta <- 0.5
params <- expand.grid(K=K,deg.max=deg.max, n=n, sigma=sigma, y.c.signal=y.c.signal, ord=ord, p0_fh=p0_fh, p0_fx=p0_fx, p_isox=p_isox, p0_beta=p0_beta)
params$J <- round(params$K/0.1)

set.seed(2)
pm <- proc.time()					   
sens.sem.km <- sens.model(reps, params,cheat=FALSE, sparse=TRUE, trace=FALSE)
proc.time() - pm
#178 mins

labs <- paste(K,J, sep="-")
xaxis <- "K-J"
yaxis <- ""
boxPlot2(sens.sem.km$ext.rate, labs, xaxis, "",title="c-node extraction rate")
boxPlot2(sens.sem.km$cancor.cs.sva, labs,xaxis,"", title="c-node overlap")
boxPlot3(sens.sem.km$cancor.hs, labs, xaxis, "method" , title="h-node overlap")
boxPlot3(sens.sem.km$r2yh, labs, xaxis, "method" , title="real r2 - est. r2 (y ~ hs)")
boxPlot3(sens.sem.km$err.fx, labs, xaxis, "method" , title="mean absolute error")
ks.plot(sens.sem.km$ks.nested, labs, xaxis, yaxis)


#############################
# sparsity of model p0_h
#############################

reps <- 100
K <- 10
J <- 1000
deg.max <- 1
n <- 100
sigma <- 1
y.c.signal <- 1
ord <- deg.max
p0_fh <- seq(0,1,length.out=10)
p0_fx <- 0.5
p_isox <- 0.25
p0_beta <- 0.5
params <- expand.grid(K=K,J=J, deg.max=deg.max, n=n, sigma=sigma, y.c.signal=y.c.signal, ord=ord, p0_fh=p0_fh, p0_fx=p0_fx, p_isox=p_isox, p0_beta=p0_beta)

set.seed(2)
pm <- proc.time()		
sens.sem.sp_fh <- sens.model(reps, params,cheat=FALSE, sparse=TRUE, trace=TRUE)
proc.time() - pm
#78 mins

labs <- as.character(round(p0_fh,2))
xaxis <- "% fh(y)=0"
yaxis <- ""
boxPlot2(sens.sem.sp_fh$ext.rate, labs, xaxis, "",title="c-node extraction rate")
boxPlot2(sens.sem.sp_fh$cancor.cs.sva, labs,xaxis,"", title="c-node overlap")
boxPlot3(sens.sem.sp_fh$cancor.hs, labs, xaxis, "method" , title="h-node overlap")
boxPlot3(sens.sem.sp_fh$r2yh, labs, xaxis, "method" , title="real r2 - est. r2 (y ~ hs)")
boxPlot3(sens.sem.sp_fh$err.fx, labs, xaxis, "method" , title="mean absolute error")
ks.plot(sens.sem.sp_fh$ks.nested, labs, xaxis, yaxis)

#############################
# sparsity of model p0_x
#############################

reps <- 100
K <- 10
J <- 1000
deg.max <- 1
n <- 100
sigma <- 1
y.c.signal <- 1
ord <- deg.max
p0_fh <- 0.5
p0_fx <- seq(0.25,1,length.out=10)
p_isox <- 0.25
p0_beta <- 0.5
params <- expand.grid(K=K,J=J, deg.max=deg.max, n=n, sigma=sigma, y.c.signal=y.c.signal, ord=ord, p0_fh=p0_fh, p0_fx=p0_fx, p_isox=p_isox, p0_beta=p0_beta)

set.seed(2)
pm <- proc.time()		
sens.sem.sp_fx <- sens.model(reps, params,cheat=FALSE, sparse=TRUE, trace=TRUE)
proc.time() - pm
#100 reps, 78 mins

labs <- as.character(round(p0_fx,2))
xaxis <- "% fx(y)=0"
yaxis <- ""
boxPlot2(sens.sem.sp_fx$ext.rate, labs, xaxis, "",title="c-node extraction rate")
boxPlot2(sens.sem.sp_fx$cancor.cs.sva, labs,xaxis,"", title="c-node overlap")
boxPlot3(sens.sem.sp_fx$cancor.hs, labs, xaxis, "method" , title="h-node overlap")
boxPlot3(sens.sem.sp_fx$r2yh, labs, xaxis, "method" , title="real r2 - est. r2 (y ~ hs)")
boxPlot3(sens.sem.sp_fx$err.fx, labs, xaxis, "method" , title="mean absolute error")
ks.plot(sens.sem.sp_fx$ks.nested, labs, xaxis, yaxis)

#############################
# sparsity of model p0_beta
#############################

reps <- 100
K <- 10
J <- 1000
deg.max <- 1
n <- 100
sigma <- 1
y.c.signal <- 1
ord <- deg.max
p0_fh <- 0.5
p0_fx <- 0.5
p_isox <- 0.25
p0_beta <- seq(0,0.99,length.out=10)
params <- expand.grid(K=K,J=J, deg.max=deg.max, n=n, sigma=sigma, y.c.signal=y.c.signal, ord=ord, p0_fh=p0_fh, p0_fx=p0_fx, p_isox=p_isox, p0_beta=p0_beta)

set.seed(2)
pm <- proc.time()		
sens.sem.sp_beta <- sens.model(reps, params,cheat=FALSE, sparse=TRUE, trace=FALSE)
proc.time() - pm
#75 mins

labs <- as.character(round(p0_beta,2))
xaxis <- "% beta=0"
yaxis <- ""
boxPlot2(mat=sens.sem.sp_beta$ext.rate, labs, xaxis, yaxis="",title="c-node extraction rate")
boxPlot2(sens.sem.sp_beta$cancor.cs.sva, labs,xaxis,"", title="c-node overlap")
boxPlot3(sens.sem.sp_beta$cancor.hs, labs, xaxis, "method" , title="h-node overlap")
boxPlot3(sens.sem.sp_beta$r2yh, labs, xaxis, "method" , title="real r2 - est. r2 (y ~ hs)")
boxPlot3(sens.sem.sp_beta$err.fx, labs, xaxis, "method" , title="mean absolute error")
ks.plot(sens.sem.sp_beta$ks.nested, labs, xaxis, yaxis)


#############################
# sparsity of model p0_isox
#############################

reps <- 100
K <- 10
J <- 1000
deg.max <- 1
n <- 100
sigma <- 1
y.c.signal <- 1
ord <- deg.max
p0_fh <- 0.5
p0_fx <-  seq(0,0.99,length.out=10)
p_isox <- p0_fx^2 ; plot(p0_fx, p_isox)
p0_beta <- 0.5
params <- expand.grid(K=K,J=J, deg.max=deg.max, n=n, sigma=sigma, y.c.signal=y.c.signal, ord=ord, p0_fh=p0_fh, p0_fx=p0_fx, p0_beta=p0_beta)
params$p_isox <- params$p0_fx^2

set.seed(2)
pm <- proc.time()		
sens.sem.sp_isox <- sens.model(reps, params,cheat=FALSE, sparse=TRUE, trace=FALSE)
proc.time() - pm
#103 mins

labs <- paste(round(p0_fx,2),round(p_isox,2),sep="-")
xaxis <- "% fx(y)=0-% p_isox=0"
yaxis <- ""
boxPlot2(sens.sem.sp_isox$ext.rate, labs, xaxis, "",title="c-node extraction rate")
boxPlot2(sens.sem.sp_isox$cancor.cs.sva, labs,xaxis,"", title="c-node overlap")
boxPlot3(sens.sem.sp_isox$cancor.hs, labs, xaxis, "method" , title="h-node overlap")
boxPlot3(sens.sem.sp_isox$r2yh, labs, xaxis, "method" , title="real r2 - est. r2 (y ~ hs)")
boxPlot3(sens.sem.sp_isox$err.fx, labs, xaxis, "method" , title="mean absoulute error")
ks.plot(sens.sem.sp_isox$ks.nested, labs, xaxis, yaxis)


#############################
# y.c.signal
#############################

reps <- 100
K <- 10
J <- 1000
deg.max <- 1
n <- 100
sigma <- 1
y.c.signal <- c(1/10, 1/5, 1/3, 1/2, 1, 2)
ord <- deg.max
p0_fh <- 0.5
p0_fx <- 0.5
p_isox <- 0.25
p0_beta <- 0.5
params <- expand.grid(K=K,J=J, deg.max=deg.max, n=n, sigma=sigma, y.c.signal=y.c.signal, ord=ord, p0_fh=p0_fh, p0_fx=p0_fx, p_isox=p_isox, p0_beta=p0_beta)

set.seed(2)
pm <- proc.time()		
sens.sem.y.c.signal <- sens.model(reps, params,cheat=FALSE, sparse=TRUE, trace=TRUE)
proc.time() - pm
#69 mins

labs <- as.character(round(y.c.signal,2))
xaxis <- "y.c.signal"
yaxis <- ""
boxPlot2(sens.sem.y.c.signal$ext.rate, labs, xaxis, "",title="c-node extraction rate")
boxPlot2(sens.sem.y.c.signal$cancor.cs.sva, labs,xaxis,"", title="c-node overlap")
boxPlot3(sens.sem.y.c.signal$cancor.hs, labs, xaxis, "method" , title="h-node overlap")
boxPlot3(sens.sem.y.c.signal$r2yh, labs, xaxis, "method" , title="real r2 - est. r2 (y ~ hs)")
boxPlot3(sens.sem.y.c.signal$err.fx, labs, xaxis, "method" , title="mean absoulute error")
ks.plot(sens.sem.y.c.signal$ks.nested, labs, xaxis, yaxis)

#############################
# nx noise
#############################

reps <- 100
K <- 10
J <- 1000
deg.max <- 1
n <- 100
sigma <- c(1/10, 1/5, 1/3, 1/2, 1, 2, 5, 10)
y.c.signal <- 1
ord <- deg.max
p0_fh <- 0.5
p0_fx <- 0.5
p_isox <- 0.25
p0_beta <- 0.5
params <- expand.grid(K=K,J=J, deg.max=deg.max, n=n, sigma=sigma, y.c.signal=y.c.signal, ord=ord, p0_fh=p0_fh, p0_fx=p0_fx, p_isox=p_isox, p0_beta=p0_beta)

set.seed(2)
pm <- proc.time()		
sens.sem.sigma <- sens.model(reps, params,cheat=FALSE, sparse=TRUE, trace=TRUE)
proc.time() - pm
#91 mins

labs <- as.character(round(sigma,2))
xaxis <- "x-node noise"
yaxis <- ""
boxPlot2(sens.sem.sigma$ext.rate, labs, xaxis, "",title="c-node extraction rate")
boxPlot2(sens.sem.sigma$cancor.cs.sva, labs,xaxis,"", title="c-node overlap")
boxPlot3(sens.sem.sigma$cancor.hs, labs, xaxis, "method" , title="h-node overlap")
boxPlot3(sens.sem.sigma$r2yh, labs, xaxis, "method" , title="real r2 - est. r2 (y ~ hs)")
boxPlot3(sens.sem.sigma$err.fx, labs, xaxis, "method" , title="mean absoulute error")
ks.plot(sens.sem.sigma$ks.nested, labs, xaxis, yaxis)

##################################
# degree of functions fx and fh
#################################
reps <- 100
K <- 10
J <- 1000
deg.max <- seq(5)
n <- 100
sigma <- 1
y.c.signal <- 1
ord <- deg.max
p0_fh <- 0.5
p0_fx <- 0.5
p_isox <- 0.25
p0_beta <- 0.5
params <- expand.grid(K=K,J=J, deg.max=deg.max, n=n, sigma=sigma, y.c.signal=y.c.signal, p0_fh=p0_fh, p0_fx=p0_fx, p_isox=p_isox, p0_beta=p0_beta)
params$ord <- params$deg.max

set.seed(2)
pm <- proc.time()		
sens.sem.deg <- sens.model(reps, params,cheat=FALSE, sparse=TRUE, trace=TRUE)
proc.time() - pm
#50 mins

labs <- as.character(round(deg.max,2))
xaxis <- "fx and fh order"
yaxis <- ""
boxPlot2(sens.sem.deg$ext.rate, labs, xaxis, "",title="c-node extraction rate")
boxPlot2(sens.sem.deg$cancor.cs.sva, labs,xaxis,"", title="c-node overlap")
boxPlot3(sens.sem.deg$cancor.hs, labs, xaxis, "method" , title="h-node overlap")
boxPlot3(sens.sem.deg$r2yh, labs, xaxis, "method" , title="real r2 - est. r2 (y ~ hs)")
boxPlot3(sens.sem.deg$err.fx, labs, xaxis, "method" , title="mean absoulute error")
ks.plot(sens.sem.deg$ks.nested, labs, xaxis, yaxis)

#############################
# number of points n
#############################

reps <- 100
K <- 10
J <- 1000
deg.max <- 1
n <- round(seq(25,1000, length.out=10),0)
sigma <- 1
y.c.signal <- 1
ord <- deg.max
p0_fh <- 0.5
p0_fx <- 0.5
p_isox <- 0.25
p0_beta <- 0.5
params <- expand.grid(K=K,J=J, deg.max=deg.max, n=n, sigma=sigma, y.c.signal=y.c.signal, ord=ord, p0_fh=p0_fh, p0_fx=p0_fx, p_isox=p_isox, p0_beta=p0_beta)

set.seed(2)
pm <- proc.time()		
sens.sem.n <- sens.model(reps, params,cheat=FALSE, sparse=TRUE, trace=TRUE)
proc.time() - pm
#31 hours!!!

labs <- as.character(round(n,2))
xaxis <- "sample size"
yaxis <- ""
boxPlot2(sens.sem.n$ext.rate, labs, xaxis, "",title="c-node extraction rate")
boxPlot2(sens.sem.n$cancor.cs.sva, labs,xaxis,"", title="c-node overlap")
boxPlot3(sens.sem.n$cancor.hs, labs, xaxis, "method" , title="h-node overlap")
boxPlot3(sens.sem.n$r2yh, labs, xaxis, "method" , title="real r2 - est. r2 (y ~ hs)")
boxPlot3(sens.sem.n$err.fx, labs, xaxis, "method" , title="mean absoulute error")
ks.plot(sens.sem.n$ks.nested, labs, xaxis, yaxis)
\end{lstlisting}

\end{appendices}


\end{document}